# First-principles calculation of the stabilities of LMP/LAMP lithium superionic conductors against sodium-ion exchange in seawater

**James R. Rustad**

**Corning Incorporated, Corning, NY 14830**

**(Oct 31, 2014)**

**Abstract-**Electronic structure calculations carried out to estimate the free energies of $Na(aq)^+$ exchange for lithium in $LiM_2(PO_4)_3$ (LMP, M=$Si^{4+}$,$Ge^{4+}$,$Ti^{4+}$,$Sn^{4+}$,$Zr^{4+}$) and $Li_{1+x}Al_xM_{2-x}(PO_4)_3$ (LAMP; M=$Ge^{4+}$,$Ti^{4+}$,$0 \le x \le 0.5$) compounds in seawater. The calculations show that resistance to sodium-ion exchange increases with decreasing cell volume. For the pure LMP compounds, only the hypothetical $LiSi_2(PO_4)_3$ is predicted to be stable against sodium ion exchange in aqueous solution. The calculations indicate that increasing the extent of $Al^{3+}$ substitution for $M^{4+}$ in the LAMP compounds increases the resistance to exchange, and that both LAGP and LATP can be stabilized against sodium exchange for $x \ge$ ~0.5 $Li_{1+x}Al_xM_{2-x}(PO_4)_3$.

## Introduction

Compounds derived from $LiM_2(PO_4)_3$ (LMP) with a NASICON-type structure have been proposed for use as solid electrolyte membranes in lithium-based batteries[1]. Such compounds include aluminum-substituted LAMP materials, including LATP $Li_{1+x}Al_xTi_{2-x}(PO_4)_3$ and LAGP $Li_{1+x}Al_xGe_{2-x}(PO_4)_3$[2] . One important application area for LAMP is in seawater-environment sensor technology where seawater serves as the cathode. The seawater cathode is separated from the lithium metal anode with a protective membrane, part of which consists of the LAMP solid electrolyte[3]. When used in a seawater environment, lithium-based solid electrolytes can be susceptible to sodium ion exchange with seawater according to the reaction:

$$L(A)MP + Na^+(aq) \leftrightarrow N(A)MP + Li^+(aq) \quad (1)$$

where N(A)MP refers to the sodium equivalent of L(A)MP. This report focuses on the role of the $M^{4+}$ ion composition, as well as the effect of aluminum substitution, in determining the stability of the L(A)MP material in seawater.

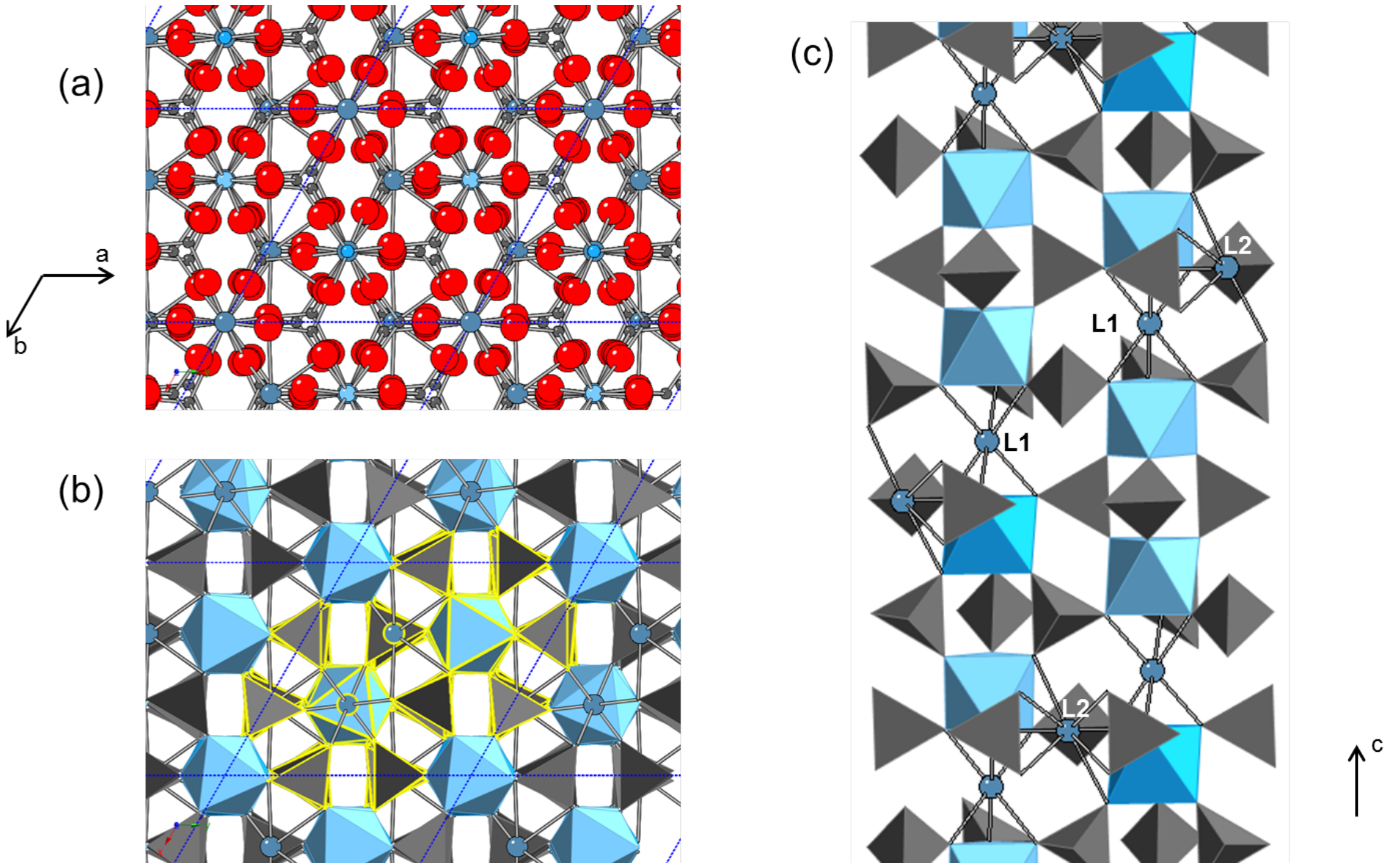

**Figure 1.** Structure of LAMP (a) Ball-and-stick representation viewed down the "c" axis. Red atoms are oxygen, grey atoms are phosphorous, light blue atoms are $M^{4+}/Al^{3+}$, and dark blue atoms are lithium. (b) Polyhedral representation. Grey tetrahedra are $PO_4$, light blue octahedra are M/AlO6, sphere+stick bonds are lithium atoms. Highlighted region is cut out and rotated 90 degrees in (c).

The structure of LMP is given in Figure 1. It consists of columns of Li-$M(PO_4)_3$M-Li-$M(PO_4)_3$M-… repeat units parallel to *z*, in the hexagonal cell convention, or along the central trigonal axis in the rhombohedral cell convention. The columns are hexagonally packed in the *x*-*y* plane and each column shares $PO_4$ tetrahedra with its neighbors. All lithium ions (L1) are six-fold coordinated. Three of the six oxygen atoms are located on the top face of the MO6 octahedron below the lithium in L1, and the other three coordinating oxygen atoms are located the bottom face of the MO6 octahedron above the lithium in L1. The LAMP compounds are derived by the coupled substitution $M^{4+} = Al^{3+} + Li^{+}$. In these compounds, aluminum occupies the MO6 octahedron and the extra lithium atom occupies the "L2" site in the empty space in near the center of the vertical stacks of phosphate tetrahedra that are surrounded by three columns of M/AlO6 octahedra. The key bond lengths in the LMP structure are given in Figure 2.

There are several crystal-chemical issues relevant to the stability of LMP compounds to sodium ion exchange. Most obvious are the relative sizes of the $Li^{+}$ and $Na^{+}$ in the six-fold coordinated environment of the alkali site in the NASICON structure. The crystal ionic radius of $Li^{+}$ is about 90 pm, while $Na^{+}$ is 116 pm, giving a typical Li-O bond length of ~2.16 Å for $Li^{+}$ versus ~2.42 Å for $Na^{+}$. The coordination environment of lithium in L1 (as shown in Figure 1) is basically a trigonal prism. This is an unusual coordination state for lithium which is, for example, predominantly tetrahedrally coordinated as $Li(H_2O)_4^{+}$ in aqueous solution. Sodium is generally six-fold coordinated as $Na(H_2O)_6^{+}$ in aqueous solution, much more akin to the L1 site in LMP. Thus it is not surprising that LMP is susceptible to sodium exchange when used in seawater

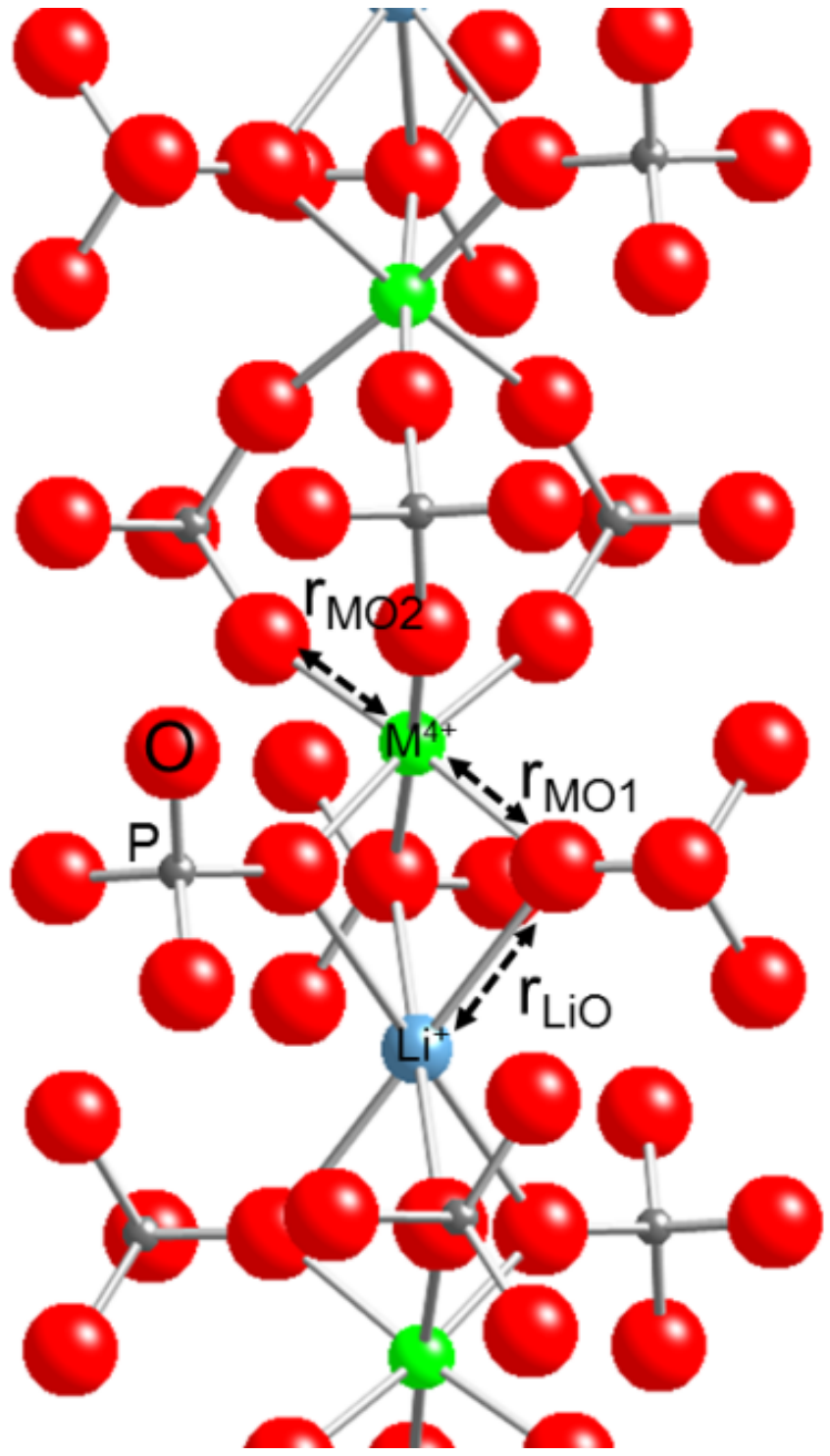

**Figure 2.** Key bond lengths in the LMP structure.

The question addressed here is what crystal-chemical factors might help stabilize Li in the L1 environment and make the LMP more resistant to sodium exchange.

## Methods

Density functional electronic structure calculations were carried out on a series of LMP/NMP compounds with systematically increasing $M^{4+}$ ionic radius Li/NaSi$_2$(PO$_4$)$_3$, Li/NaGe$_2$(PO$_4$)$_3$, Li/NaTi$_2$(PO$_4$)$_3$, Li/NaSn$_2$(PO$_4$)$_3$, Li/NaZr$_2$(PO$_4$)$_3$ and Li/NaPb$_2$(PO$_4$)$_3$, as well as LiCl and NaCl. The electronic energies of these compounds were evaluated using the CASTEP code[4] as implemented in Materials Studio 6.1. Calculations were carried out on rhombohedral cells in both the local density approximation (LDA) and the generalized gradient approximation (GGA) with the PBE exchange-correlation functional.[5] Norm-conserving pseudopotentials were used with a plane-wave energy cutoff of 750 eV and a 2×2×2 k-point mesh. The atom positions and lattice parameters, all starting from the LiTi$_2$(PO$_4$)$_3$ structure[6], were optimized (CIF files for all optimized structures are given in the Supporting Information). For Li/NaGe$_2$(PO$_4$)$_3$, Li/NaTi$_2$(PO$_4$)$_3$ vibrational calculations were done on the optimized structure (within the LDA) to evaluate the vibrational contribution to the free energies of each of the compounds at room temperature. The free energies were used to compute the energy for the reaction:

$$\mathrm{LMP + NaCl_{xtl} = NMP + LiCl_{xtl}} \qquad (2)$$

These free energies can be related to Reaction 1 because the enthalpy and entropy of solution of $NaCl_{xtl}$ ($\Delta H_{sol}$=+3.9 kJ/mol $\Delta S_{sol}$=+43.4 J/K/mol at 298.15 K) and $LiCl_{xtl}$ ($\Delta H_{sol}$ =-37 kJ/mol, $\Delta S_{sol}$ =+10.6 J/K/mol at 298.15 K) are well-known (see, for example, the CRC handbook).

In addition, calculations are carried out on candidate LATP and LAGP structures (as well as their Na-analogs) to assess the influence of aluminum substitution on LAMP stability. The structures of these materials are not known precisely, but a reasonable guess can be made at the optimal structure using the general principle that the aluminum ions should be as far apart as possible, and that the associated Li in the $M^{4+} = Al^{3+} + Li^{+}$ coupled substitution should be placed in L2 sites as close as possible to the aluminum ions. These structures were run in a hexagonal unit cell. Since there are, conveniently, 12 M ions per unit cell, one aluminum can be placed in each of three MO6 columns in the unit cell, with the z coordinates offset to maximize the distance between them in the unit cell. The L2 sites are arranged trigonally around the MO6. One of the three equivalent L2 sites associated with the aluminum-substituted MO6 octahedra is chosen arbitrarily. The other L2 sites are chosen in the same direction for the other two MO6 octahedra within the unit cell. This gives a distance of 8.551 angstroms between each substituted site and a composition $Li_{1.5}Al_{0.5}M_{1.5}(PO_4)_3$ (there are 9 lithium atoms, 3 aluminum atoms, 9 M atoms, 18 phosphorous atoms and 72 oxygen atoms per hexagonal unit cell). The structure just described is illustrated in Figure 3. Compositions with 0.333 and 0.167 aluminum atoms per formula unit (i.e. 2 and 1 aluminum atoms per unit cell) are generated by replacing one or two $Al^{3+}$-$Li^{+}$ pair(s) with $Ti^{4+}$ or $Ge^{4+}$. (CIF files for optimized structures are given in the Supporting Information). Future work could explore whether there are more optimal arrangements, but this procedure is likely to yield a reasonable preliminary initial structure.

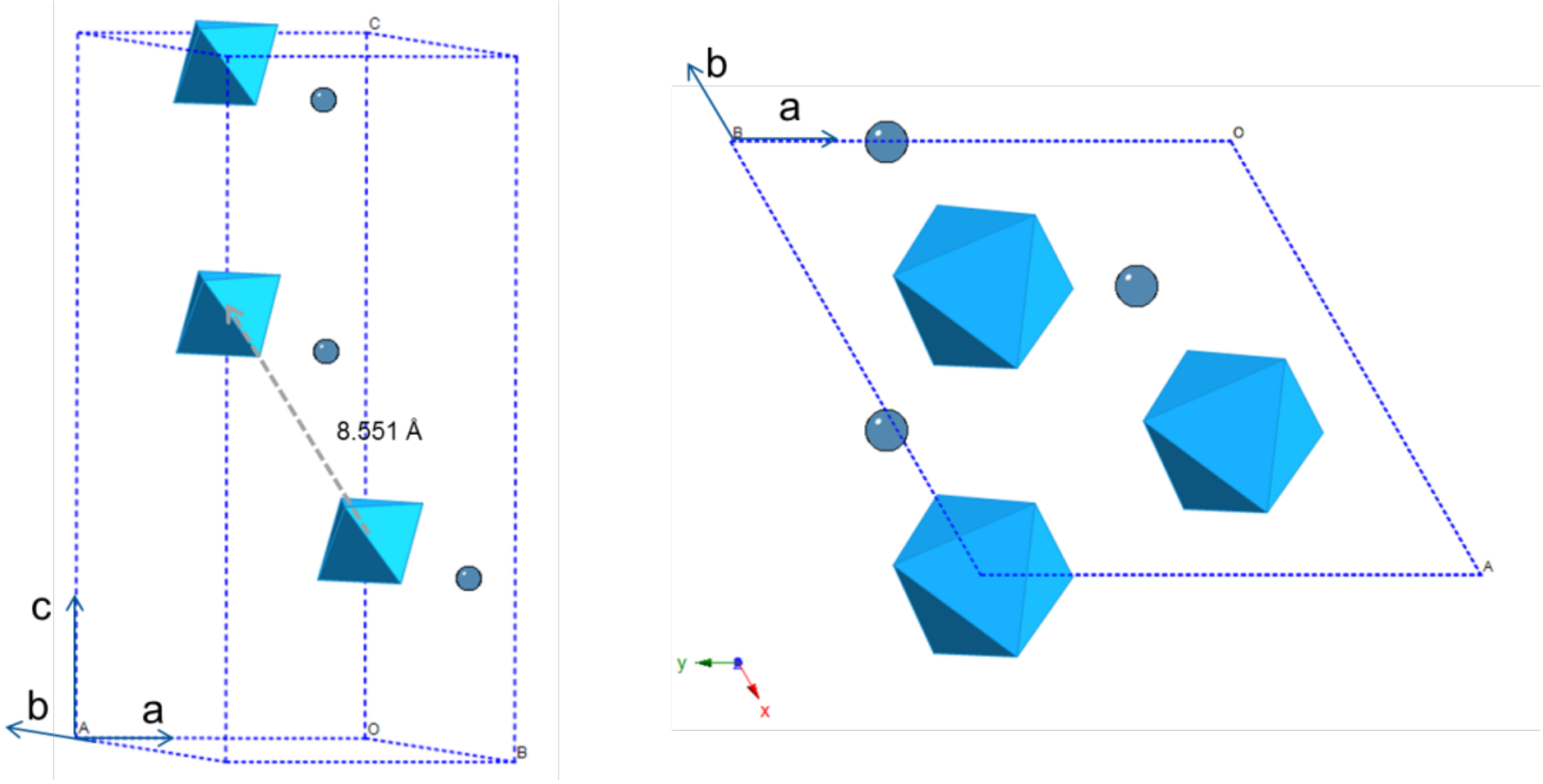

**Figure 3.** Arrangement of Al-substituted MO6 sites and associated Li+ placements in L2 sites for LATP and LAGP (distances given for LATP structure). Side view (left), top view (right).

## Results

Structural results are given in Tables 1 and 2 for the LMP and NMP compounds, respectively. Agreement with X-ray measurements is as expected with the GGA overestimating cell parameters/bond lengths and the LDA underestimating them. The major exceptions are the tin phases, for which the cell parameters are strongly underestimated even in the GGA. The most likely explanation is a poor pseudopotential for tin in the version of CASTEP supplied with Materials Studio. The calculations for tin are probably not very reliable.

**Table 1. Structural Parameters for LMP compounds (Ångstrom units)**

| | a | c | volume | $r_{LiO}$ | $r_{MO1}$ | $r_{MO2}$ |
|---|---|---|---|---|---|---|
| $LiSi_2(PO_4)_3$ | 8.1402 | 20.4018 | 1170.8 | 2.259 | 1.809 | 1.769 |
| $LiGe_2(PO_4)_3$ | 8.3815 | 20.9744 | 1276.0 | 2.296 | 1.917 | 1.879 |
| $LiGe_2(PO_4)_3$[a] | 8.1864 | 20.5083 | 1190.3 | 2.253 | 1.852 | 1.819 |
| $LiGe_2(PO_4)_3$[b] | 8.275 | 20.47 | 1213.9 | 2.209 | 1.873 | 1.835 |
| $LiTi_2(PO_4)_3$ | 8.5938 | 21.4053 | 1369.1 | 2.376 | 2.003 | 1.912 |
| $LiTi_2(PO_4)_3$[a] | 8.4626 | 20.9981 | 1302.3 | 2.335 | 1.956 | 1.888 |
| $LiTi_2(PO_4)_3$[c] | 8.5145 | 20.8633 | 1309.9 | 2.267 | 1.977 | 1.881 |
| $LiSn_2(PO_4)_3$ | 8.4725 | 21.3981 | 1330.2 | 2.374 | 1.942 | 1.913 |
| $LiSn_2(PO_4)_3$[d] | 8.6334 | 21.5330 | 1389.9 | 2.340 | 2.027 | 1.990 |
| $LiZr_2(PO_4)_3$ | 8.9288 | 22.6231 | 1562.0 | 2.526 | 2.137 | 2.074 |
| $LiZr_2(PO_4)_3$[e] | 8.847 | 22.24 | 1507.5 | 2.501 | 2.082 | 2.039 |

[a]computed in the LDA; [b]Ref 7; [c]Ref 6;[d]Ref 8;[e]Ref. 9

**Table 2. Structural parameters for NMP compounds (Ångstrom units).**

| | a | c | volume | $r_{LiO}$ | $r_{MO1}$ | $r_{MO2}$ |
|---|---|---|---|---|---|---|
| $NaSi_2(PO_4)_3$ | 8.1021 | 21.2172 | 1206.2 | 2.431 | 1.814 | 1.774 |
| $NaGe_2(PO_4)_3$ | 8.3423 | 21.8897 | 1319.3 | 2.487 | 1.919 | 1.884 |
| $NaGe_2(PO_4)_3$[a] | 8.1530 | 21.2751 | 1224.7 | 2.420 | 1.851 | 1.823 |
| $NaGe_2(PO_4)_3$[b] | 8.109 | 21.536 | 1226.39 | 2.515 | 1.879 | 1.851 |
| $NaTi_2(PO_4)_3$ | 8.5817 | 22.2259 | 1417.5 | 2.551 | 1.999 | 1.923 |
| $NaTi_2(PO_4)_3$[a] | 8.446 | 21.6971 | 1340.4 | 2.471 | 1.955 | 1.895 |
| $NaTi_2(PO_4)_3$[c] | 8.4854 | 21.7994 | 1359.32 | 2.472 | 1.983 | 1.912 |
| $NaSn_2(PO_4)_3$ | 8.4397 | 22.1985 | 1369.3 | 2.541 | 1.943 | 1.918 |
| $NaSn_2(PO_4)_3$[d] | 8.51324 | 22.5106 | 1412.89 | 2.508 | 2.021 | 1.991 |
| $NaZr_2(PO_4)_3$ | 8.9236 | 23.3138 | 1607.8 | 2.669 | 2.136 | 2.080 |
| $NaZr_2(PO_4)_3$[e] | 8.8043 | 22.7595 | 1527.79 | 2.534 | 2.083 | 2.041 |

[a]computed in the LDA [b] Ref. 10; [c]Ref. 11; [d]Ref 12; [e]Ref 13

For LATP and LAGP, major changes were seen in optimizing the initial structure, while the NATP and NAGP structures were maintained. Running the NATP composition in the optimized LATP structure optimized back to the structure described above, so the NT(G)P structure obtained from simple optimization is likely to be a low-energy structure for NAT(G)P. In the LATP and LAGP structures, the Li in the L1 position immediately above the aluminum substituted MO6 site moves much closer to the top three oxygen atoms of the AlO6 octahedron, breaking its bonds with the bottom three oxygen atoms of the TiO6/GeO6 octahedron (illustrated for LATP in Figure 4). Second, the Li in the L1 site between two TiO6/GeO6 octahedra is displaced into an L2-like site. The structure in Figure 4b is unusual because of the broken bonds, and one suspects that it might not be stable (vibrational analysis was not carried out because of the large size of the hexagonal cell). On the other hand, published structures for LAGTP compounds give the same site for $Al^{3+}$, $Ge^{4+}$ and $Ti^{4+}$.[14] One possible solution would be to postulate clustering of Al into chains terminated by L1 vacancies, giving a v-M[$(PO_4)_3$Al-Li-Al]$_n$$(PO_4)_3$M-v structure along the columns parallel to *z* to minimize the mismatch between the $M^{4+}$ side and the $Al^{3+}$ side of the $M^{4+}(O_3)$---Li-$(O_3)Al^{3+}$ L1 bonding environments (as shown in Figure 4b). Working out the energetics for these structures theoretically would require further study with large supercells. Because of the dramatic change in the lithium bonding upon changing $M^{4+}$ to $Al^{3+}$, the structure was also run using VASP[15] with the same PBE exchange-correlation functional and projector-augmented-wave pseudopotentials[16] and an energy cutoff of 520 eV. The Li-O distances to the AlO6 octahedron were reduced, but the effect is not as great as seen in Figure 4.

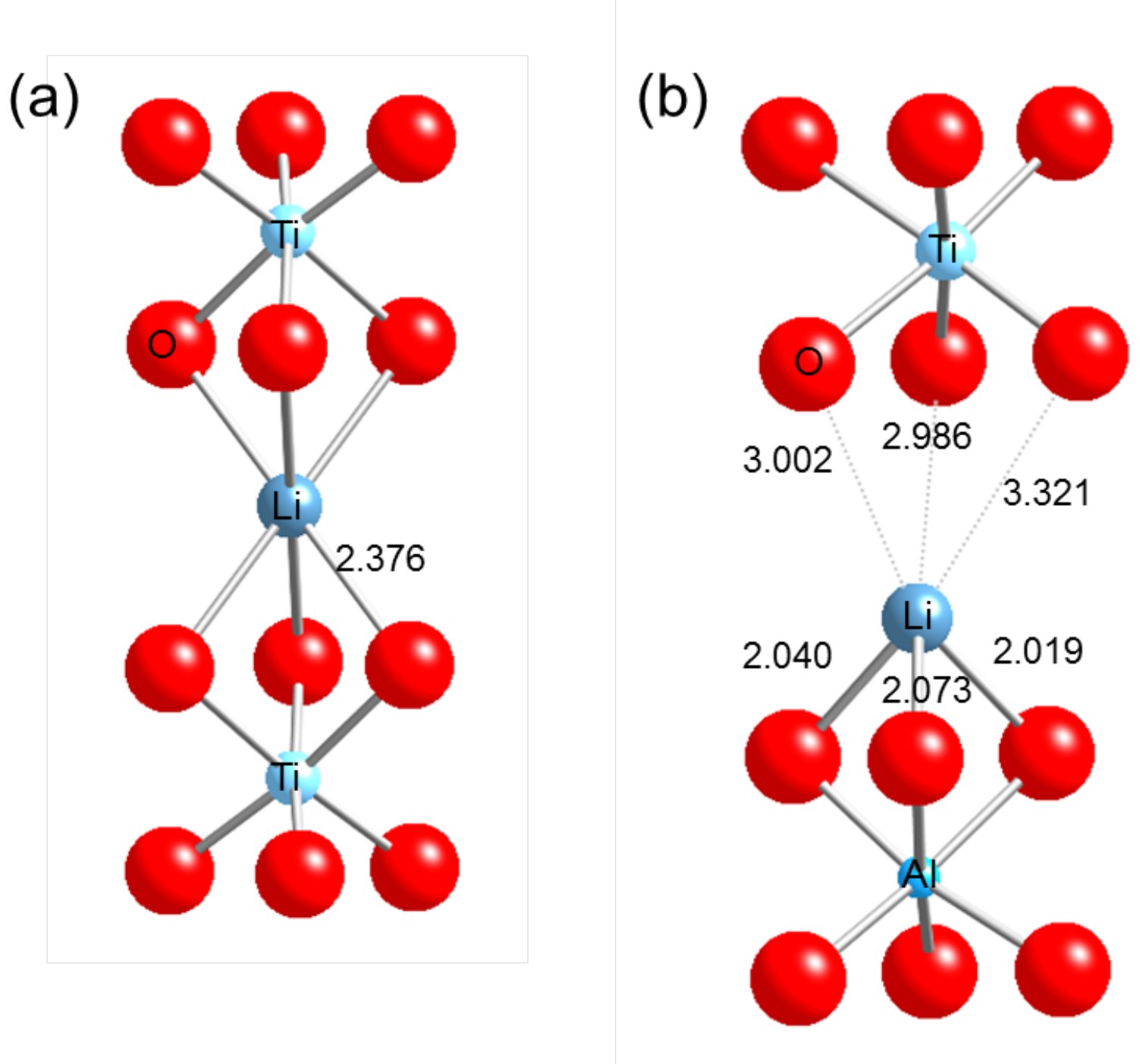

**Figure 4.** Structural changes in LTP (a) upon aluminum substitution at the MO6 site to make LATP (b). Bond lengths are given in angstroms (all Li-O bond lengths equivalent for LTP).

Energies calculated for the LMP/LAMP and NMP/NAMP compounds, along with energies calculated for the Na-Li exchange reaction are given in Table 3. Vibrational contributions for Li/$NaTi_2(PO_4)_3$ and Li/$NaGe_2(PO_4)_3$ are given in Table 4. The vibrational contributions to the exchange reaction energies for these compounds are small and are ignored hereafter. The results in Table 3 are plotted in Figure 5.

**Table 3.** Data and reaction energies for Na exchange for LMP and LAMP phases.

| | Li phase energy (eV) | NaCl | Na phase energy (eV) | LiCl | $DE_e$ (eV) | $DE_e$ (kJ/mol) | DG (kJ/mol) |
|---|---|---|---|---|---|---|---|
| $LiSi_2(PO_4)_3$ | -5947.424 | -1566.336 | -7095.638 | -417.753 | 0.369 | 35.6 | 4.5 |
| $LiGe_2(PO_4)_3$ | -5939.850 | -1566.336 | -7088.202 | -417.753 | 0.231 | 22.3 | -8.8 |
| $LiTi_2(PO_4)_3$ | -8884.005 | -1566.336 | -10032.525 | -417.753 | 0.063 | 6.1 | -25.0 |
| $LiSn_2(PO_4)_3$ | -5926.216 | -1566.336 | -7074.683 | -417.753 | 0.116 | 11.2 | -19.9 |
| $LiZr_2(PO_4)_3$ | -8284.608 | -1566.336 | -9433.336 | -417.753 | -0.145 | -14.0 | -45.1 |
| $LiPb_2(PO_4)_3{}^{*}$ | -16279.397 | -1566.368 | -17428.096 | -417.769 | -0.101 | -9.7 | -40.8 |
| $LiGe_2(PO_4)_3{}^{\dagger}$ | -5944.3452 | -1562.114 | -7087.8952 | -418.279 | 0.286 | 27.6 | -3.5 |
| $Li_{1.5}Al_{0.5}Ge_{1.5}(PO_4)_3$ | -35537.615 | -1566.336 | -45870.875 | -417.753 | 0.443 | 42.8 | 11.7 |
| $Li_{1.333}Al_{0.333}Ge_{1.667}(PO_4)_3$ | -35571.500 | -1566.336 | -44756.914 | -417.753 | 0.406 | 39.2 | 8.1 |
| $Li_{1.167}Al_{0.167}Ge_{1.833}(PO_4)_3$ | -35605.305 | -1566.336 | -43643.030 | -417.753 | 0.336 | 32.5 | 1.4 |
| $Li_{1.5}Al_{0.5}Ti_{1.5}(PO_4)_3$ | -48786.454 | -1566.336 | -59120.263 | -417.753 | 0.382 | 36.9 | 5.8 |
| $Li_{1.333}Al_{0.333}Ti_{1.667}(PO_4)_3$ | -50292.284 | -1566.336 | -59478.508 | -417.753 | 0.301 | 29.0 | -2.1 |
| $Li_{1.167}Al_{0.167}Ti_{1.833}(PO_4)_3$ | -51798.345 | -1566.336 | -59836.802 | -417.753 | 0.232 | 22.4 | -8.7 |

*calculation fo $LiPb_2(PO_4)_3$ was done with a plane-wave cutoff of 820 eV; all others done at 750 eV; †computed with LDA

**Table 4.** Vibrational contributions to free energy for Na exchange for LTP and LGP phases.

| | ZPE LMP (eV) | G LMP (eV) | ZPE NaCl (eV) | G NaCl (eV) | ZPE NMP (eV) | G NMP (eV) | ZPE LiCl (eV) | G LiCl (eV) | $DG_{tot}$ (kJ/mol) |
|---|---|---|---|---|---|---|---|---|---|
| M=Ti | 1.629 | -0.359 | 0.057 | -0.108 | 1.632 | -0.403 | 0.075 | -0.085 | -0.020 |
| M=Ge | 1.726 | -0.359 | 0.057 | -0.108 | 1.709 | -0.381 | 0.075 | -0.085 | 0.230 |

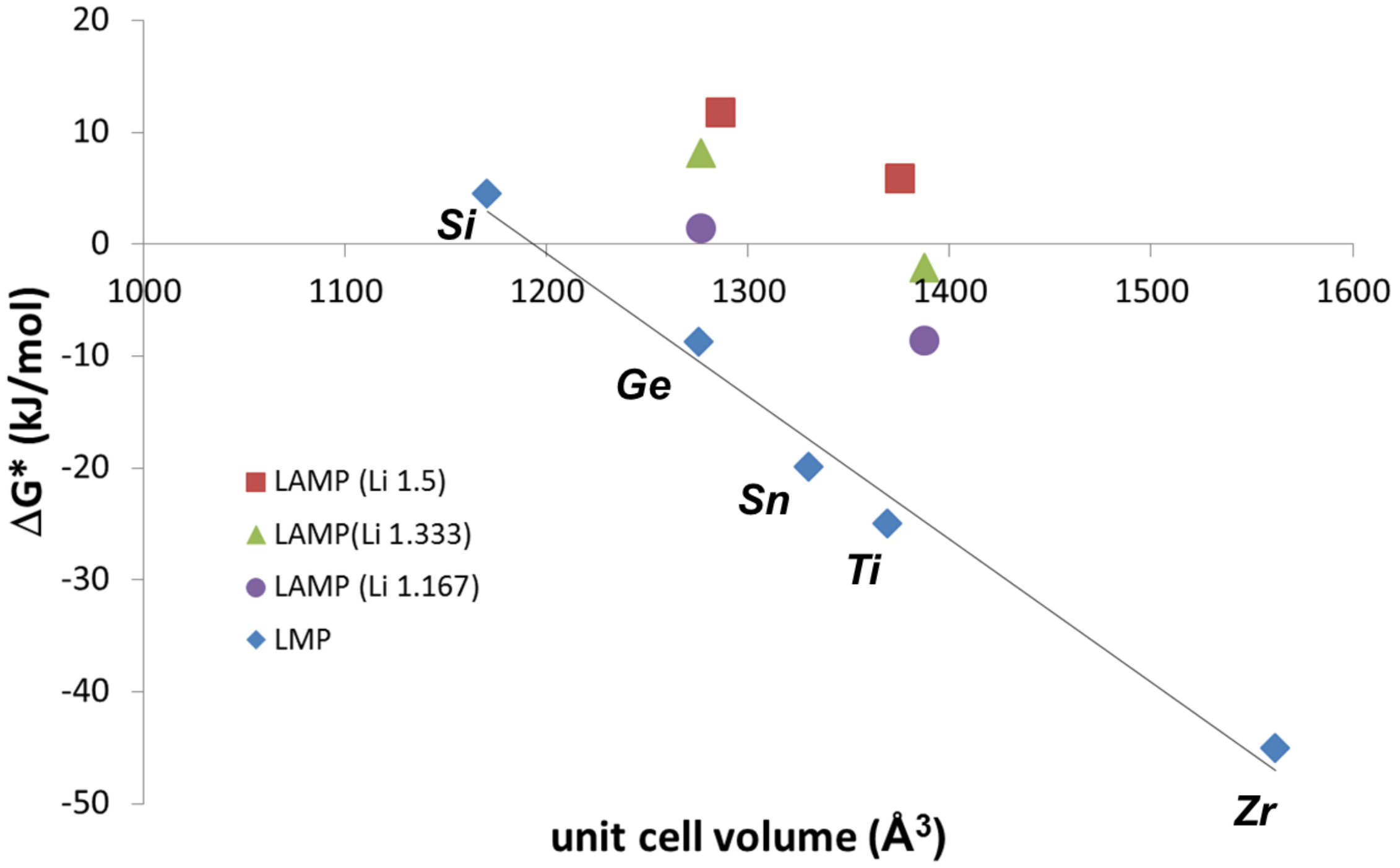

**Figure 5.** Free energies calculated for the $Na^+$-$Li^+$ exchange reaction for LMP (blue diamonds) and LATP, LAGP compounds (purple circles, green triangles, and red squares).

For the LMP compounds there is a strong correlation of the favorability of the exchange reaction with the volume of the unit cell, and, hence, the ionic radius of the $M^{4+}$ cation. The larger the cell, the more favorable the reaction. This relationship can be understood as simply indicating the overall “fit” of the alkali cation into the L1 site. A larger cell is better able to accommodate the larger $Na^+$ ion. Although the relative ΔG values are probably more robust than the absolute values, Figure 5 also suggests that for the Al-free compositions, both LGP and LTP would be unstable to ion exchange in seawater. The only Al-free composition that would be predicted to be stable is the hypothetical $LiSi_2(PO_4)_3$ compound, which, to our knowledge, has not been made and is probably unstable with respect to phases with four-fold coordinated $Si^{4+}$.[17]

Figure 5 indicates a strong dependence of the calculated reaction energy of the LAMP compounds on the extent of $Al^{3+}$ substitution. The higher the aluminum content of the LAMP, the greater is its resistance to $Na^+$-$Li^+$ exchange. This effect is apparently not simply related to the effect of Al on the unit cell volume, as this influence is small. This observed effect is probably the increased effective attraction between the oxygen atoms on the AlO6 octahedron and the Li ion. The $Na^+$ ion is not able to take as much advantage of the increased basicity of the oxide ion upon substitution of $Al^{3+}$ for $Ti^{4+}$ or $Ge^{4+}$. This is the same interaction that drives the structural response shown in Figure 4. Whatever the ultimate cause, Figure 5 clearly shows aluminum substitution is a strong factor in improving the stability of the LAGP/LATP compounds against ion exchange in sodium-rich environments.

Typical values of $x$ in $Li_{1+x}M_{2-x}Al_x(PO_4)_3$ compounds are around 0.4. Figure 5 suggests that it may be possible to stabilize LATP against sodium ion exchange by increasing $x$ to 0.5 or beyond. Figure 6 shows a significant degree of nonlinearity suggesting, for example, that changing $x$ from 0.5 to 0.6 would not give as much improvement as changing $x$ from 0.4 to 0.5.

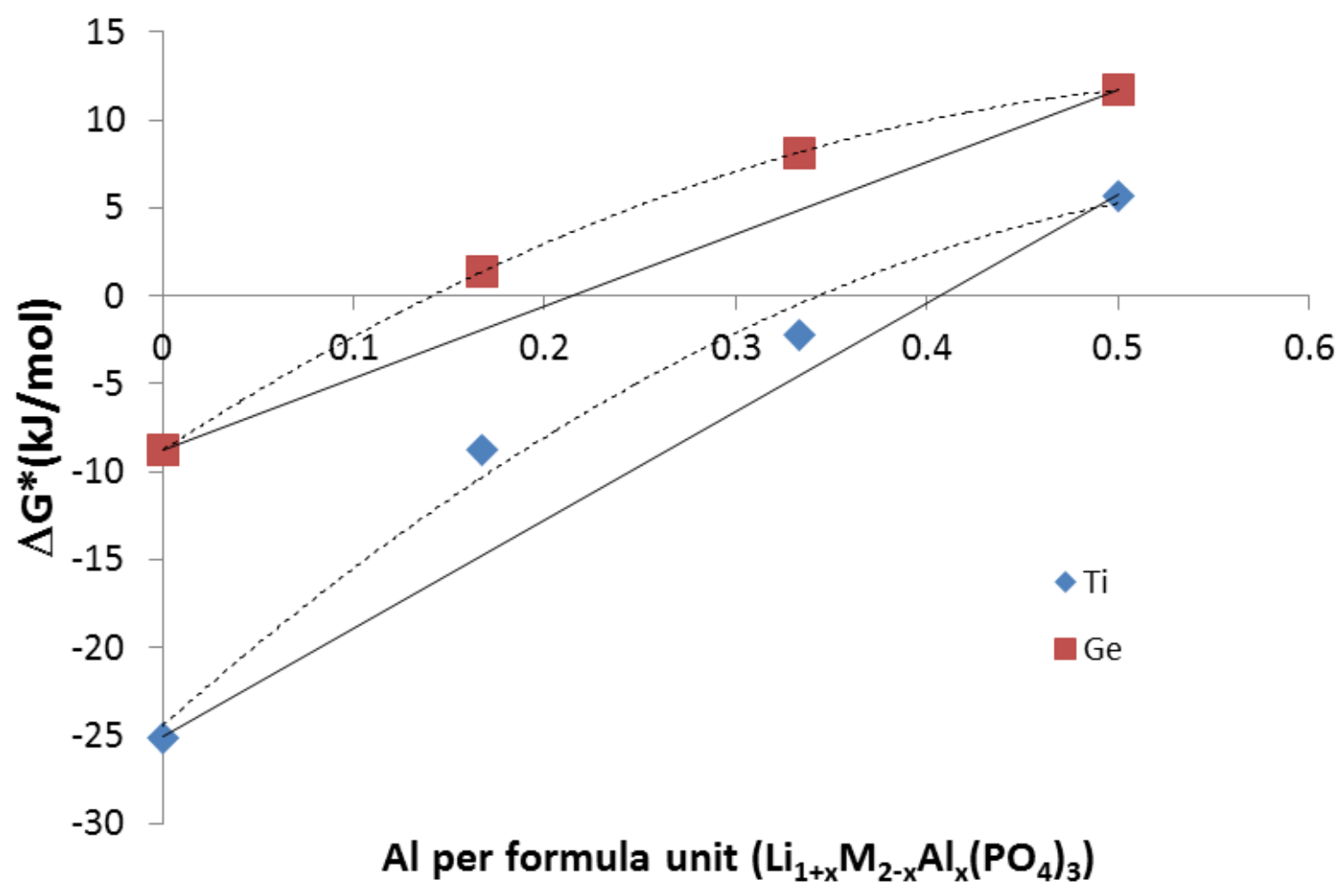

**Figure 6.** Free energy of $Na^+$-$Li^+$ exchange reaction for LATP and LAGP compounds as a function of substituted aluminum. Dashed lines represent a quadratic fit to the calculated points.

## Conclusions

In summary, the calculations presented here indicate the following:

(1) Resistance to $Na^+$(aq)-$Li^+$ ion exchange in LMP compounds is inversely correlated to unit cell volume with smaller unit cell volumes being more resistant to sodium exchange. Thus L(A)GP is more resistant to sodium exchange than L(A)TP.

(2) Increasing $Al^{3+}$ substitution in $Li_{1+x}M_{2-x}Al_x(PO_4)_3$ compounds increases the resistance to $Na^+$(aq)-$Li^+$ ion exchange in the LAMP compounds. This is not driven by systematic changes in volume but rather seems to be driven chemically by improved $AlO_3 \equiv Li$ association in the substituted compound. The calculations suggest that LATP can be made stable to ion exchange if $x > 0.4$.

**Acknowledgement-** The author thanks Michael E. Badding and Xinyuan (Shelly) Liu, of Corning Incorporated, for useful discussions and suggestions.

---

[1] Aono H; Sugimoto E; Sadaoka Y; Imanaka N; Adachi G "Ionic conductivity of solid electrolytes based on lithium titanium phosphate" *Journal of the Electrochemical Society* **137**, 1023-1027, 1990.

[2] Kumar B; Thomas D; Kumar J "Space-charge-mediated superionic transport in lithium ion conduction glass-ceramics" *Journal of the Electrochemical Society* **156**, A506-A513, 2009

[3] "Seawater battery sparks sub dreams" *NewScientist* 25 April 2011 Magazine issue 2861.

[4] Clark SJ; Segall MD; Pickard CJ; Hasnip PJ; Probert MJ; Refson K, Payne MC. "First principles methods using CASTEP" *Zeitschrift fur Kristallographie* **220**, 567-570, 2005.

[5] Perdew JP; Burke K; Ernzerhof M "Generalized gradient approximation made simple" *Physical Review Letters* **77**, 3865-368, 1996.

[6] Pinus I-Y; Khoroshilov AV; Gavrichev KS; Tarasov VP; Yaroslavtsev AB "On cationic mobility in NASICON phosphates $LiTi_2(PO_4)_3$ and $Li_{0.9}Ti_{1.9}Nb_{0.1}(PO_4)_3$" *Journal of Materials Chemistry* **12**, 2971-2978, 2002.

[7] Alami M; Brochu R; Soubeyroux JL; Gravereau P, le Flem G; Hagenmuller P J. "Structure and thermal expansion of $LiGe_2(PO_4)_3$" *Journal of Solid State Chemistry* **90**, 185-193, 1991.

[8] Morin, E.; Angenault J; Couturier JC; Quarton M; He H; Klinowski J "Phase transition and crystal structures of $LiSn_2(PO_4)_3$" *European Journal of Solid State Inorganic Chemistry* 34, 947-958, 1997.

[9] Petit D; Colomban P; Collin G; Boilot JP "Fast ion transport in $LiZr_2(PO_4)_3$: Structure and conductivity" *Materials Research Bulletin* **21**, 365-371, 1986.

[10] Zhao D; Xie S; Hu J-M; Zhang H; Zhang W-l; Yang S-L; Cheng W-D "Structure determination, electronic and optical properties of $NaGe_2P_3O_{12}$ and $Cs_2GeP_4O_{13}$" *Journal of Molecular Structure* **922**, 127-134, 2009

[11] Rodrigo JL; Carrasco P; Alamo J "Thermal expansion of $NaTi_2(PO_4)_3$ studied by Rietveld method from X-ray diffraction data" *Materials Research Bulletin* **24**, 611-618. 1989.

[12] Alamo J; Rodrigo JL "High temperature neutron diffraction study of sodium di-tin tri-phosphate" *Materials Research Bulletin* **27**, 1091-1098, 1992.

[13] Hagman L-O; Kierkegaard P "The crystal structure of $NaMe_2(PO_4)_3$; Me = Ge, Ti Zr" *Acta Chemica Scandinavica* **22**, 1822-1832, 1968.

[14] Maldonado-Manso P;Morsli K;Mouahid FE;Zahir M;Bruque S;Losilla ER "New NASICON $Li_{1+x}Al_x Ge_yTi_{2-x-y}(PO_4)_3$ compositions with high lithium ionic conductivity" *Annales de Chimie* (Paris) **32**, 561-572, 2007 (structure located in ICSD database, original reference not examined).

[15] Kresse G and Hafner J "Ab initio molecular dynamics for liquid metals" Physical Review B, 47, 558-, 1993; Kresse G and Hafner J "Ab initio molecular-dynamics simulation of the liquid-metal-amorphous-semiconductor transition in germanium" Physical Review B, 49, 14251-, 1994; Kresse G and Furthmueller J. "Efficiency of ab-initio total energy calculations for metals and semiconductors using a plane-wave basis set" Computational Materials Science 6,15-, 1996; Kresse G and Furthmueller J "Efficient iterative

schemes for ab initio total-energy calculations using a plane-wave basis set" Physical Review B, 54, 11169-, 1996

[16] Bloechl PE "Projector augmented-wave method" Physical Review B, 50,17953-, 1994; Kresse G and Jourbert D "From ultrasoft pseudopotentials to the projector augmented-wave method" Physical Review B, 59, 1758-, 1999

[17] Concerning comparisons with VASP, it is noted that energies for Reaction 2 are quite different in the Materials Genome Project database (-2 kJ/mol (M=Ge), -4 kJ/mol (M=Sn), -7 kJ/mol (M=Ti), ~+1.9 kJ/mol (M=Zr (with $LiZr_2(PO_4)_3$ taken in the R-3c structure). The issue requires further study. Since the trend with increasing volume seems reasonable, questions arise concerning how the corrections applied in the MGP to normalize the formation energies are affecting these results. On the other hand, the PAW approach is probably overall more reliable than the norm-conserving pseudopotentials applied here. The whole set of structures should probably be redone directly with VASP to check the results.

**Supporting Information**

CIF files for optimized structures

```
LAMP STRUCTURES
------------------------------------------------------------
lagp1_opt.cif
data_LAGP_1AL
_audit_creation_date              2014-10-29
_audit_creation_method            'Materials Studio'
_symmetry_space_group_name_H-M    'P1'
_symmetry_Int_Tables_number       1
_symmetry_cell_setting            triclinic
loop_
_symmetry_equiv_pos_as_xyz
  x,y,z
_cell_length_a                    8.3746
_cell_length_b                    8.3710
_cell_length_c                    21.0167
_cell_angle_alpha                 89.9277
_cell_angle_beta                  90.0215
_cell_angle_gamma                 119.9413
loop_
_atom_site_label
_atom_site_type_symbol
_atom_site_fract_x
_atom_site_fract_y
_atom_site_fract_z
_atom_site_U_iso_or_equiv
_atom_site_adp_type
_atom_site_occupancy
Ge107  Ge    0.00440   0.00677   0.64069   0.00000  Uiso   1.00
Ge108  Ge    0.67091   0.33943   0.97361   0.00000  Uiso   1.00
Al109  Al    0.33957   0.67188   0.30978   0.00000  Uiso   1.00
Li91   Li   -0.01100  -0.00575  -0.00201   0.00000  Uiso   1.00
Li92   Li    0.48428   0.17494   0.35029   0.00000  Uiso   1.00
Li93   Li    0.33225   0.66805   0.66610   0.00000  Uiso   1.00
Li94   Li    0.01724   0.01006   0.49934   0.00000  Uiso   1.00
Li95   Li    0.67367   0.32872   0.83065   0.00000  Uiso   1.00
Li96   Li    0.33328   0.66762   0.17336   0.00000  Uiso   1.00
Li97   Li   -0.11062   0.53530   0.30523   0.00000  Uiso   1.00
O1     O     0.17364   0.98068   0.19231   0.00000  Uiso   1.00
O10    O     0.49699   0.35420   0.41567   0.00000  Uiso   1.00
O11    O     0.35863   0.87225   0.85461   0.00000  Uiso   1.00
O12    O     0.17747   0.70011   0.74901   0.00000  Uiso   1.00
O13    O     0.79951   0.83679   0.18935   0.00000  Uiso   1.00
O14    O     0.97165   0.81493   0.08340   0.00000  Uiso   1.00
O15    O     0.47592   0.16095   0.52263   0.00000  Uiso   1.00
O16    O     0.64032   0.14898   0.41669   0.00000  Uiso   1.00
O17    O     0.13891   0.49347   0.85524   0.00000  Uiso   1.00
O18    O     0.30951   0.48445   0.74912   0.00000  Uiso   1.00
O19    O     0.98180   0.17205   0.30293   0.00000  Uiso   1.00
O2     O     0.19509   0.16018   0.08283   0.00000  Uiso   1.00
O20    O     0.15636   0.19638   0.41444   0.00000  Uiso   1.00
O21    O     0.65584   0.51925   0.64106   0.00000  Uiso   1.00
O22    O     0.83435   0.52459   0.74683   0.00000  Uiso   1.00
O23    O     0.32273   0.85194   0.97417   0.00000  Uiso   1.00
O24    O     0.50238   0.85700   0.08028   0.00000  Uiso   1.00
O25    O     0.22299   0.03492   0.31167   0.00000  Uiso   1.00
O26    O     0.03827   0.85249   0.41658   0.00000  Uiso   1.00
O27    O     0.86599   0.35272   0.64120   0.00000  Uiso   1.00
O28    O     0.69401   0.17600   0.74787   0.00000  Uiso   1.00
O29    O     0.53410   0.68554   0.97513   0.00000  Uiso   1.00
O3     O     0.85389   0.32651   0.52130   0.00000  Uiso   1.00
O30    O     0.35717   0.50525   0.08137   0.00000  Uiso   1.00
O31    O     0.84958   0.79870   0.30588   0.00000  Uiso   1.00
O32    O     0.80889   0.97274   0.40937   0.00000  Uiso   1.00
O33    O     0.48819   0.14009   0.64218   0.00000  Uiso   1.00
O34    O     0.48462   0.31586   0.74710   0.00000  Uiso   1.00
O35    O     0.15507   0.47366   0.97483   0.00000  Uiso   1.00
O36    O     0.15221   0.65168   0.08030   0.00000  Uiso   1.00
O37    O     0.82438   0.02407   0.80805   0.00000  Uiso   1.00
O38    O     0.81525   0.84522   0.91440   0.00000  Uiso   1.00
O39    O     0.49121   0.35555   0.14169   0.00000  Uiso   1.00
O4     O     0.85601   0.50833   0.41639   0.00000  Uiso   1.00
O40    O     0.48674   0.17105   0.24610   0.00000  Uiso   1.00
O41    O     0.16224   0.68802   0.47349   0.00000  Uiso   1.00
O42    O     0.14976   0.51303   0.58201   0.00000  Uiso   1.00
O43    O     0.98317   0.80532   0.80782   0.00000  Uiso   1.00
O44    O     0.16355   0.97522   0.91361   0.00000  Uiso   1.00
O45    O     0.65430   0.13644   0.14186   0.00000  Uiso   1.00
O46    O     0.84721   0.32760   0.24592   0.00000  Uiso   1.00
O47    O     0.32022   0.46782   0.47980   0.00000  Uiso   1.00
O48    O     0.49988   0.64888   0.58301   0.00000  Uiso   1.00
O49    O     0.20199   0.18351   0.80726   0.00000  Uiso   1.00
O5     O     0.51641   0.65283   0.85540   0.00000  Uiso   1.00
O50    O     0.03291   0.19322   0.91355   0.00000  Uiso   1.00
O51    O     0.86793   0.51305   0.13929   0.00000  Uiso   1.00
O52    O     0.70425   0.53050   0.24373   0.00000  Uiso   1.00
O53    O     0.54045   0.85093   0.47624   0.00000  Uiso   1.00
O54    O     0.36229   0.85994   0.58081   0.00000  Uiso   1.00
O55    O     0.02175   0.82604   0.68842   0.00000  Uiso   1.00
O56    O     0.84386   0.81904   0.58244   0.00000  Uiso   1.00
```

```
O57    O     0.68595   0.15878   0.02203   0.00000  Uiso   1.00
O58    O     0.50741   0.15261   0.91558   0.00000  Uiso   1.00
O59    O     0.34716   0.50484   0.36109   0.00000  Uiso   1.00
O6     O     0.52571   0.83201   0.74901   0.00000  Uiso   1.00
O60    O     0.18512   0.48507   0.25400   0.00000  Uiso   1.00
O61    O     0.80718   0.99007   0.68842   0.00000  Uiso   1.00
O62    O     0.97904   0.16815   0.58174   0.00000  Uiso   1.00
O63    O     0.47572   0.32474   0.02199   0.00000  Uiso   1.00
O64    O     0.64775   0.50120   0.91530   0.00000  Uiso   1.00
O65    O     0.13072   0.65469   0.35328   0.00000  Uiso   1.00
O66    O     0.30968   0.82758   0.25174   0.00000  Uiso   1.00
O67    O     0.18543   0.20493   0.68783   0.00000  Uiso   1.00
O68    O     0.19161   0.03272   0.58192   0.00000  Uiso   1.00
O69    O     0.85425   0.53868   0.02010   0.00000  Uiso   1.00
O7     O     0.03234   0.21498   0.18489   0.00000  Uiso   1.00
O70    O     0.85607   0.35985   0.91516   0.00000  Uiso   1.00
O71    O     0.49671   0.86479   0.35809   0.00000  Uiso   1.00
O72    O     0.53186   0.69989   0.25889   0.00000  Uiso   1.00
O8     O     0.84644   0.03472   0.08072   0.00000  Uiso   1.00
O9     O     0.68492   0.53869   0.52130   0.00000  Uiso   1.00
P73    P     0.29345  -0.00147   0.24826   0.00000  Uiso   1.00
P74    P     0.95904   0.34005   0.58147   0.00000  Uiso   1.00
P75    P     0.62486   0.67107   0.91500   0.00000  Uiso   1.00
P76    P     0.01545   0.29823   0.24665   0.00000  Uiso   1.00
P77    P     0.67126   0.62829   0.58193   0.00000  Uiso   1.00
P78    P     0.33775   0.95953   0.91466   0.00000  Uiso   1.00
P79    P     0.71768   0.72065   0.24869   0.00000  Uiso   1.00
P80    P     0.38328   0.05177   0.58185   0.00000  Uiso   1.00
P81    P     0.04945   0.38350   0.91477   0.00000  Uiso   1.00
P82    P     0.71602   0.00507   0.74828   0.00000  Uiso   1.00
P83    P     0.38267   0.33679   0.08178   0.00000  Uiso   1.00
P84    P     0.05204   0.67578   0.41538   0.00000  Uiso   1.00
P85    P     0.00397   0.71730   0.74801   0.00000  Uiso   1.00
P86    P     0.67140   0.04949   0.08101   0.00000  Uiso   1.00
P87    P     0.33134   0.38792   0.41648   0.00000  Uiso   1.00
P88    P     0.29217   0.29371   0.74790   0.00000  Uiso   1.00
P89    P     0.95855   0.62631   0.08064   0.00000  Uiso   1.00
P90    P     0.61868   0.95423   0.41430   0.00000  Uiso   1.00
Ge98   Ge    0.00317   0.00695   0.14076   0.00000  Uiso   1.00
Ge99   Ge    0.67167   0.34254   0.47405   0.00000  Uiso   1.00
Ge100  Ge    0.33768   0.67241   0.80762   0.00000  Uiso   1.00
Ge101  Ge    0.00627   0.00652   0.35567   0.00000  Uiso   1.00
Ge102  Ge    0.67034   0.33835   0.68892   0.00000  Uiso   1.00
Ge103  Ge    0.33699   0.67032   0.02284   0.00000  Uiso   1.00
Ge104  Ge    0.00432   0.00469   0.85543   0.00000  Uiso   1.00
Ge105  Ge    0.67142   0.33510   0.18791   0.00000  Uiso   1.00
Ge106  Ge    0.34033   0.67228   0.52351   0.00000  Uiso   1.00
lagp2_opt.cif
data_lagp2
_audit_creation_date              2014-10-31
_audit_creation_method            'Materials Studio'
_symmetry_space_group_name_H-M    'P1'
_symmetry_Int_Tables_number       1
_symmetry_cell_setting            triclinic
loop_
_symmetry_equiv_pos_as_xyz
  x,y,z
_cell_length_a                    8.3674
_cell_length_b                    8.3603
_cell_length_c                    21.0416
_cell_angle_alpha                 89.8809
_cell_angle_beta                  90.0271
_cell_angle_gamma                 119.8746
loop_
_atom_site_label
_atom_site_type_symbol
_atom_site_fract_x
_atom_site_fract_y
_atom_site_fract_z
_atom_site_U_iso_or_equiv
_atom_site_adp_type
_atom_site_occupancy
Ge108  Ge    0.00488   0.00801   0.63990   0.00000  Uiso   1.00
Al109  Al    0.67477   0.34099   0.97572   0.00000  Uiso   1.00
Al110  Al    0.34028   0.67457   0.30943   0.00000  Uiso   1.00
Li91   Li   -0.17957  -0.15722   0.01595   0.00000  Uiso   1.00
Li92   Li    0.48588   0.17589   0.34939   0.00000  Uiso   1.00
Li93   Li    0.32031   0.65726   0.66448   0.00000  Uiso   1.00
Li94   Li    0.02068   0.00911   0.50004   0.00000  Uiso   1.00
Li95   Li    0.67209   0.33480   0.83957   0.00000  Uiso   1.00
Li96   Li    0.34716   0.67558   0.17486   0.00000  Uiso   1.00
Li97   Li   -0.11365   0.53738   0.30633   0.00000  Uiso   1.00
Li98   Li    0.22925   0.20606   0.96964   0.00000  Uiso   1.00
O1     O     0.17899   0.98996   0.19093   0.00000  Uiso   1.00
O10    O     0.49750   0.35423   0.41514   0.00000  Uiso   1.00
O11    O     0.36842   0.88374   0.85103   0.00000  Uiso   1.00
O12    O     0.18135   0.70318   0.74722   0.00000  Uiso   1.00
O13    O     0.80353   0.83847   0.18899   0.00000  Uiso   1.00
O14    O     0.97315   0.81629   0.08385   0.00000  Uiso   1.00
O15    O     0.47786   0.16221   0.52226   0.00000  Uiso   1.00
O16    O     0.64304   0.15144   0.41631   0.00000  Uiso   1.00
O17    O     0.13396   0.50568   0.85585   0.00000  Uiso   1.00
O18    O     0.30480   0.48240   0.75030   0.00000  Uiso   1.00
O19    O     0.98309   0.17202   0.30173   0.00000  Uiso   1.00
O2     O     0.19337   0.17177   0.08234   0.00000  Uiso   1.00
```

```
O20    O     0.15692   0.19922   0.41326   0.00000  Uiso   1.00
O21    O     0.65862   0.51952   0.64013   0.00000  Uiso   1.00
O22    O     0.83758   0.52458   0.74581   0.00000  Uiso   1.00
O23    O     0.31746   0.83901   0.96889   0.00000  Uiso   1.00
O24    O     0.49306   0.86479   0.08014   0.00000  Uiso   1.00
O25    O     0.22325   0.03626   0.31084   0.00000  Uiso   1.00
O26    O     0.04177   0.85537   0.41614   0.00000  Uiso   1.00
O27    O     0.86602   0.35103   0.64030   0.00000  Uiso   1.00
O28    O     0.69126   0.17274   0.74736   0.00000  Uiso   1.00
O29    O     0.55796   0.70140   0.97797   0.00000  Uiso   1.00
O3     O     0.85647   0.32917   0.52029   0.00000  Uiso   1.00
O30    O     0.37151   0.51790   0.08303   0.00000  Uiso   1.00
O31    O     0.85071   0.79883   0.30559   0.00000  Uiso   1.00
O32    O     0.80880   0.97239   0.40808   0.00000  Uiso   1.00
O33    O     0.48859   0.13999   0.64171   0.00000  Uiso   1.00
O34    O     0.48623   0.32064   0.74557   0.00000  Uiso   1.00
O35    O     0.18461   0.46583   0.97214   0.00000  Uiso   1.00
O36    O     0.14438   0.64260   0.07495   0.00000  Uiso   1.00
O37    O     0.82746   0.02418   0.80703   0.00000  Uiso   1.00
O38    O     0.82087   0.83947   0.91206   0.00000  Uiso   1.00
O39    O     0.49744   0.35308   0.13922   0.00000  Uiso   1.00
O4     O     0.85700   0.51105   0.41562   0.00000  Uiso   1.00
O40    O     0.48868   0.17401   0.24640   0.00000  Uiso   1.00
O41    O     0.16306   0.68794   0.47293   0.00000  Uiso   1.00
O42    O     0.15102   0.51300   0.58168   0.00000  Uiso   1.00
O43    O     0.98647   0.80202   0.80825   0.00000  Uiso   1.00
O44    O     0.18160   0.99429   0.91197   0.00000  Uiso   1.00
O45    O     0.65883   0.13400   0.14614   0.00000  Uiso   1.00
O46    O     0.85081   0.33366   0.24740   0.00000  Uiso   1.00
O47    O     0.32142   0.46840   0.47933   0.00000  Uiso   1.00
O48    O     0.50151   0.64936   0.58226   0.00000  Uiso   1.00
O49    O     0.20349   0.17855   0.80515   0.00000  Uiso   1.00
O5     O     0.50768   0.64766   0.85871   0.00000  Uiso   1.00
O50    O     0.04014   0.19866   0.90942   0.00000  Uiso   1.00
O51    O     0.87528   0.51663   0.14115   0.00000  Uiso   1.00
O52    O     0.70063   0.53095   0.24345   0.00000  Uiso   1.00
O53    O     0.54142   0.85191   0.47534   0.00000  Uiso   1.00
O54    O     0.36308   0.86016   0.57997   0.00000  Uiso   1.00
O55    O     0.02237   0.82920   0.68872   0.00000  Uiso   1.00
O56    O     0.84593   0.81943   0.58182   0.00000  Uiso   1.00
O57    O     0.68212   0.17511   0.02761   0.00000  Uiso   1.00
O58    O     0.51962   0.15299   0.92029   0.00000  Uiso   1.00
O59    O     0.35130   0.50931   0.36089   0.00000  Uiso   1.00
O6     O     0.52967   0.82726   0.74933   0.00000  Uiso   1.00
O60    O     0.18927   0.48612   0.25370   0.00000  Uiso   1.00
O61    O     0.80750   0.99069   0.68741   0.00000  Uiso   1.00
O62    O     0.97912   0.16761   0.58039   0.00000  Uiso   1.00
O63    O     0.46684   0.32401   0.01904   0.00000  Uiso   1.00
O64    O     0.64540   0.49547   0.91753   0.00000  Uiso   1.00
O65    O     0.13193   0.65577   0.35274   0.00000  Uiso   1.00
O66    O     0.30852   0.82923   0.25054   0.00000  Uiso   1.00
O67    O     0.18549   0.20844   0.68642   0.00000  Uiso   1.00
O68    O     0.19277   0.03333   0.58142   0.00000  Uiso   1.00
O69    O     0.83255   0.53556   0.02339   0.00000  Uiso   1.00
O7     O     0.02895   0.21821   0.18379   0.00000  Uiso   1.00
O70    O     0.86654   0.36689   0.92514   0.00000  Uiso   1.00
O71    O     0.49618   0.86857   0.35749   0.00000  Uiso   1.00
O72    O     0.53462   0.70676   0.25802   0.00000  Uiso   1.00
O8     O     0.83444   0.02429   0.08029   0.00000  Uiso   1.00
O9     O     0.68669   0.53998   0.52045   0.00000  Uiso   1.00
P73    P     0.29490   0.00208   0.24738   0.00000  Uiso   1.00
P74    P     0.95973   0.34020   0.58058   0.00000  Uiso   1.00
P75    P     0.62814   0.66611   0.91448   0.00000  Uiso   1.00
P76    P     0.01730   0.30117   0.24638   0.00000  Uiso   1.00
P77    P     0.67299   0.62868   0.58127   0.00000  Uiso   1.00
P78    P     0.35057   0.96571   0.91294   0.00000  Uiso   1.00
P79    P     0.71936   0.72335   0.24818   0.00000  Uiso   1.00
P80    P     0.38475   0.05246   0.58132   0.00000  Uiso   1.00
P81    P     0.05291   0.38857   0.91505   0.00000  Uiso   1.00
P82    P     0.71676   0.00390   0.74765   0.00000  Uiso   1.00
P83    P     0.38700   0.34220   0.08124   0.00000  Uiso   1.00
P84    P     0.05377   0.67739   0.41482   0.00000  Uiso   1.00
P85    P     0.00602   0.71752   0.74724   0.00000  Uiso   1.00
P86    P     0.66745   0.05581   0.08202   0.00000  Uiso   1.00
P87    P     0.33280   0.38974   0.41581   0.00000  Uiso   1.00
P88    P     0.29221   0.29402   0.74678   0.00000  Uiso   1.00
P89    P     0.95385   0.62287   0.07985   0.00000  Uiso   1.00
P90    P     0.61962   0.95588   0.41352   0.00000  Uiso   1.00
Ge99   Ge    0.00500   0.01236   0.13985   0.00000  Uiso   1.00
Ge100  Ge    0.67294   0.34392   0.47336   0.00000  Uiso   1.00
Ge101  Ge    0.33777   0.67507   0.80727   0.00000  Uiso   1.00
Ge102  Ge    0.00764   0.00790   0.35506   0.00000  Uiso   1.00
Ge103  Ge    0.67119   0.33762   0.68879   0.00000  Uiso   1.00
Ge104  Ge    0.34164   0.67372   0.02243   0.00000  Uiso   1.00
Ge105  Ge    0.00610   0.00230   0.85392   0.00000  Uiso   1.00
Ge106  Ge    0.67614   0.33741   0.18841   0.00000  Uiso   1.00
Ge107  Ge    0.34221   0.67319   0.52279   0.00000  Uiso   1.00
lagp3_opt.cif
data_LAGP
_audit_creation_date              2014-10-29
_audit_creation_method            'Materials Studio'
_symmetry_space_group_name_H-M    'P1'
_symmetry_Int_Tables_number       1
_symmetry_cell_setting            triclinic
loop_
```

```
_symmetry_equiv_pos_as_xyz
  x,y,z
_cell_length_a                    8.3462
_cell_length_b                    8.3302
_cell_length_c                    21.3493
_cell_angle_alpha                 89.7929
_cell_angle_beta                  90.0539
_cell_angle_gamma                 119.8857
loop_
_atom_site_label
_atom_site_type_symbol
_atom_site_fract_x
_atom_site_fract_y
_atom_site_fract_z
_atom_site_U_iso_or_equiv
_atom_site_adp_type
_atom_site_occupancy
Al109  Al    0.00673   0.00919   0.64400   0.00000  Uiso   1.00
Al110  Al    0.67346   0.34241   0.97731   0.00000  Uiso   1.00
Al111  Al    0.34007   0.67571   0.31076   0.00000  Uiso   1.00
Li91   Li   -0.19350  -0.17678   0.02312   0.00000  Uiso   1.00
Li92   Li    0.47302   0.15643   0.35657   0.00000  Uiso   1.00
Li93   Li    0.13973   0.48984   0.68984   0.00000  Uiso   1.00
Li94   Li    0.02433   0.01673   0.51463   0.00000  Uiso   1.00
Li95   Li    0.69110   0.35001   0.84793   0.00000  Uiso   1.00
Li96   Li    0.35737   0.68282   0.18152   0.00000  Uiso   1.00
Li97   Li   -0.08915   0.55893   0.29475   0.00000  Uiso   1.00
Li98   Li    0.57724  -0.10780   0.62817   0.00000  Uiso   1.00
Li99   Li    0.24465   0.22610   0.96121   0.00000  Uiso   1.00
O1     O     0.16779   0.98025   0.19424   0.00000  Uiso   1.00
O10    O     0.50459   0.36147   0.41444   0.00000  Uiso   1.00
O11    O     0.37239   0.89438   0.84964   0.00000  Uiso   1.00
O12    O     0.17141   0.69503   0.74762   0.00000  Uiso   1.00
O13    O     0.79741   0.85382   0.19079   0.00000  Uiso   1.00
O14    O     0.96327   0.80869   0.08945   0.00000  Uiso   1.00
O15    O     0.46405   0.18728   0.52413   0.00000  Uiso   1.00
O16    O     0.63002   0.14207   0.42290   0.00000  Uiso   1.00
O17    O     0.13079   0.52091   0.85739   0.00000  Uiso   1.00
O18    O     0.29655   0.47540   0.75614   0.00000  Uiso   1.00
O19    O     0.98762   0.17027   0.29896   0.00000  Uiso   1.00
O2     O     0.19810   0.16425   0.08543   0.00000  Uiso   1.00
O20    O     0.16271   0.19617   0.41058   0.00000  Uiso   1.00
O21    O     0.65473   0.50391   0.63225   0.00000  Uiso   1.00
O22    O     0.82953   0.52962   0.74382   0.00000  Uiso   1.00
O23    O     0.32116   0.83699   0.96546   0.00000  Uiso   1.00
O24    O     0.49612   0.86297   0.07705   0.00000  Uiso   1.00
O25    O     0.22850   0.03667   0.31125   0.00000  Uiso   1.00
O26    O     0.03028   0.84538   0.41345   0.00000  Uiso   1.00
O27    O     0.89514   0.36983   0.64460   0.00000  Uiso   1.00
O28    O     0.69691   0.17872   0.74675   0.00000  Uiso   1.00
O29    O     0.56184   0.70308   0.97786   0.00000  Uiso   1.00
O3     O     0.83459   0.31400   0.52759   0.00000  Uiso   1.00
O30    O     0.36348   0.51195   0.07999   0.00000  Uiso   1.00
O31    O     0.85537   0.79761   0.30347   0.00000  Uiso   1.00
O32    O     0.81164   0.98345   0.40280   0.00000  Uiso   1.00
O33    O     0.52196   0.13116   0.63679   0.00000  Uiso   1.00
O34    O     0.47834   0.31694   0.73603   0.00000  Uiso   1.00
O35    O     0.18883   0.46452   0.97001   0.00000  Uiso   1.00
O36    O     0.14489   0.65012   0.06928   0.00000  Uiso   1.00
O37    O     0.84317   0.03186   0.80160   0.00000  Uiso   1.00
O38    O     0.81872   0.84853   0.90924   0.00000  Uiso   1.00
O39    O     0.50990   0.36532   0.13493   0.00000  Uiso   1.00
O4     O     0.86468   0.49765   0.41881   0.00000  Uiso   1.00
O40    O     0.48532   0.18158   0.24257   0.00000  Uiso   1.00
O41    O     0.17632   0.69856   0.46839   0.00000  Uiso   1.00
O42    O     0.15198   0.51515   0.57601   0.00000  Uiso   1.00
O43    O     0.98318   0.79203   0.81218   0.00000  Uiso   1.00
O44    O     0.18239   0.99647   0.91213   0.00000  Uiso   1.00
O45    O     0.64974   0.12536   0.14546   0.00000  Uiso   1.00
O46    O     0.84897   0.32960   0.24551   0.00000  Uiso   1.00
O47    O     0.31644   0.45865   0.47893   0.00000  Uiso   1.00
O48    O     0.51547   0.66294   0.57898   0.00000  Uiso   1.00
O49    O     0.21658   0.16730   0.80444   0.00000  Uiso   1.00
O5     O     0.50135   0.64751   0.86085   0.00000  Uiso   1.00
O50    O     0.04678   0.20781   0.90341   0.00000  Uiso   1.00
O51    O     0.88322   0.50068   0.13781   0.00000  Uiso   1.00
O52    O     0.71306   0.54067   0.23693   0.00000  Uiso   1.00
O53    O     0.54989   0.83416   0.47128   0.00000  Uiso   1.00
O54    O     0.37969   0.87415   0.57032   0.00000  Uiso   1.00
O55    O     0.01811   0.84666   0.69575   0.00000  Uiso   1.00
O56    O     0.85467   0.82343   0.58812   0.00000  Uiso   1.00
O57    O     0.68506   0.17996   0.02903   0.00000  Uiso   1.00
O58    O     0.52150   0.15678   0.92141   0.00000  Uiso   1.00
O59    O     0.35159   0.51328   0.36247   0.00000  Uiso   1.00
O6     O     0.53147   0.83097   0.75199   0.00000  Uiso   1.00
O60    O     0.18821   0.48979   0.25484   0.00000  Uiso   1.00
O61    O     0.79581   0.98748   0.68428   0.00000  Uiso   1.00
O62    O     0.98294   0.16808   0.58421   0.00000  Uiso   1.00
O63    O     0.46239   0.32063   0.01759   0.00000  Uiso   1.00
O64    O     0.64963   0.50132   0.91740   0.00000  Uiso   1.00
O65    O     0.12902   0.65370   0.35104   0.00000  Uiso   1.00
O66    O     0.31610   0.83444   0.25099   0.00000  Uiso   1.00
O67    O     0.15974   0.20760   0.69070   0.00000  Uiso   1.00
O68    O     0.20075   0.03502   0.59378   0.00000  Uiso   1.00
O69    O     0.82632   0.54083   0.02401   0.00000  Uiso   1.00
```

```
O7     O     0.03895   0.22722   0.18312   0.00000  Uiso   1.00
O70    O     0.86745   0.36828   0.92702   0.00000  Uiso   1.00
O71    O     0.49286   0.87405   0.35749   0.00000  Uiso   1.00
O72    O     0.53413   0.70160   0.26050   0.00000  Uiso   1.00
O8     O     0.83801   0.02814   0.08102   0.00000  Uiso   1.00
O9     O     0.70542   0.56082   0.51641   0.00000  Uiso   1.00
P73    P     0.29662   0.00500   0.24743   0.00000  Uiso   1.00
P74    P     0.96335   0.33850   0.58078   0.00000  Uiso   1.00
P75    P     0.63005   0.67184   0.91401   0.00000  Uiso   1.00
P76    P     0.01909   0.30211   0.24559   0.00000  Uiso   1.00
P77    P     0.68573   0.63567   0.57892   0.00000  Uiso   1.00
P78    P     0.35252   0.96900   0.91216   0.00000  Uiso   1.00
P79    P     0.72140   0.72678   0.24734   0.00000  Uiso   1.00
P80    P     0.38803   0.06026   0.58068   0.00000  Uiso   1.00
P81    P     0.05487   0.39373   0.91389   0.00000  Uiso   1.00
P82    P     0.71976   0.00690   0.74651   0.00000  Uiso   1.00
P83    P     0.38637   0.34014   0.07982   0.00000  Uiso   1.00
P84    P     0.05302   0.67347   0.41327   0.00000  Uiso   1.00
P85    P     0.00135   0.72241   0.74833   0.00000  Uiso   1.00
P86    P     0.66804   0.05572   0.08160   0.00000  Uiso   1.00
P87    P     0.33459   0.38896   0.41510   0.00000  Uiso   1.00
P88    P     0.28624   0.28746   0.74612   0.00000  Uiso   1.00
P89    P     0.95283   0.62069   0.07942   0.00000  Uiso   1.00
P90    P     0.61955   0.95406   0.41290   0.00000  Uiso   1.00
Ge100  Ge    0.00509   0.01490   0.14150   0.00000  Uiso   1.00
Ge101  Ge    0.67172   0.34841   0.47485   0.00000  Uiso   1.00
Ge102  Ge    0.33849   0.68187   0.80812   0.00000  Uiso   1.00
Ge103  Ge    0.00923   0.00662   0.35385   0.00000  Uiso   1.00
Ge104  Ge    0.67597   0.34009   0.68716   0.00000  Uiso   1.00
Ge105  Ge    0.34263   0.67336   0.02044   0.00000  Uiso   1.00
Ge106  Ge    0.01056   0.00242   0.85238   0.00000  Uiso   1.00
Ge107  Ge    0.67715   0.33566   0.18570   0.00000  Uiso   1.00
Ge108  Ge    0.34377   0.66905   0.51912   0.00000  Uiso   1.00
latp1_opt.cif
data_LATP_1AL
_audit_creation_date              2014-10-29
_audit_creation_method            'Materials Studio'
_symmetry_space_group_name_H-M    'P1'
_symmetry_Int_Tables_number       1
_symmetry_cell_setting            triclinic
loop_
_symmetry_equiv_pos_as_xyz
  x,y,z
_cell_length_a                    8.5807
_cell_length_b                    8.5720
_cell_length_c                    21.7420
_cell_angle_alpha                 89.6057
_cell_angle_beta                  90.4079
_cell_angle_gamma                 119.8156
loop_
_atom_site_label
_atom_site_type_symbol
_atom_site_fract_x
_atom_site_fract_y
_atom_site_fract_z
_atom_site_U_iso_or_equiv
_atom_site_adp_type
_atom_site_occupancy
Ti109  Ti    0.00429   0.00256   0.64442   0.00000  Uiso   1.00
Ti110  Ti    0.66731   0.33801   0.97451   0.00000  Uiso   1.00
Al111  Al    0.33324   0.66714   0.31156   0.00000  Uiso   1.00
Li91   Li   -0.14466  -0.12281   0.01487   0.00000  Uiso   1.00
Li92   Li    0.45070   0.15409   0.36057   0.00000  Uiso   1.00
Li93   Li    0.45686   0.77976   0.65369   0.00000  Uiso   1.00
Li94   Li    0.16188   0.18699   0.51953   0.00000  Uiso   1.00
Li95   Li    0.53415   0.17037   0.81715   0.00000  Uiso   1.00
Li96   Li    0.35492   0.67694   0.18672   0.00000  Uiso   1.00
Li97   Li   -0.08217   0.55654   0.28885   0.00000  Uiso   1.00
O1     O     0.16323   0.97280   0.19813   0.00000  Uiso   1.00
O10    O     0.49546   0.36165   0.41563   0.00000  Uiso   1.00
O11    O     0.36908   0.88679   0.85215   0.00000  Uiso   1.00
O12    O     0.16751   0.69421   0.75547   0.00000  Uiso   1.00
O13    O     0.79481   0.83499   0.19238   0.00000  Uiso   1.00
O14    O     0.96545   0.80672   0.08723   0.00000  Uiso   1.00
O15    O     0.46884   0.17023   0.52519   0.00000  Uiso   1.00
O16    O     0.61853   0.13190   0.41975   0.00000  Uiso   1.00
O17    O     0.14332   0.50928   0.86190   0.00000  Uiso   1.00
O18    O     0.31296   0.46772   0.76114   0.00000  Uiso   1.00
O19    O     0.99967   0.17932   0.30268   0.00000  Uiso   1.00
O2     O     0.20051   0.16710   0.09246   0.00000  Uiso   1.00
O20    O     0.16635   0.19878   0.41195   0.00000  Uiso   1.00
O21    O     0.63994   0.49789   0.63688   0.00000  Uiso   1.00
O22    O     0.83893   0.53416   0.73872   0.00000  Uiso   1.00
O23    O     0.32824   0.83827   0.96737   0.00000  Uiso   1.00
O24    O     0.49747   0.86402   0.08078   0.00000  Uiso   1.00
O25    O     0.22514   0.02666   0.31119   0.00000  Uiso   1.00
O26    O     0.03022   0.83504   0.40853   0.00000  Uiso   1.00
O27    O     0.86629   0.35099   0.63948   0.00000  Uiso   1.00
O28    O     0.69118   0.16329   0.74816   0.00000  Uiso   1.00
O29    O     0.54807   0.69534   0.98017   0.00000  Uiso   1.00
O3     O     0.83324   0.31341   0.52600   0.00000  Uiso   1.00
O30    O     0.36112   0.49996   0.08079   0.00000  Uiso   1.00
O31    O     0.84032   0.80733   0.30382   0.00000  Uiso   1.00
O32    O     0.79287   0.97620   0.40088   0.00000  Uiso   1.00
O33    O     0.49860   0.13425   0.63938   0.00000  Uiso   1.00
```

```
O34    O     0.47336   0.29621   0.74562   0.00000  Uiso   1.00
O35    O     0.17262   0.47017   0.97340   0.00000  Uiso   1.00
O36    O     0.13841   0.64657   0.07419   0.00000  Uiso   1.00
O37    O     0.82307   0.00732   0.80518   0.00000  Uiso   1.00
O38    O     0.81214   0.83756   0.91682   0.00000  Uiso   1.00
O39    O     0.50244   0.35567   0.13858   0.00000  Uiso   1.00
O4     O     0.86385   0.50514   0.42220   0.00000  Uiso   1.00
O40    O     0.47783   0.15452   0.24083   0.00000  Uiso   1.00
O41    O     0.17503   0.68888   0.46380   0.00000  Uiso   1.00
O42    O     0.13440   0.51031   0.57246   0.00000  Uiso   1.00
O43    O     0.97370   0.79403   0.80905   0.00000  Uiso   1.00
O44    O     0.16934   0.96713   0.91052   0.00000  Uiso   1.00
O45    O     0.66269   0.14334   0.14062   0.00000  Uiso   1.00
O46    O     0.85767   0.33007   0.24666   0.00000  Uiso   1.00
O47    O     0.31389   0.45633   0.47864   0.00000  Uiso   1.00
O48    O     0.50637   0.65297   0.58088   0.00000  Uiso   1.00
O49    O     0.19767   0.16008   0.80531   0.00000  Uiso   1.00
O5     O     0.51069   0.66505   0.86457   0.00000  Uiso   1.00
O50    O     0.03478   0.20299   0.90557   0.00000  Uiso   1.00
O51    O     0.87065   0.51214   0.14013   0.00000  Uiso   1.00
O52    O     0.70379   0.53727   0.24300   0.00000  Uiso   1.00
O53    O     0.53178   0.83452   0.46971   0.00000  Uiso   1.00
O54    O     0.36628   0.86810   0.57557   0.00000  Uiso   1.00
O55    O     0.03107   0.84189   0.69658   0.00000  Uiso   1.00
O56    O     0.83694   0.80315   0.59268   0.00000  Uiso   1.00
O57    O     0.67830   0.15768   0.02605   0.00000  Uiso   1.00
O58    O     0.50139   0.14544   0.91766   0.00000  Uiso   1.00
O59    O     0.34320   0.50918   0.36370   0.00000  Uiso   1.00
O6     O     0.53001   0.82951   0.74747   0.00000  Uiso   1.00
O60    O     0.18840   0.48390   0.25526   0.00000  Uiso   1.00
O61    O     0.80790   0.99294   0.68959   0.00000  Uiso   1.00
O62    O     0.98873   0.18022   0.58325   0.00000  Uiso   1.00
O63    O     0.46725   0.31367   0.02512   0.00000  Uiso   1.00
O64    O     0.63799   0.50400   0.92166   0.00000  Uiso   1.00
O65    O     0.11755   0.63789   0.35033   0.00000  Uiso   1.00
O66    O     0.30271   0.81987   0.25323   0.00000  Uiso   1.00
O67    O     0.17938   0.20487   0.69240   0.00000  Uiso   1.00
O68    O     0.20142   0.03926   0.58724   0.00000  Uiso   1.00
O69    O     0.83466   0.53225   0.02665   0.00000  Uiso   1.00
O7     O     0.02822   0.20996   0.18979   0.00000  Uiso   1.00
O70    O     0.86999   0.36591   0.92619   0.00000  Uiso   1.00
O71    O     0.48064   0.85909   0.35759   0.00000  Uiso   1.00
O72    O     0.52743   0.69659   0.26005   0.00000  Uiso   1.00
O8     O     0.83144   0.02314   0.08333   0.00000  Uiso   1.00
O9     O     0.68381   0.53791   0.52379   0.00000  Uiso   1.00
P73    P     0.29071  -0.00883   0.25054   0.00000  Uiso   1.00
P74    P     0.95536   0.33906   0.58115   0.00000  Uiso   1.00
P75    P     0.62299   0.67291   0.91962   0.00000  Uiso   1.00
P76    P     0.02032   0.30148   0.24847   0.00000  Uiso   1.00
P77    P     0.66841   0.62352   0.58289   0.00000  Uiso   1.00
P78    P     0.34033   0.95512   0.91339   0.00000  Uiso   1.00
P79    P     0.71148   0.71959   0.24933   0.00000  Uiso   1.00
P80    P     0.38645   0.05520   0.58201   0.00000  Uiso   1.00
P81    P     0.05378   0.38592   0.91727   0.00000  Uiso   1.00
P82    P     0.71597  -0.00185   0.74892   0.00000  Uiso   1.00
P83    P     0.38272   0.33390   0.08282   0.00000  Uiso   1.00
P84    P     0.04923   0.66827   0.41081   0.00000  Uiso   1.00
P85    P     0.00316   0.71675   0.75125   0.00000  Uiso   1.00
P86    P     0.66684   0.04800   0.08228   0.00000  Uiso   1.00
P87    P     0.33127   0.38845   0.41509   0.00000  Uiso   1.00
P88    P     0.28888   0.28194   0.75069   0.00000  Uiso   1.00
P89    P     0.95160   0.62229   0.08248   0.00000  Uiso   1.00
P90    P     0.60479   0.94800   0.41126   0.00000  Uiso   1.00
Ti100  Ti    0.00445   0.00767   0.14466   0.00000  Uiso   1.00
Ti101  Ti    0.67169   0.34528   0.47488   0.00000  Uiso   1.00
Ti102  Ti    0.33201   0.66357   0.80935   0.00000  Uiso   1.00
Ti103  Ti   -0.00502   0.00229   0.35707   0.00000  Uiso   1.00
Ti104  Ti    0.67530   0.33887   0.68904   0.00000  Uiso   1.00
Ti105  Ti    0.33443   0.66658   0.02489   0.00000  Uiso   1.00
Ti106  Ti    0.00409  -0.00201   0.85899   0.00000  Uiso   1.00
Ti107  Ti    0.66879   0.33098   0.18811   0.00000  Uiso   1.00
Ti108  Ti    0.33526   0.67165   0.51976   0.00000  Uiso   1.00
latp2_opt.cif
data_latp2
_audit_creation_date              2014-10-29
_audit_creation_method            'Materials Studio'
_symmetry_space_group_name_H-M    'P1'
_symmetry_Int_Tables_number       1
_symmetry_cell_setting            triclinic
loop_
_symmetry_equiv_pos_as_xyz
  x,y,z
_cell_length_a                    8.5290
_cell_length_b                    8.5194
_cell_length_c                    22.0258
_cell_angle_alpha                 89.7009
_cell_angle_beta                  90.5005
_cell_angle_gamma                 119.8781
loop_
_atom_site_label
_atom_site_type_symbol
_atom_site_fract_x
_atom_site_fract_y
_atom_site_fract_z
_atom_site_U_iso_or_equiv
```

```
_atom_site_adp_type
_atom_site_occupancy
Ti108  Ti    0.00124   0.00482   0.64452   0.00000  Uiso   1.00
Al109  Al    0.67097   0.34016   0.98108   0.00000  Uiso   1.00
Al110  Al    0.33588   0.67103   0.31288   0.00000  Uiso   1.00
Li91   Li   -0.22226  -0.19148   0.03502   0.00000  Uiso   1.00
Li92   Li    0.44190   0.14036   0.36610   0.00000  Uiso   1.00
Li93   Li    0.13949   0.49761   0.69226   0.00000  Uiso   1.00
Li94   Li    0.02800   0.02612   0.51898   0.00000  Uiso   1.00
Li95   Li    0.69342   0.34839   0.86102   0.00000  Uiso   1.00
Li96   Li    0.36141   0.68306   0.19295   0.00000  Uiso   1.00
Li97   Li   -0.08399   0.56506   0.29012   0.00000  Uiso   1.00
Li98   Li    0.25031   0.22951   0.95818   0.00000  Uiso   1.00
O1     O     0.16438   0.97796   0.19731   0.00000  Uiso   1.00
O10    O     0.49885   0.36176   0.41458   0.00000  Uiso   1.00
O11    O     0.37260   0.88878   0.85585   0.00000  Uiso   1.00
O12    O     0.16956   0.69275   0.75168   0.00000  Uiso   1.00
O13    O     0.79807   0.84887   0.19181   0.00000  Uiso   1.00
O14    O     0.96057   0.80298   0.08963   0.00000  Uiso   1.00
O15    O     0.46907   0.18225   0.52256   0.00000  Uiso   1.00
O16    O     0.62704   0.13521   0.42034   0.00000  Uiso   1.00
O17    O     0.13287   0.51556   0.86051   0.00000  Uiso   1.00
O18    O     0.30026   0.47452   0.75762   0.00000  Uiso   1.00
O19    O     0.99945   0.18010   0.29891   0.00000  Uiso   1.00
O2     O     0.20487   0.17334   0.09045   0.00000  Uiso   1.00
O20    O     0.16883   0.20324   0.40706   0.00000  Uiso   1.00
O21    O     0.66570   0.51288   0.63128   0.00000  Uiso   1.00
O22    O     0.83777   0.53667   0.74044   0.00000  Uiso   1.00
O23    O     0.33214   0.84688   0.96688   0.00000  Uiso   1.00
O24    O     0.50469   0.87300   0.07447   0.00000  Uiso   1.00
O25    O     0.23238   0.03568   0.30906   0.00000  Uiso   1.00
O26    O     0.03628   0.84025   0.40732   0.00000  Uiso   1.00
O27    O     0.89438   0.36697   0.64185   0.00000  Uiso   1.00
O28    O     0.69724   0.16969   0.74054   0.00000  Uiso   1.00
O29    O     0.56932   0.70494   0.97768   0.00000  Uiso   1.00
O3     O     0.83309   0.31697   0.52832   0.00000  Uiso   1.00
O30    O     0.37009   0.50799   0.07594   0.00000  Uiso   1.00
O31    O     0.85144   0.80533   0.30068   0.00000  Uiso   1.00
O32    O     0.79961   0.97805   0.39755   0.00000  Uiso   1.00
O33    O     0.50944   0.14671   0.63321   0.00000  Uiso   1.00
O34    O     0.47502   0.31814   0.73492   0.00000  Uiso   1.00
O35    O     0.18528   0.47261   0.96953   0.00000  Uiso   1.00
O36    O     0.13441   0.64774   0.06584   0.00000  Uiso   1.00
O37    O     0.84781   0.03847   0.80149   0.00000  Uiso   1.00
O38    O     0.81505   0.83096   0.90390   0.00000  Uiso   1.00
O39    O     0.51700   0.36374   0.13095   0.00000  Uiso   1.00
O4     O     0.87070   0.50611   0.42071   0.00000  Uiso   1.00
O40    O     0.48013   0.16482   0.23649   0.00000  Uiso   1.00
O41    O     0.18208   0.69519   0.46259   0.00000  Uiso   1.00
O42    O     0.14505   0.50582   0.57166   0.00000  Uiso   1.00
O43    O     0.99031   0.80711   0.80632   0.00000  Uiso   1.00
O44    O     0.19514   1.00388   0.91133   0.00000  Uiso   1.00
O45    O     0.65980   0.13811   0.14121   0.00000  Uiso   1.00
O46    O     0.86197   0.33787   0.24450   0.00000  Uiso   1.00
O47    O     0.32056   0.46947   0.47322   0.00000  Uiso   1.00
O48    O     0.51391   0.65043   0.57397   0.00000  Uiso   1.00
O49    O     0.21270   0.17232   0.80280   0.00000  Uiso   1.00
O5     O     0.49745   0.64177   0.86656   0.00000  Uiso   1.00
O50    O     0.04317   0.21132   0.90499   0.00000  Uiso   1.00
O51    O     0.88047   0.50445   0.13623   0.00000  Uiso   1.00
O52    O     0.70695   0.54404   0.23624   0.00000  Uiso   1.00
O53    O     0.54663   0.83770   0.46813   0.00000  Uiso   1.00
O54    O     0.36930   0.87725   0.56671   0.00000  Uiso   1.00
O55    O     0.03144   0.84126   0.69416   0.00000  Uiso   1.00
O56    O     0.84926   0.81720   0.58431   0.00000  Uiso   1.00
O57    O     0.68902   0.18184   0.02889   0.00000  Uiso   1.00
O58    O     0.52679   0.15779   0.92359   0.00000  Uiso   1.00
O59    O     0.35290   0.51201   0.36058   0.00000  Uiso   1.00
O6     O     0.54094   0.83716   0.76043   0.00000  Uiso   1.00
O60    O     0.19401   0.49045   0.25518   0.00000  Uiso   1.00
O61    O     0.80112   0.97611   0.69045   0.00000  Uiso   1.00
O62    O     0.97753   0.16978   0.58292   0.00000  Uiso   1.00
O63    O     0.45726   0.30988   0.01843   0.00000  Uiso   1.00
O64    O     0.64482   0.49561   0.92167   0.00000  Uiso   1.00
O65    O     0.12427   0.64320   0.34965   0.00000  Uiso   1.00
O66    O     0.30896   0.82840   0.25253   0.00000  Uiso   1.00
O67    O     0.16642   0.21282   0.69205   0.00000  Uiso   1.00
O68    O     0.20219   0.04014   0.58804   0.00000  Uiso   1.00
O69    O     0.81828   0.53428   0.02609   0.00000  Uiso   1.00
O7     O     0.03675   0.22147   0.18773   0.00000  Uiso   1.00
O70    O     0.86908   0.37139   0.92858   0.00000  Uiso   1.00
O71    O     0.48328   0.86506   0.35779   0.00000  Uiso   1.00
O72    O     0.53512   0.70561   0.26032   0.00000  Uiso   1.00
O8     O     0.83529   0.02980   0.08255   0.00000  Uiso   1.00
O9     O     0.70090   0.54940   0.51873   0.00000  Uiso   1.00
P73    P     0.29540  -0.00078   0.24879   0.00000  Uiso   1.00
P74    P     0.96008   0.33990   0.58034   0.00000  Uiso   1.00
P75    P     0.63046   0.66612   0.91755   0.00000  Uiso   1.00
P76    P     0.02493   0.30776   0.24682   0.00000  Uiso   1.00
P77    P     0.68034   0.63118   0.57731   0.00000  Uiso   1.00
P78    P     0.35848   0.97460   0.91463   0.00000  Uiso   1.00
P79    P     0.71815   0.72744   0.24731   0.00000  Uiso   1.00
P80    P     0.38842   0.06104   0.57745   0.00000  Uiso   1.00
P81    P     0.05286   0.39458   0.91596   0.00000  Uiso   1.00
P82    P     0.72014   0.00536   0.74791   0.00000  Uiso   1.00
```

```
P83    P     0.38895   0.33882   0.07886   0.00000  Uiso   1.00
P84    P     0.05686   0.67171   0.41040   0.00000  Uiso   1.00
P85    P     0.00599   0.71965   0.74875   0.00000  Uiso   1.00
P86    P     0.67360   0.06141   0.08057   0.00000  Uiso   1.00
P87    P     0.33628   0.39370   0.41317   0.00000  Uiso   1.00
P88    P     0.28925   0.29241   0.74785   0.00000  Uiso   1.00
P89    P     0.94558   0.61852   0.07936   0.00000  Uiso   1.00
P90    P     0.61120   0.94987   0.41080   0.00000  Uiso   1.00
Ti99   Ti    0.00826   0.01533   0.14409   0.00000  Uiso   1.00
Ti100  Ti    0.67439   0.34874   0.47315   0.00000  Uiso   1.00
Ti101  Ti    0.34375   0.68118   0.81280   0.00000  Uiso   1.00
Ti102  Ti    0.00361   0.00885   0.35447   0.00000  Uiso   1.00
Ti103  Ti    0.66863   0.33800   0.68529   0.00000  Uiso   1.00
Ti104  Ti    0.33725   0.67733   0.02296   0.00000  Uiso   1.00
Ti105  Ti    0.00744   0.00452   0.85346   0.00000  Uiso   1.00
Ti106  Ti    0.67548   0.33543   0.18593   0.00000  Uiso   1.00
Ti107  Ti    0.34649   0.67467   0.51984   0.00000  Uiso   1.00
latp3_opt.cif
data_LATP_1\(2)
_audit_creation_date              2014-10-29
_audit_creation_method            'Materials Studio'
_symmetry_space_group_name_H-M    'P1'
_symmetry_Int_Tables_number       1
_symmetry_cell_setting            triclinic
loop_
_symmetry_equiv_pos_as_xyz
  x,y,z
_cell_length_a                    8.5002
_cell_length_b                    8.4898
_cell_length_c                    21.9699
_cell_angle_alpha                 89.7149
_cell_angle_beta                  90.5159
_cell_angle_gamma                 119.8224
loop_
_atom_site_label
_atom_site_type_symbol
_atom_site_fract_x
_atom_site_fract_y
_atom_site_fract_z
_atom_site_U_iso_or_equiv
_atom_site_adp_type
_atom_site_occupancy
Al109  Al    0.00461   0.00696   0.64618   0.00000  Uiso   1.00
Al110  Al    0.67115   0.34022   0.97952   0.00000  Uiso   1.00
Al111  Al    0.33787   0.67341   0.31293   0.00000  Uiso   1.00
Li91   Li   -0.21938  -0.19040   0.03217   0.00000  Uiso   1.00
Li92   Li    0.44710   0.14292   0.36573   0.00000  Uiso   1.00
Li93   Li    0.11320   0.47566   0.69911   0.00000  Uiso   1.00
Li94   Li    0.02815   0.01661   0.52583   0.00000  Uiso   1.00
Li95   Li    0.69481   0.34992   0.85923   0.00000  Uiso   1.00
Li96   Li    0.36127   0.68299   0.19258   0.00000  Uiso   1.00
Li97   Li   -0.08167   0.56716   0.28931   0.00000  Uiso   1.00
Li98   Li    0.58496  -0.09960   0.62264   0.00000  Uiso   1.00
Li99   Li    0.25147   0.23384   0.95597   0.00000  Uiso   1.00
O1     O     0.16326   0.97633   0.19767   0.00000  Uiso   1.00
O10    O     0.50260   0.36294   0.41529   0.00000  Uiso   1.00
O11    O     0.37123   0.88909   0.85405   0.00000  Uiso   1.00
O12    O     0.16894   0.69603   0.74853   0.00000  Uiso   1.00
O13    O     0.79716   0.84969   0.19190   0.00000  Uiso   1.00
O14    O     0.96062   0.80325   0.08922   0.00000  Uiso   1.00
O15    O     0.46422   0.18318   0.52532   0.00000  Uiso   1.00
O16    O     0.62710   0.13642   0.42268   0.00000  Uiso   1.00
O17    O     0.13074   0.51643   0.85854   0.00000  Uiso   1.00
O18    O     0.29392   0.46991   0.75598   0.00000  Uiso   1.00
O19    O     0.99950   0.17990   0.29899   0.00000  Uiso   1.00
O2     O     0.20489   0.17151   0.09007   0.00000  Uiso   1.00
O20    O     0.17104   0.20578   0.40723   0.00000  Uiso   1.00
O21    O     0.66617   0.51339   0.63230   0.00000  Uiso   1.00
O22    O     0.83753   0.53917   0.74046   0.00000  Uiso   1.00
O23    O     0.33255   0.84662   0.96555   0.00000  Uiso   1.00
O24    O     0.50439   0.87249   0.07370   0.00000  Uiso   1.00
O25    O     0.23314   0.03659   0.30976   0.00000  Uiso   1.00
O26    O     0.03557   0.84019   0.40863   0.00000  Uiso   1.00
O27    O     0.89958   0.36994   0.64309   0.00000  Uiso   1.00
O28    O     0.70222   0.17368   0.74190   0.00000  Uiso   1.00
O29    O     0.56616   0.70323   0.97634   0.00000  Uiso   1.00
O3     O     0.83058   0.31008   0.53096   0.00000  Uiso   1.00
O30    O     0.36872   0.50686   0.07517   0.00000  Uiso   1.00
O31    O     0.85250   0.80576   0.30108   0.00000  Uiso   1.00
O32    O     0.80044   0.98006   0.39809   0.00000  Uiso   1.00
O33    O     0.51911   0.13904   0.63447   0.00000  Uiso   1.00
O34    O     0.46675   0.31317   0.73131   0.00000  Uiso   1.00
O35    O     0.18597   0.47244   0.96766   0.00000  Uiso   1.00
O36    O     0.13378   0.64675   0.06464   0.00000  Uiso   1.00
O37    O     0.85029   0.03003   0.79697   0.00000  Uiso   1.00
O38    O     0.81387   0.83188   0.90352   0.00000  Uiso   1.00
O39    O     0.51723   0.36355   0.13022   0.00000  Uiso   1.00
O4     O     0.87180   0.50486   0.42352   0.00000  Uiso   1.00
O40    O     0.48024   0.16500   0.23661   0.00000  Uiso   1.00
O41    O     0.18408   0.69685   0.46369   0.00000  Uiso   1.00
O42    O     0.14735   0.49841   0.57024   0.00000  Uiso   1.00
O43    O     0.99281   0.80430   0.80741   0.00000  Uiso   1.00
O44    O     0.19565   1.00535   0.91098   0.00000  Uiso   1.00
O45    O     0.65980   0.13770   0.14067   0.00000  Uiso   1.00
O46    O     0.86212   0.33829   0.24442   0.00000  Uiso   1.00
```

```
O47    O     0.32646   0.47087   0.47424   0.00000  Uiso   1.00
O48    O     0.52923   0.67202   0.57773   0.00000  Uiso   1.00
O49    O     0.21358   0.17078   0.80232   0.00000  Uiso   1.00
O5     O     0.49720   0.64353   0.86418   0.00000  Uiso   1.00
O50    O     0.03974   0.21093   0.90294   0.00000  Uiso   1.00
O51    O     0.88057   0.50423   0.13566   0.00000  Uiso   1.00
O52    O     0.70671   0.54431   0.23628   0.00000  Uiso   1.00
O53    O     0.54726   0.83746   0.46909   0.00000  Uiso   1.00
O54    O     0.37337   0.87766   0.56970   0.00000  Uiso   1.00
O55    O     0.02327   0.84908   0.69471   0.00000  Uiso   1.00
O56    O     0.86193   0.82439   0.58887   0.00000  Uiso   1.00
O57    O     0.68972   0.18221   0.02797   0.00000  Uiso   1.00
O58    O     0.52837   0.15765   0.92218   0.00000  Uiso   1.00
O59    O     0.35631   0.51554   0.36152   0.00000  Uiso   1.00
O6     O     0.53802   0.83834   0.75670   0.00000  Uiso   1.00
O60    O     0.19491   0.49087   0.25561   0.00000  Uiso   1.00
O61    O     0.78966   0.97525   0.68405   0.00000  Uiso   1.00
O62    O     0.97545   0.16126   0.58677   0.00000  Uiso   1.00
O63    O     0.45648   0.30875   0.01732   0.00000  Uiso   1.00
O64    O     0.64217   0.49473   0.91998   0.00000  Uiso   1.00
O65    O     0.12328   0.64212   0.35076   0.00000  Uiso   1.00
O66    O     0.30881   0.82791   0.25336   0.00000  Uiso   1.00
O67    O     0.14962   0.20179   0.69198   0.00000  Uiso   1.00
O68    O     0.20220   0.04071   0.59441   0.00000  Uiso   1.00
O69    O     0.81660   0.53495   0.02528   0.00000  Uiso   1.00
O7     O     0.03793   0.22228   0.18747   0.00000  Uiso   1.00
O70    O     0.86885   0.37410   0.92770   0.00000  Uiso   1.00
O71    O     0.48326   0.86813   0.35870   0.00000  Uiso   1.00
O72    O     0.53537   0.70705   0.26112   0.00000  Uiso   1.00
O8     O     0.83589   0.02952   0.08175   0.00000  Uiso   1.00
O9     O     0.70483   0.55578   0.52079   0.00000  Uiso   1.00
P73    P     0.29527  -0.00052   0.24911   0.00000  Uiso   1.00
P74    P     0.96213   0.33296   0.58252   0.00000  Uiso   1.00
P75    P     0.62867   0.66632   0.91574   0.00000  Uiso   1.00
P76    P     0.02544   0.30806   0.24667   0.00000  Uiso   1.00
P77    P     0.69242   0.64161   0.57999   0.00000  Uiso   1.00
P78    P     0.35883   0.97491   0.91325   0.00000  Uiso   1.00
P79    P     0.71846   0.72826   0.24754   0.00000  Uiso   1.00
P80    P     0.38521   0.06167   0.58091   0.00000  Uiso   1.00
P81    P     0.05184   0.39505   0.91416   0.00000  Uiso   1.00
P82    P     0.72185   0.00465   0.74460   0.00000  Uiso   1.00
P83    P     0.38859   0.33799   0.07788   0.00000  Uiso   1.00
P84    P     0.05551   0.67136   0.41135   0.00000  Uiso   1.00
P85    P     0.00666   0.72775   0.74645   0.00000  Uiso   1.00
P86    P     0.67349   0.06107   0.07969   0.00000  Uiso   1.00
P87    P     0.34011   0.39436   0.41325   0.00000  Uiso   1.00
P88    P     0.27844   0.28485   0.74519   0.00000  Uiso   1.00
P89    P     0.94533   0.61821   0.07848   0.00000  Uiso   1.00
P90    P     0.61195   0.95144   0.41192   0.00000  Uiso   1.00
Ti100  Ti    0.00786   0.01466   0.14335   0.00000  Uiso   1.00
Ti101  Ti    0.67478   0.34814   0.47674   0.00000  Uiso   1.00
Ti102  Ti    0.34122   0.68146   0.80997   0.00000  Uiso   1.00
Ti103  Ti    0.00444   0.01016   0.35478   0.00000  Uiso   1.00
Ti104  Ti    0.67096   0.34346   0.68808   0.00000  Uiso   1.00
Ti105  Ti    0.33779   0.67695   0.02132   0.00000  Uiso   1.00
Ti106  Ti    0.00872   0.00228   0.85196   0.00000  Uiso   1.00
Ti107  Ti    0.67549   0.33556   0.18522   0.00000  Uiso   1.00
Ti108  Ti    0.34237   0.66891   0.51874   0.00000  Uiso   1.00
```

LMP STRUCTURES

------------------------------------------------------------

```
LiGe2PO4_3_hex.cif
data_LiGe2PO4_3_hex
_audit_creation_date              2014-09-09
_audit_creation_method            'Materials Studio'
_symmetry_space_group_name_H-M    'R-3C'
_symmetry_Int_Tables_number       167
_symmetry_cell_setting            trigonal
loop_
_symmetry_equiv_pos_as_xyz
  x,y,z
  -y,x-y,z
  -x+y,-x,z
  y,x,-z+1/2
  x-y,-y,-z+1/2
  -x,-x+y,-z+1/2
  -x,-y,-z
  y,-x+y,-z
  x-y,x,-z
  -y,-x,z+1/2
  -x+y,y,z+1/2
  x,x-y,z+1/2
  x+2/3,y+1/3,z+1/3
  -y+2/3,x-y+1/3,z+1/3
  -x+y+2/3,-x+1/3,z+1/3
  y+2/3,x+1/3,-z+5/6
  x-y+2/3,-y+1/3,-z+5/6
  -x+2/3,-x+y+1/3,-z+5/6
  -x+2/3,-y+1/3,-z+1/3
  y+2/3,-x+y+1/3,-z+1/3
  x-y+2/3,x+1/3,-z+1/3
  -y+2/3,-x+1/3,z+5/6
  -x+y+2/3,y+1/3,z+5/6
  x+2/3,x-y+1/3,z+5/6
  x+1/3,y+2/3,z+2/3
```

```
  -y+1/3,x-y+2/3,z+2/3
  -x+y+1/3,-x+2/3,z+2/3
  y+1/3,x+2/3,-z+1/6
  x-y+1/3,-y+2/3,-z+1/6
  -x+1/3,-x+y+2/3,-z+1/6
  -x+1/3,-y+2/3,-z+2/3
  y+1/3,-x+y+2/3,-z+2/3
  x-y+1/3,x+2/3,-z+2/3
  -y+1/3,-x+2/3,z+1/6
  -x+y+1/3,y+2/3,z+1/6
  x+1/3,x-y+2/3,z+1/6
_cell_length_a                    8.3815
_cell_length_b                    8.3815
_cell_length_c                    20.9744
_cell_angle_alpha                 90.0000
_cell_angle_beta                  90.0000
_cell_angle_gamma                 120.0000
loop_
_atom_site_label
_atom_site_type_symbol
_atom_site_fract_x
_atom_site_fract_y
_atom_site_fract_z
_atom_site_U_iso_or_equiv
_atom_site_adp_type
_atom_site_occupancy
O1     O     0.18058  -0.01784   0.19010   0.01119  Uani   1.00
O2     O     0.18665   0.16169   0.08406   0.00853  Uani   1.00
Li1    Li    0.00000   0.00000   0.00000   0.03181  Uani   1.00
Ge1    Ge    0.00000   0.00000   0.14267   0.00335  Uani   1.00
P1     P     0.28849   0.00000   0.25000   0.00386  Uani   1.00
LiGe2PO4_3_LDA_hex.cif
data_LiGe2PO4_3_LDA_hex
_audit_creation_date              2014-09-09
_audit_creation_method            'Materials Studio'
_symmetry_space_group_name_H-M    'R-3C'
_symmetry_Int_Tables_number       167
_symmetry_cell_setting            trigonal
loop_
_symmetry_equiv_pos_as_xyz
  x,y,z
  -y,x-y,z
  -x+y,-x,z
  y,x,-z+1/2
  x-y,-y,-z+1/2
  -x,-x+y,-z+1/2
  -x,-y,-z
  y,-x+y,-z
  x-y,x,-z
  -y,-x,z+1/2
  -x+y,y,z+1/2
  x,x-y,z+1/2
  x+2/3,y+1/3,z+1/3
  -y+2/3,x-y+1/3,z+1/3
  -x+y+2/3,-x+1/3,z+1/3
  y+2/3,x+1/3,-z+5/6
  x-y+2/3,-y+1/3,-z+5/6
  -x+2/3,-x+y+1/3,-z+5/6
  -x+2/3,-y+1/3,-z+1/3
  y+2/3,-x+y+1/3,-z+1/3
  x-y+2/3,x+1/3,-z+1/3
  -y+2/3,-x+1/3,z+5/6
  -x+y+2/3,y+1/3,z+5/6
  x+2/3,x-y+1/3,z+5/6
  x+1/3,y+2/3,z+2/3
  -y+1/3,x-y+2/3,z+2/3
  -x+y+1/3,-x+2/3,z+2/3
  y+1/3,x+2/3,-z+1/6
  x-y+1/3,-y+2/3,-z+1/6
  -x+1/3,-x+y+2/3,-z+1/6
  -x+1/3,-y+2/3,-z+2/3
  y+1/3,-x+y+2/3,-z+2/3
  x-y+1/3,x+2/3,-z+2/3
  -y+1/3,-x+2/3,z+1/6
  -x+y+1/3,y+2/3,z+1/6
  x+1/3,x-y+2/3,z+1/6
_cell_length_a                    8.1864
_cell_length_b                    8.1864
_cell_length_c                    20.5083
_cell_angle_alpha                 90.0000
_cell_angle_beta                  90.0000
_cell_angle_gamma                 120.0000
loop_
_atom_site_label
_atom_site_type_symbol
_atom_site_fract_x
_atom_site_fract_y
_atom_site_fract_z
_atom_site_U_iso_or_equiv
_atom_site_adp_type
_atom_site_occupancy
O1     O     0.17768  -0.01756   0.18988   0.01068  Uani   1.00
O2     O     0.18650   0.16038   0.08481   0.00815  Uani   1.00
Li1    Li    0.00000   0.00000   0.00000   0.03034  Uani   1.00
Ge1    Ge    0.00000   0.00000   0.14205   0.00320  Uani   1.00
```

```
P1     P     0.28721   0.00000   0.25000   0.00369  Uani   1.00
LiSi2PO4_3_hex.cif
data_LiSi2(PO4)_3_hex
_audit_creation_date              2014-09-09
_audit_creation_method            'Materials Studio'
_symmetry_space_group_name_H-M    'R-3C'
_symmetry_Int_Tables_number       167
_symmetry_cell_setting            trigonal
loop_
_symmetry_equiv_pos_as_xyz
  x,y,z
  -y,x-y,z
  -x+y,-x,z
  y,x,-z+1/2
  x-y,-y,-z+1/2
  -x,-x+y,-z+1/2
  -x,-y,-z
  y,-x+y,-z
  x-y,x,-z
  -y,-x,z+1/2
  -x+y,y,z+1/2
  x,x-y,z+1/2
  x+2/3,y+1/3,z+1/3
  -y+2/3,x-y+1/3,z+1/3
  -x+y+2/3,-x+1/3,z+1/3
  y+2/3,x+1/3,-z+5/6
  x-y+2/3,-y+1/3,-z+5/6
  -x+2/3,-x+y+1/3,-z+5/6
  -x+2/3,-y+1/3,-z+1/3
  y+2/3,-x+y+1/3,-z+1/3
  x-y+2/3,x+1/3,-z+1/3
  -y+2/3,-x+1/3,z+5/6
  -x+y+2/3,y+1/3,z+5/6
  x+2/3,x-y+1/3,z+5/6
  x+1/3,y+2/3,z+2/3
  -y+1/3,x-y+2/3,z+2/3
  -x+y+1/3,-x+2/3,z+2/3
  y+1/3,x+2/3,-z+1/6
  x-y+1/3,-y+2/3,-z+1/6
  -x+1/3,-x+y+2/3,-z+1/6
  -x+1/3,-y+2/3,-z+2/3
  y+1/3,-x+y+2/3,-z+2/3
  x-y+1/3,x+2/3,-z+2/3
  -y+1/3,-x+2/3,z+1/6
  -x+y+1/3,y+2/3,z+1/6
  x+1/3,x-y+2/3,z+1/6
_cell_length_a                    8.1402
_cell_length_b                    8.1402
_cell_length_c                    20.4018
_cell_angle_alpha                 90.0000
_cell_angle_beta                  90.0000
_cell_angle_gamma                 120.0000
loop_
_atom_site_label
_atom_site_type_symbol
_atom_site_fract_x
_atom_site_fract_y
_atom_site_fract_z
_atom_site_U_iso_or_equiv
_atom_site_adp_type
_atom_site_occupancy
O1     O     0.16955  -0.02353   0.18902   0.01056  Uani   1.00
O2     O     0.18598   0.15722   0.08644   0.00806  Uani   1.00
Li1    Li    0.00000   0.00000   0.00000   0.03001  Uani   1.00
Si1    Si    0.00000   0.00000   0.14193   0.00316  Uani   1.00
P1     P     0.28418  -0.00000   0.25000   0.00364  Uani   1.00
LiSn2PO4_3_hex.cif
data_LiSn2PO4_3_hex
_audit_creation_date              2014-09-09
_audit_creation_method            'Materials Studio'
_symmetry_space_group_name_H-M    'R-3C'
_symmetry_Int_Tables_number       167
_symmetry_cell_setting            trigonal
loop_
_symmetry_equiv_pos_as_xyz
  x,y,z
  -y,x-y,z
  -x+y,-x,z
  y,x,-z+1/2
  x-y,-y,-z+1/2
  -x,-x+y,-z+1/2
  -x,-y,-z
  y,-x+y,-z
  x-y,x,-z
  -y,-x,z+1/2
  -x+y,y,z+1/2
  x,x-y,z+1/2
  x+2/3,y+1/3,z+1/3
  -y+2/3,x-y+1/3,z+1/3
  -x+y+2/3,-x+1/3,z+1/3
  y+2/3,x+1/3,-z+5/6
  x-y+2/3,-y+1/3,-z+5/6
  -x+2/3,-x+y+1/3,-z+5/6
  -x+2/3,-y+1/3,-z+1/3
  y+2/3,-x+y+1/3,-z+1/3
```

```
  x-y+2/3,x+1/3,-z+1/3
  -y+2/3,-x+1/3,z+5/6
  -x+y+2/3,y+1/3,z+5/6
  x+2/3,x-y+1/3,z+5/6
  x+1/3,y+2/3,z+2/3
  -y+1/3,x-y+2/3,z+2/3
  -x+y+1/3,-x+2/3,z+2/3
  y+1/3,x+2/3,-z+1/6
  x-y+1/3,-y+2/3,-z+1/6
  -x+1/3,-x+y+2/3,-z+1/6
  -x+1/3,-y+2/3,-z+2/3
  y+1/3,-x+y+2/3,-z+2/3
  x-y+1/3,x+2/3,-z+2/3
  -y+1/3,-x+2/3,z+1/6
  -x+y+1/3,y+2/3,z+1/6
  x+1/3,x-y+2/3,z+1/6
_cell_length_a                    8.4725
_cell_length_b                    8.4725
_cell_length_c                    21.3981
_cell_angle_alpha                 90.0000
_cell_angle_beta                  90.0000
_cell_angle_gamma                 120.0000
loop_
_atom_site_label
_atom_site_type_symbol
_atom_site_fract_x
_atom_site_fract_y
_atom_site_fract_z
_atom_site_U_iso_or_equiv
_atom_site_adp_type
_atom_site_occupancy
O1     O     0.18025   0.98297   0.69195   0.01151  Uani   1.00
O2     O     0.18832   0.16430   0.58584   0.00877  Uani   1.00
Li1    Li    0.33333   0.66667   0.16667   0.03257  Uani   1.00
Sn1    Sn    0.33333   0.66667   0.30995   0.00343  Uani   1.00
P1     P     0.28898   1.00000   0.75000   0.00396  Uani   1.00
LiTi2PO4_3_hex.cif
data_LiTi2PO4_3_hex
_audit_creation_date              2014-09-09
_audit_creation_method            'Materials Studio'
_symmetry_space_group_name_H-M    'R-3C'
_symmetry_Int_Tables_number       167
_symmetry_cell_setting            trigonal
loop_
_symmetry_equiv_pos_as_xyz
  x,y,z
  -y,x-y,z
  -x+y,-x,z
  y,x,-z+1/2
  x-y,-y,-z+1/2
  -x,-x+y,-z+1/2
  -x,-y,-z
  y,-x+y,-z
  x-y,x,-z
  -y,-x,z+1/2
  -x+y,y,z+1/2
  x,x-y,z+1/2
  x+2/3,y+1/3,z+1/3
  -y+2/3,x-y+1/3,z+1/3
  -x+y+2/3,-x+1/3,z+1/3
  y+2/3,x+1/3,-z+5/6
  x-y+2/3,-y+1/3,-z+5/6
  -x+2/3,-x+y+1/3,-z+5/6
  -x+2/3,-y+1/3,-z+1/3
  y+2/3,-x+y+1/3,-z+1/3
  x-y+2/3,x+1/3,-z+1/3
  -y+2/3,-x+1/3,z+5/6
  -x+y+2/3,y+1/3,z+5/6
  x+2/3,x-y+1/3,z+5/6
  x+1/3,y+2/3,z+2/3
  -y+1/3,x-y+2/3,z+2/3
  -x+y+1/3,-x+2/3,z+2/3
  y+1/3,x+2/3,-z+1/6
  x-y+1/3,-y+2/3,-z+1/6
  -x+1/3,-x+y+2/3,-z+1/6
  -x+1/3,-y+2/3,-z+2/3
  y+1/3,-x+y+2/3,-z+2/3
  x-y+1/3,x+2/3,-z+2/3
  -y+1/3,-x+2/3,z+1/6
  -x+y+1/3,y+2/3,z+1/6
  x+1/3,x-y+2/3,z+1/6
_cell_length_a                    8.5938
_cell_length_b                    8.5938
_cell_length_c                    21.4053
_cell_angle_alpha                 90.0000
_cell_angle_beta                  90.0000
_cell_angle_gamma                 120.0000
loop_
_atom_site_label
_atom_site_type_symbol
_atom_site_fract_x
_atom_site_fract_y
_atom_site_fract_z
_atom_site_U_iso_or_equiv
_atom_site_adp_type
```

```
_atom_site_occupancy
O1     O     0.18107   0.99030   0.19209   0.01191  Uiso   1.00
O2     O     0.19238   0.16615   0.08400   0.00874  Uiso   1.00
Li1    Li    0.00000   0.00000   0.00000   0.06079  Uiso   1.00
Ti1    Ti    0.00000   0.00000   0.14311   0.00887  Uiso   1.00
P1     P     0.28847  -0.00000   0.25000   0.00633  Uiso   1.00
LiTi2PO4_3_LDA_hex.cif
data_LiTi2PO4_3_lda_hex
_audit_creation_date              2014-09-09
_audit_creation_method            'Materials Studio'
_symmetry_space_group_name_H-M    'R-3C'
_symmetry_Int_Tables_number       167
_symmetry_cell_setting            trigonal
loop_
_symmetry_equiv_pos_as_xyz
  x,y,z
  -y,x-y,z
  -x+y,-x,z
  y,x,-z+1/2
  x-y,-y,-z+1/2
  -x,-x+y,-z+1/2
  -x,-y,-z
  y,-x+y,-z
  x-y,x,-z
  -y,-x,z+1/2
  -x+y,y,z+1/2
  x,x-y,z+1/2
  x+2/3,y+1/3,z+1/3
  -y+2/3,x-y+1/3,z+1/3
  -x+y+2/3,-x+1/3,z+1/3
  y+2/3,x+1/3,-z+5/6
  x-y+2/3,-y+1/3,-z+5/6
  -x+2/3,-x+y+1/3,-z+5/6
  -x+2/3,-y+1/3,-z+1/3
  y+2/3,-x+y+1/3,-z+1/3
  x-y+2/3,x+1/3,-z+1/3
  -y+2/3,-x+1/3,z+5/6
  -x+y+2/3,y+1/3,z+5/6
  x+2/3,x-y+1/3,z+5/6
  x+1/3,y+2/3,z+2/3
  -y+1/3,x-y+2/3,z+2/3
  -x+y+1/3,-x+2/3,z+2/3
  y+1/3,x+2/3,-z+1/6
  x-y+1/3,-y+2/3,-z+1/6
  -x+1/3,-x+y+2/3,-z+1/6
  -x+1/3,-y+2/3,-z+2/3
  y+1/3,-x+y+2/3,-z+2/3
  x-y+1/3,x+2/3,-z+2/3
  -y+1/3,-x+2/3,z+1/6
  -x+y+1/3,y+2/3,z+1/6
  x+1/3,x-y+2/3,z+1/6
_cell_length_a                    8.4626
_cell_length_b                    8.4626
_cell_length_c                    20.9981
_cell_angle_alpha                 90.0000
_cell_angle_beta                  90.0000
_cell_angle_gamma                 120.0000
loop_
_atom_site_label
_atom_site_type_symbol
_atom_site_fract_x
_atom_site_fract_y
_atom_site_fract_z
_atom_site_U_iso_or_equiv
_atom_site_adp_type
_atom_site_occupancy
O1     O     0.18164   0.99104   0.19186   0.01191  Uiso   1.00
O2     O     0.19170   0.16636   0.08413   0.00874  Uiso   1.00
Li1    Li    0.00000   0.00000   0.00000   0.06079  Uiso   1.00
Ti1    Ti    0.00000   0.00000   0.14241   0.00887  Uiso   1.00
P1     P     0.28869   0.00000   0.25000   0.00633  Uiso   1.00
LiZr2PO4_3_hex.cif
data_LiZr2(PO4)_3_hex
_audit_creation_date              2014-09-09
_audit_creation_method            'Materials Studio'
_symmetry_space_group_name_H-M    'R-3C'
_symmetry_Int_Tables_number       167
_symmetry_cell_setting            trigonal
loop_
_symmetry_equiv_pos_as_xyz
  x,y,z
  -y,x-y,z
  -x+y,-x,z
  y,x,-z+1/2
  x-y,-y,-z+1/2
  -x,-x+y,-z+1/2
  -x,-y,-z
  y,-x+y,-z
  x-y,x,-z
  -y,-x,z+1/2
  -x+y,y,z+1/2
  x,x-y,z+1/2
  x+2/3,y+1/3,z+1/3
  -y+2/3,x-y+1/3,z+1/3
  -x+y+2/3,-x+1/3,z+1/3
```

```
  y+2/3,x+1/3,-z+5/6
  x-y+2/3,-y+1/3,-z+5/6
  -x+2/3,-x+y+1/3,-z+5/6
  -x+2/3,-y+1/3,-z+1/3
  y+2/3,-x+y+1/3,-z+1/3
  x-y+2/3,x+1/3,-z+1/3
  -y+2/3,-x+1/3,z+5/6
  -x+y+2/3,y+1/3,z+5/6
  x+2/3,x-y+1/3,z+5/6
  x+1/3,y+2/3,z+2/3
  -y+1/3,x-y+2/3,z+2/3
  -x+y+1/3,-x+2/3,z+2/3
  y+1/3,x+2/3,-z+1/6
  x-y+1/3,-y+2/3,-z+1/6
  -x+1/3,-x+y+2/3,-z+1/6
  -x+1/3,-y+2/3,-z+2/3
  y+1/3,-x+y+2/3,-z+2/3
  x-y+1/3,x+2/3,-z+2/3
  -y+1/3,-x+2/3,z+1/6
  -x+y+1/3,y+2/3,z+1/6
  x+1/3,x-y+2/3,z+1/6
_cell_length_a                    8.9288
_cell_length_b                    8.9288
_cell_length_c                    22.6231
_cell_angle_alpha                 90.0000
_cell_angle_beta                  90.0000
_cell_angle_gamma                 120.0000
loop_
_atom_site_label
_atom_site_type_symbol
_atom_site_fract_x
_atom_site_fract_y
_atom_site_fract_z
_atom_site_U_iso_or_equiv
_atom_site_adp_type
_atom_site_occupancy
O1     O     0.18931   0.99217   0.19522   0.01191  Uiso   1.00
O2     O     0.19536   0.17246   0.08446   0.00874  Uiso   1.00
Li1    Li    0.00000   0.00000   0.00000   0.06079  Uiso   1.00
Zr1    Zr    0.00000   0.00000   0.14443   0.00887  Uiso   1.00
P1     P     0.29175  -0.00000   0.25000   0.00633  Uiso   1.00

NAMP STRUCTURES
------------------------------------------------------------
nagp1_opt.cif
data_NAGP_1AL
_audit_creation_date              2014-10-29
_audit_creation_method            'Materials Studio'
_symmetry_space_group_name_H-M    'P1'
_symmetry_Int_Tables_number       1
_symmetry_cell_setting            triclinic
loop_
_symmetry_equiv_pos_as_xyz
  x,y,z
_cell_length_a                    8.3481
_cell_length_b                    8.3534
_cell_length_c                    21.8506
_cell_angle_alpha                 89.9996
_cell_angle_beta                  89.9728
_cell_angle_gamma                 119.9781
loop_
_atom_site_label
_atom_site_type_symbol
_atom_site_fract_x
_atom_site_fract_y
_atom_site_fract_z
_atom_site_U_iso_or_equiv
_atom_site_adp_type
_atom_site_occupancy
Ge107  Ge   -0.00034   0.00145   0.64562   0.00000  Uiso   1.00
Ge108  Ge    0.66400   0.33337   0.97877   0.00000  Uiso   1.00
Al109  Al    0.34209   0.66986   0.31730   0.00000  Uiso   1.00
Na91   Na    0.00181   0.00291  -0.00057   0.00000  Uiso   1.00
Na92   Na    0.63244   0.30172   0.33479   0.00000  Uiso   1.00
Na93   Na    0.33373   0.66904   0.66608   0.00000  Uiso   1.00
Na94   Na   -0.00063   0.00063   0.49965   0.00000  Uiso   1.00
Na95   Na    0.66610   0.33460   0.83202   0.00000  Uiso   1.00
Na96   Na    0.35652   0.67010   0.16898   0.00000  Uiso   1.00
Na97   Na   -0.00005   0.63532   0.26115   0.00000  Uiso   1.00
O1     O     0.15120   0.94955   0.19710   0.00000  Uiso   1.00
O10    O     0.50799   0.36830   0.42291   0.00000  Uiso   1.00
O11    O     0.37009   0.87280   0.85960   0.00000  Uiso   1.00
O12    O     0.17322   0.69942   0.75710   0.00000  Uiso   1.00
O13    O     0.77939   0.83491   0.19181   0.00000  Uiso   1.00
O14    O     0.96706   0.80923   0.09156   0.00000  Uiso   1.00
O15    O     0.45959   0.16853   0.52561   0.00000  Uiso   1.00
O16    O     0.63439   0.14079   0.42430   0.00000  Uiso   1.00
O17    O     0.12383   0.50013   0.85902   0.00000  Uiso   1.00
O18    O     0.30267   0.47590   0.75782   0.00000  Uiso   1.00
O19    O     0.96080   0.16601   0.30582   0.00000  Uiso   1.00
O2     O     0.19458   0.15880   0.09301   0.00000  Uiso   1.00
O20    O     0.16428   0.19056   0.41051   0.00000  Uiso   1.00
O21    O     0.63479   0.50565   0.63975   0.00000  Uiso   1.00
O22    O     0.83023   0.52381   0.74232   0.00000  Uiso   1.00
O23    O     0.30068   0.84030   0.97364   0.00000  Uiso   1.00
```

```
O24    O     0.49741   0.85537   0.07622   0.00000  Uiso   1.00
O25    O     0.21476   0.04797   0.30810   0.00000  Uiso   1.00
O26    O     0.02336   0.83562   0.40877   0.00000  Uiso   1.00
O27    O     0.87452   0.37083   0.64075   0.00000  Uiso   1.00
O28    O     0.69080   0.16960   0.74188   0.00000  Uiso   1.00
O29    O     0.54046   0.70396   0.97444   0.00000  Uiso   1.00
O3     O     0.83240   0.29612   0.52699   0.00000  Uiso   1.00
O30    O     0.35320   0.50031   0.07499   0.00000  Uiso   1.00
O31    O     0.84683   0.79932   0.30253   0.00000  Uiso   1.00
O32    O     0.81133   0.97521   0.40848   0.00000  Uiso   1.00
O33    O     0.49981   0.13083   0.63905   0.00000  Uiso   1.00
O34    O     0.47682   0.30850   0.74120   0.00000  Uiso   1.00
O35    O     0.16532   0.46460   0.97249   0.00000  Uiso   1.00
O36    O     0.14329   0.64327   0.07481   0.00000  Uiso   1.00
O37    O     0.83628   0.04347   0.80555   0.00000  Uiso   1.00
O38    O     0.80673   0.84195   0.90806   0.00000  Uiso   1.00
O39    O     0.50770   0.38120   0.13676   0.00000  Uiso   1.00
O4     O     0.86211   0.49295   0.42286   0.00000  Uiso   1.00
O40    O     0.47135   0.17883   0.23828   0.00000  Uiso   1.00
O41    O     0.17001   0.70722   0.47159   0.00000  Uiso   1.00
O42    O     0.14176   0.50880   0.57512   0.00000  Uiso   1.00
O43    O     0.96102   0.79449   0.80657   0.00000  Uiso   1.00
O44    O     0.15906   0.96919   0.90912   0.00000  Uiso   1.00
O45    O     0.62956   0.12934   0.13900   0.00000  Uiso   1.00
O46    O     0.82799   0.30819   0.24272   0.00000  Uiso   1.00
O47    O     0.29786   0.45929   0.47557   0.00000  Uiso   1.00
O48    O     0.49473   0.63790   0.57646   0.00000  Uiso   1.00
O49    O     0.20933   0.16862   0.80690   0.00000  Uiso   1.00
O5     O     0.49604   0.62639   0.86103   0.00000  Uiso   1.00
O50    O     0.03050   0.19330   0.90847   0.00000  Uiso   1.00
O51    O     0.88059   0.50768   0.14261   0.00000  Uiso   1.00
O52    O     0.69515   0.52927   0.24038   0.00000  Uiso   1.00
O53    O     0.54635   0.83761   0.47524   0.00000  Uiso   1.00
O54    O     0.36297   0.86009   0.57484   0.00000  Uiso   1.00
O55    O     0.03556   0.83250   0.69284   0.00000  Uiso   1.00
O56    O     0.83819   0.81267   0.58993   0.00000  Uiso   1.00
O57    O     0.69744   0.16163   0.02504   0.00000  Uiso   1.00
O58    O     0.50265   0.14603   0.92260   0.00000  Uiso   1.00
O59    O     0.35753   0.50027   0.36172   0.00000  Uiso   1.00
O6     O     0.52642   0.82786   0.75840   0.00000  Uiso   1.00
O60    O     0.16701   0.47979   0.25945   0.00000  Uiso   1.00
O61    O     0.79295   0.96346   0.69223   0.00000  Uiso   1.00
O62    O     0.97512   0.16586   0.59032   0.00000  Uiso   1.00
O63    O     0.45527   0.29519   0.02427   0.00000  Uiso   1.00
O64    O     0.64358   0.50048   0.92405   0.00000  Uiso   1.00
O65    O     0.13666   0.64319   0.35752   0.00000  Uiso   1.00
O66    O     0.31364   0.83890   0.25985   0.00000  Uiso   1.00
O67    O     0.16564   0.20531   0.69360   0.00000  Uiso   1.00
O68    O     0.19061   0.02935   0.59081   0.00000  Uiso   1.00
O69    O     0.82839   0.53353   0.02902   0.00000  Uiso   1.00
O7     O     0.03797   0.20903   0.19256   0.00000  Uiso   1.00
O70    O     0.85582   0.36110   0.92447   0.00000  Uiso   1.00
O71    O     0.49864   0.86219   0.36211   0.00000  Uiso   1.00
O72    O     0.52221   0.69459   0.26395   0.00000  Uiso   1.00
O8     O     0.84153   0.03522   0.08953   0.00000  Uiso   1.00
O9     O     0.70537   0.54204   0.52562   0.00000  Uiso   1.00
P73    P     0.28822   0.00361   0.25057   0.00000  Uiso   1.00
P74    P     0.95362   0.33569   0.58298   0.00000  Uiso   1.00
P75    P     0.61913   0.66858   0.91659   0.00000  Uiso   1.00
P76    P     0.00269   0.28888   0.25027   0.00000  Uiso   1.00
P77    P     0.66775   0.62155   0.58268   0.00000  Uiso   1.00
P78    P     0.33283   0.95418   0.91605   0.00000  Uiso   1.00
P79    P     0.70756   0.71828   0.25004   0.00000  Uiso   1.00
P80    P     0.38121   0.04974   0.58233   0.00000  Uiso   1.00
P81    P     0.04663   0.38213   0.91592   0.00000  Uiso   1.00
P82    P     0.71428   0.00126   0.74938   0.00000  Uiso   1.00
P83    P     0.37930   0.33348   0.08205   0.00000  Uiso   1.00
P84    P     0.05074   0.66795   0.41487   0.00000  Uiso   1.00
P85    P    -0.00027   0.71510   0.74953   0.00000  Uiso   1.00
P86    P     0.66631   0.04764   0.08216   0.00000  Uiso   1.00
P87    P     0.33358   0.38468   0.41671   0.00000  Uiso   1.00
P88    P     0.28599   0.28676   0.74968   0.00000  Uiso   1.00
P89    P     0.95284   0.62123   0.08371   0.00000  Uiso   1.00
P90    P     0.61808   0.95085   0.41667   0.00000  Uiso   1.00
Ge98   Ge   -0.00480   0.00058   0.14722   0.00000  Uiso   1.00
Ge99   Ge    0.66760   0.33562   0.47886   0.00000  Uiso   1.00
Ge100  Ge    0.33204   0.66702   0.81272   0.00000  Uiso   1.00
Ge101  Ge    0.00366   0.00516   0.35494   0.00000  Uiso   1.00
Ge102  Ge    0.66778   0.33478   0.68576   0.00000  Uiso   1.00
Ge103  Ge    0.33314   0.66806   0.01951   0.00000  Uiso   1.00
Ge104  Ge    0.00057   0.00173   0.85244   0.00000  Uiso   1.00
Ge105  Ge    0.66471   0.33358   0.18459   0.00000  Uiso   1.00
Ge106  Ge    0.33563   0.66859   0.51944   0.00000  Uiso   1.00
nagp2_opt.cif
data_nagp2
_audit_creation_date              2014-10-31
_audit_creation_method            'Materials Studio'
_symmetry_space_group_name_H-M    'P1'
_symmetry_Int_Tables_number       1
_symmetry_cell_setting            triclinic
loop_
_symmetry_equiv_pos_as_xyz
  x,y,z
_cell_length_a                    8.3543
_cell_length_b                    8.3629
```

```
_cell_length_c                    21.8078
_cell_angle_alpha                 90.0094
_cell_angle_beta                  89.9222
_cell_angle_gamma                 119.9340
loop_
_atom_site_label
_atom_site_type_symbol
_atom_site_fract_x
_atom_site_fract_y
_atom_site_fract_z
_atom_site_U_iso_or_equiv
_atom_site_adp_type
_atom_site_occupancy
Ge108  Ge   -0.00240   0.00033   0.64479   0.00000  Uiso   1.00
Al109  Al    0.67351   0.33601   0.98308   0.00000  Uiso   1.00
Al110  Al    0.34178   0.67008   0.31635   0.00000  Uiso   1.00
Na91   Na   -0.02924  -0.02743   0.00091   0.00000  Uiso   1.00
Na92   Na    0.63548   0.30472   0.33386   0.00000  Uiso   1.00
Na93   Na    0.33249   0.66759   0.66522   0.00000  Uiso   1.00
Na94   Na    0.00168   0.00075   0.49862   0.00000  Uiso   1.00
Na95   Na    0.69093   0.33626   0.83488   0.00000  Uiso   1.00
Na96   Na    0.35842   0.66978   0.16905   0.00000  Uiso   1.00
Na97   Na    0.00099   0.63588   0.25741   0.00000  Uiso   1.00
Na98   Na    0.33440   0.30326   0.92416   0.00000  Uiso   1.00
O1     O     0.15398   0.95203   0.19613   0.00000  Uiso   1.00
O10    O     0.50790   0.36830   0.42242   0.00000  Uiso   1.00
O11    O     0.37001   0.87387   0.85941   0.00000  Uiso   1.00
O12    O     0.17442   0.70102   0.75609   0.00000  Uiso   1.00
O13    O     0.78159   0.83493   0.19132   0.00000  Uiso   1.00
O14    O     0.96547   0.80732   0.09089   0.00000  Uiso   1.00
O15    O     0.45932   0.16766   0.52535   0.00000  Uiso   1.00
O16    O     0.63449   0.14087   0.42382   0.00000  Uiso   1.00
O17    O     0.11353   0.50055   0.85841   0.00000  Uiso   1.00
O18    O     0.30122   0.47609   0.75819   0.00000  Uiso   1.00
O19    O     0.96240   0.16637   0.30539   0.00000  Uiso   1.00
O2     O     0.19592   0.15734   0.09036   0.00000  Uiso   1.00
O20    O     0.16443   0.19070   0.41013   0.00000  Uiso   1.00
O21    O     0.63608   0.50779   0.64024   0.00000  Uiso   1.00
O22    O     0.83077   0.52234   0.74288   0.00000  Uiso   1.00
O23    O     0.29538   0.83478   0.97301   0.00000  Uiso   1.00
O24    O     0.49792   0.85573   0.07747   0.00000  Uiso   1.00
O25    O     0.21424   0.04699   0.30777   0.00000  Uiso   1.00
O26    O     0.02300   0.83551   0.40854   0.00000  Uiso   1.00
O27    O     0.87424   0.36990   0.64114   0.00000  Uiso   1.00
O28    O     0.68704   0.16702   0.74168   0.00000  Uiso   1.00
O29    O     0.54722   0.71393   0.97503   0.00000  Uiso   1.00
O3     O     0.83106   0.29573   0.52717   0.00000  Uiso   1.00
O30    O     0.35258   0.49975   0.07520   0.00000  Uiso   1.00
O31    O     0.84696   0.79887   0.30230   0.00000  Uiso   1.00
O32    O     0.81105   0.97524   0.40799   0.00000  Uiso   1.00
O33    O     0.49905   0.13126   0.63910   0.00000  Uiso   1.00
O34    O     0.47624   0.30933   0.74141   0.00000  Uiso   1.00
O35    O     0.17933   0.46610   0.96938   0.00000  Uiso   1.00
O36    O     0.14414   0.64308   0.07541   0.00000  Uiso   1.00
O37    O     0.84013   0.04674   0.80356   0.00000  Uiso   1.00
O38    O     0.80510   0.84537   0.90545   0.00000  Uiso   1.00
O39    O     0.50592   0.37702   0.13636   0.00000  Uiso   1.00
O4     O     0.86226   0.49315   0.42280   0.00000  Uiso   1.00
O40    O     0.47244   0.17910   0.23876   0.00000  Uiso   1.00
O41    O     0.17018   0.70753   0.47112   0.00000  Uiso   1.00
O42    O     0.14086   0.50879   0.57490   0.00000  Uiso   1.00
O43    O     0.96347   0.79573   0.80591   0.00000  Uiso   1.00
O44    O     0.16143   0.97513   0.90955   0.00000  Uiso   1.00
O45    O     0.63285   0.12706   0.14126   0.00000  Uiso   1.00
O46    O     0.82944   0.30924   0.24310   0.00000  Uiso   1.00
O47    O     0.29835   0.45947   0.47532   0.00000  Uiso   1.00
O48    O     0.49420   0.63786   0.57649   0.00000  Uiso   1.00
O49    O     0.21396   0.17531   0.80971   0.00000  Uiso   1.00
O5     O     0.48562   0.61640   0.86384   0.00000  Uiso   1.00
O50    O     0.02719   0.19531   0.90753   0.00000  Uiso   1.00
O51    O     0.88311   0.51061   0.14422   0.00000  Uiso   1.00
O52    O     0.69331   0.52858   0.24039   0.00000  Uiso   1.00
O53    O     0.54667   0.83857   0.47522   0.00000  Uiso   1.00
O54    O     0.36152   0.85967   0.57500   0.00000  Uiso   1.00
O55    O     0.03133   0.82919   0.69178   0.00000  Uiso   1.00
O56    O     0.83800   0.81249   0.58892   0.00000  Uiso   1.00
O57    O     0.68683   0.16384   0.02705   0.00000  Uiso   1.00
O58    O     0.50052   0.14673   0.92518   0.00000  Uiso   1.00
O59    O     0.35732   0.50058   0.36124   0.00000  Uiso   1.00
O6     O     0.52758   0.82579   0.75958   0.00000  Uiso   1.00
O60    O     0.16832   0.47985   0.25887   0.00000  Uiso   1.00
O61    O     0.78878   0.96146   0.69082   0.00000  Uiso   1.00
O62    O     0.97603   0.16678   0.59030   0.00000  Uiso   1.00
O63    O     0.46582   0.30795   0.02279   0.00000  Uiso   1.00
O64    O     0.64800   0.50626   0.92704   0.00000  Uiso   1.00
O65    O     0.13577   0.64288   0.35682   0.00000  Uiso   1.00
O66    O     0.31341   0.83875   0.25936   0.00000  Uiso   1.00
O67    O     0.16051   0.19979   0.69598   0.00000  Uiso   1.00
O68    O     0.19097   0.03101   0.59099   0.00000  Uiso   1.00
O69    O     0.82812   0.52365   0.03061   0.00000  Uiso   1.00
O7     O     0.03859   0.20959   0.19175   0.00000  Uiso   1.00
O70    O     0.85584   0.36263   0.93062   0.00000  Uiso   1.00
O71    O     0.49759   0.86151   0.36193   0.00000  Uiso   1.00
O72    O     0.52343   0.69663   0.26350   0.00000  Uiso   1.00
O8     O     0.84201   0.03665   0.08836   0.00000  Uiso   1.00
```

```
O9     O     0.70321   0.53982   0.52566   0.00000  Uiso   1.00
P73    P     0.28905   0.00412   0.25011   0.00000  Uiso   1.00
P74    P     0.95322   0.33591   0.58283   0.00000  Uiso   1.00
P75    P     0.62201   0.67055   0.91747   0.00000  Uiso   1.00
P76    P     0.00361   0.28930   0.24973   0.00000  Uiso   1.00
P77    P     0.66707   0.62135   0.58239   0.00000  Uiso   1.00
P78    P     0.33592   0.95573   0.91679   0.00000  Uiso   1.00
P79    P     0.70810   0.71850   0.24956   0.00000  Uiso   1.00
P80    P     0.38087   0.04962   0.58216   0.00000  Uiso   1.00
P81    P     0.04049   0.38461   0.91671   0.00000  Uiso   1.00
P82    P     0.71249  -0.00006   0.74847   0.00000  Uiso   1.00
P83    P     0.38169   0.33331   0.08049   0.00000  Uiso   1.00
P84    P     0.05049   0.66808   0.41427   0.00000  Uiso   1.00
P85    P    -0.00040   0.71414   0.74865   0.00000  Uiso   1.00
P86    P     0.66622   0.05031   0.08221   0.00000  Uiso   1.00
P87    P     0.33331   0.38451   0.41612   0.00000  Uiso   1.00
P88    P     0.28621   0.28781   0.75027   0.00000  Uiso   1.00
P89    P     0.95154   0.61854   0.08369   0.00000  Uiso   1.00
P90    P     0.61806   0.95086   0.41618   0.00000  Uiso   1.00
Ge99   Ge   -0.00319   0.00181   0.14591   0.00000  Uiso   1.00
Ge100  Ge    0.66700   0.33502   0.47808   0.00000  Uiso   1.00
Ge101  Ge    0.32880   0.66700   0.81342   0.00000  Uiso   1.00
Ge102  Ge    0.00360   0.00473   0.35405   0.00000  Uiso   1.00
Ge103  Ge    0.66694   0.33468   0.68559   0.00000  Uiso   1.00
Ge104  Ge    0.33578   0.67134   0.02139   0.00000  Uiso   1.00
Ge105  Ge   -0.00258  -0.00034   0.85100   0.00000  Uiso   1.00
Ge106  Ge    0.66498   0.33299   0.18431   0.00000  Uiso   1.00
Ge107  Ge    0.33552   0.66884   0.51866   0.00000  Uiso   1.00
nagp3_opt.cif
data_NAGP_input
_audit_creation_date              2014-10-29
_audit_creation_method            'Materials Studio'
_symmetry_space_group_name_H-M    'P1'
_symmetry_Int_Tables_number       1
_symmetry_cell_setting            triclinic
loop_
_symmetry_equiv_pos_as_xyz
  x,y,z
_cell_length_a                    8.3652
_cell_length_b                    8.3727
_cell_length_c                    21.7382
_cell_angle_alpha                 89.9869
_cell_angle_beta                  89.8892
_cell_angle_gamma                 119.8713
loop_
_atom_site_label
_atom_site_type_symbol
_atom_site_fract_x
_atom_site_fract_y
_atom_site_fract_z
_atom_site_U_iso_or_equiv
_atom_site_adp_type
_atom_site_occupancy
Al109  Al    0.00671   0.00320   0.64894   0.00000  Uiso   1.00
Al110  Al    0.67334   0.33652   0.98231   0.00000  Uiso   1.00
Al111  Al    0.34001   0.66991   0.31571   0.00000  Uiso   1.00
Na91   Na   -0.02678  -0.02530  -0.00013   0.00000  Uiso   1.00
Na92   Na    0.64001   0.30815   0.33309   0.00000  Uiso   1.00
Na93   Na    0.30675   0.64150   0.66645   0.00000  Uiso   1.00
Na94   Na    0.02539   0.00326   0.50140   0.00000  Uiso   1.00
Na95   Na    0.69219   0.33659   0.83481   0.00000  Uiso   1.00
Na96   Na    0.35887   0.66997   0.16805   0.00000  Uiso   1.00
Na97   Na    0.00146   0.63685   0.25450   0.00000  Uiso   1.00
Na98   Na    0.66818  -0.02981   0.58782   0.00000  Uiso   1.00
Na99   Na    0.33490   0.30351   0.92110   0.00000  Uiso   1.00
O1     O     0.15613   0.95396   0.19600   0.00000  Uiso   1.00
O10    O     0.50759   0.36836   0.42117   0.00000  Uiso   1.00
O11    O     0.36959   0.87434   0.85833   0.00000  Uiso   1.00
O12    O     0.17430   0.70164   0.75440   0.00000  Uiso   1.00
O13    O     0.78319   0.83303   0.19117   0.00000  Uiso   1.00
O14    O     0.96744   0.80829   0.09035   0.00000  Uiso   1.00
O15    O     0.44988   0.16634   0.52448   0.00000  Uiso   1.00
O16    O     0.63416   0.14169   0.42369   0.00000  Uiso   1.00
O17    O     0.11650   0.49965   0.85775   0.00000  Uiso   1.00
O18    O     0.30080   0.47493   0.75696   0.00000  Uiso   1.00
O19    O     0.96315   0.16777   0.30592   0.00000  Uiso   1.00
O2     O     0.19463   0.15835   0.08971   0.00000  Uiso   1.00
O20    O     0.16382   0.18960   0.41071   0.00000  Uiso   1.00
O21    O     0.62987   0.50118   0.63922   0.00000  Uiso   1.00
O22    O     0.83053   0.52289   0.74395   0.00000  Uiso   1.00
O23    O     0.29662   0.83456   0.97254   0.00000  Uiso   1.00
O24    O     0.49716   0.85624   0.07729   0.00000  Uiso   1.00
O25    O     0.21352   0.04633   0.30815   0.00000  Uiso   1.00
O26    O     0.02066   0.83429   0.40889   0.00000  Uiso   1.00
O27    O     0.88015   0.37968   0.64142   0.00000  Uiso   1.00
O28    O     0.68733   0.16755   0.74212   0.00000  Uiso   1.00
O29    O     0.54677   0.71308   0.97471   0.00000  Uiso   1.00
O3     O     0.82283   0.28727   0.52929   0.00000  Uiso   1.00
O30    O     0.35399   0.50092   0.07547   0.00000  Uiso   1.00
O31    O     0.84643   0.79871   0.30268   0.00000  Uiso   1.00
O32    O     0.81012   0.97487   0.40879   0.00000  Uiso   1.00
O33    O     0.51301   0.13207   0.63599   0.00000  Uiso   1.00
O34    O     0.47684   0.30822   0.74207   0.00000  Uiso   1.00
O35    O     0.17975   0.46544   0.96927   0.00000  Uiso   1.00
O36    O     0.14342   0.64152   0.07539   0.00000  Uiso   1.00
```

```
O37    O     0.83589   0.04022   0.80361   0.00000  Uiso   1.00
O38    O     0.80698   0.84462   0.90618   0.00000  Uiso   1.00
O39    O     0.50266   0.37358   0.13694   0.00000  Uiso   1.00
O4     O     0.86134   0.49174   0.42309   0.00000  Uiso   1.00
O40    O     0.47364   0.17801   0.23958   0.00000  Uiso   1.00
O41    O     0.16934   0.70698   0.47034   0.00000  Uiso   1.00
O42    O     0.14032   0.51133   0.57290   0.00000  Uiso   1.00
O43    O     0.96814   0.79387   0.80812   0.00000  Uiso   1.00
O44    O     0.16322   0.97671   0.90999   0.00000  Uiso   1.00
O45    O     0.63463   0.12717   0.14151   0.00000  Uiso   1.00
O46    O     0.82987   0.31002   0.24336   0.00000  Uiso   1.00
O47    O     0.30149   0.46053   0.47492   0.00000  Uiso   1.00
O48    O     0.49654   0.64339   0.57672   0.00000  Uiso   1.00
O49    O     0.21536   0.17922   0.81123   0.00000  Uiso   1.00
O5     O     0.48947   0.62056   0.86257   0.00000  Uiso   1.00
O50    O     0.02462   0.19408   0.90783   0.00000  Uiso   1.00
O51    O     0.88204   0.51260   0.14461   0.00000  Uiso   1.00
O52    O     0.69120   0.52740   0.24121   0.00000  Uiso   1.00
O53    O     0.54881   0.84610   0.47802   0.00000  Uiso   1.00
O54    O     0.35800   0.86075   0.57457   0.00000  Uiso   1.00
O55    O     0.01976   0.83100   0.69347   0.00000  Uiso   1.00
O56    O     0.83497   0.81319   0.59146   0.00000  Uiso   1.00
O57    O     0.68647   0.16443   0.02687   0.00000  Uiso   1.00
O58    O     0.50167   0.14646   0.92478   0.00000  Uiso   1.00
O59    O     0.35298   0.49775   0.36027   0.00000  Uiso   1.00
O6     O     0.52800   0.82498   0.75629   0.00000  Uiso   1.00
O60    O     0.16828   0.47979   0.25819   0.00000  Uiso   1.00
O61    O     0.79951   0.97534   0.68928   0.00000  Uiso   1.00
O62    O     0.97992   0.17194   0.59323   0.00000  Uiso   1.00
O63    O     0.46602   0.30867   0.02261   0.00000  Uiso   1.00
O64    O     0.64654   0.50528   0.92657   0.00000  Uiso   1.00
O65    O     0.13274   0.64209   0.35601   0.00000  Uiso   1.00
O66    O     0.31326   0.83860   0.25993   0.00000  Uiso   1.00
O67    O     0.16095   0.18991   0.69721   0.00000  Uiso   1.00
O68    O     0.19094   0.03213   0.59688   0.00000  Uiso   1.00
O69    O     0.82749   0.52326   0.03059   0.00000  Uiso   1.00
O7     O     0.03630   0.20767   0.19174   0.00000  Uiso   1.00
O70    O     0.85765   0.36549   0.93024   0.00000  Uiso   1.00
O71    O     0.49417   0.85657   0.36401   0.00000  Uiso   1.00
O72    O     0.52435   0.69893   0.26366   0.00000  Uiso   1.00
O8     O     0.84094   0.03511   0.08785   0.00000  Uiso   1.00
O9     O     0.70296   0.54109   0.52504   0.00000  Uiso   1.00
P73    P     0.28947   0.00430   0.25035   0.00000  Uiso   1.00
P74    P     0.95614   0.33764   0.58364   0.00000  Uiso   1.00
P75    P     0.62277   0.67097   0.91693   0.00000  Uiso   1.00
P76    P     0.00337   0.28941   0.24961   0.00000  Uiso   1.00
P77    P     0.67006   0.62279   0.58292   0.00000  Uiso   1.00
P78    P     0.33673   0.95609   0.91619   0.00000  Uiso   1.00
P79    P     0.70785   0.71809   0.24967   0.00000  Uiso   1.00
P80    P     0.37452   0.05140   0.58297   0.00000  Uiso   1.00
P81    P     0.04120   0.38473   0.91625   0.00000  Uiso   1.00
P82    P     0.71454   0.00008   0.74689   0.00000  Uiso   1.00
P83    P     0.38119   0.33343   0.08025   0.00000  Uiso   1.00
P84    P     0.04787   0.66682   0.41365   0.00000  Uiso   1.00
P85    P    -0.00083   0.71653   0.74845   0.00000  Uiso   1.00
P86    P     0.66577   0.04990   0.08185   0.00000  Uiso   1.00
P87    P     0.33246   0.38321   0.41524   0.00000  Uiso   1.00
P88    P     0.28509   0.28536   0.75003   0.00000  Uiso   1.00
P89    P     0.95170   0.61869   0.08339   0.00000  Uiso   1.00
P90    P     0.61845   0.95212   0.41679   0.00000  Uiso   1.00
Ge100  Ge   -0.00303   0.00176   0.14512   0.00000  Uiso   1.00
Ge101  Ge    0.66363   0.33510   0.47844   0.00000  Uiso   1.00
Ge102  Ge    0.33029   0.66839   0.81171   0.00000  Uiso   1.00
Ge103  Ge    0.00256   0.00417   0.35402   0.00000  Uiso   1.00
Ge104  Ge    0.66922   0.33750   0.68731   0.00000  Uiso   1.00
Ge105  Ge    0.33585   0.67082   0.02061   0.00000  Uiso   1.00
Ge106  Ge   -0.00252  -0.00118   0.85073   0.00000  Uiso   1.00
Ge107  Ge    0.66414   0.33218   0.18409   0.00000  Uiso   1.00
Ge108  Ge    0.33086   0.66556   0.51747   0.00000  Uiso   1.00
natp1_opt.cif
data_NATP_1AL
_audit_creation_date              2014-10-29
_audit_creation_method            'Materials Studio'
_symmetry_space_group_name_H-M    'P1'
_symmetry_Int_Tables_number       1
_symmetry_cell_setting            triclinic
loop_
_symmetry_equiv_pos_as_xyz
  x,y,z
_cell_length_a                    8.5657
_cell_length_b                    8.5697
_cell_length_c                    22.1713
_cell_angle_alpha                 90.0571
_cell_angle_beta                  89.9253
_cell_angle_gamma                 119.9323
loop_
_atom_site_label
_atom_site_type_symbol
_atom_site_fract_x
_atom_site_fract_y
_atom_site_fract_z
_atom_site_U_iso_or_equiv
_atom_site_adp_type
_atom_site_occupancy
Ti109  Ti    0.00074   0.00076   0.64514   0.00000  Uiso   1.00
```

| | | | | | | | |
|---|---|---|---|---|---|---|---|
| Ti110 | Ti | 0.66612 | 0.33458 | 0.97909 | 0.00000 | Uiso | 1.00 |
| Al111 | Al | 0.34172 | 0.66767 | 0.31648 | 0.00000 | Uiso | 1.00 |
| Na91 | Na | 0.00157 | -0.00010 | 0.00025 | 0.00000 | Uiso | 1.00 |
| Na92 | Na | 0.62861 | 0.29757 | 0.33469 | 0.00000 | Uiso | 1.00 |
| Na93 | Na | 0.33318 | 0.66760 | 0.66496 | 0.00000 | Uiso | 1.00 |
| Na94 | Na | -0.00096 | -0.00163 | 0.50017 | 0.00000 | Uiso | 1.00 |
| Na95 | Na | 0.66701 | 0.33419 | 0.83199 | 0.00000 | Uiso | 1.00 |
| Na96 | Na | 0.35782 | 0.67109 | 0.16957 | 0.00000 | Uiso | 1.00 |
| Na97 | Na | -0.00452 | 0.63414 | 0.26222 | 0.00000 | Uiso | 1.00 |
| O1 | O | 0.16021 | 0.96269 | 0.19835 | 0.00000 | Uiso | 1.00 |
| O10 | O | 0.50058 | 0.36466 | 0.42134 | 0.00000 | Uiso | 1.00 |
| O11 | O | 0.35952 | 0.86485 | 0.86099 | 0.00000 | Uiso | 1.00 |
| O12 | O | 0.16672 | 0.69560 | 0.75685 | 0.00000 | Uiso | 1.00 |
| O13 | O | 0.78775 | 0.82992 | 0.19481 | 0.00000 | Uiso | 1.00 |
| O14 | O | 0.96895 | 0.80541 | 0.09184 | 0.00000 | Uiso | 1.00 |
| O15 | O | 0.47012 | 0.16414 | 0.52651 | 0.00000 | Uiso | 1.00 |
| O16 | O | 0.63515 | 0.13480 | 0.42245 | 0.00000 | Uiso | 1.00 |
| O17 | O | 0.13578 | 0.49562 | 0.86097 | 0.00000 | Uiso | 1.00 |
| O18 | O | 0.30517 | 0.47180 | 0.75691 | 0.00000 | Uiso | 1.00 |
| O19 | O | 0.97869 | 0.17458 | 0.30610 | 0.00000 | Uiso | 1.00 |
| O2 | O | 0.19582 | 0.16522 | 0.09287 | 0.00000 | Uiso | 1.00 |
| O20 | O | 0.16828 | 0.19827 | 0.40928 | 0.00000 | Uiso | 1.00 |
| O21 | O | 0.64214 | 0.50511 | 0.63786 | 0.00000 | Uiso | 1.00 |
| O22 | O | 0.83395 | 0.52938 | 0.74253 | 0.00000 | Uiso | 1.00 |
| O23 | O | 0.30922 | 0.84073 | 0.97254 | 0.00000 | Uiso | 1.00 |
| O24 | O | 0.50178 | 0.86249 | 0.07758 | 0.00000 | Uiso | 1.00 |
| O25 | O | 0.20699 | 0.02990 | 0.30793 | 0.00000 | Uiso | 1.00 |
| O26 | O | 0.02854 | 0.83112 | 0.40834 | 0.00000 | Uiso | 1.00 |
| O27 | O | 0.86375 | 0.35859 | 0.63811 | 0.00000 | Uiso | 1.00 |
| O28 | O | 0.69485 | 0.16658 | 0.74259 | 0.00000 | Uiso | 1.00 |
| O29 | O | 0.53023 | 0.69339 | 0.97269 | 0.00000 | Uiso | 1.00 |
| O3 | O | 0.83865 | 0.30835 | 0.52653 | 0.00000 | Uiso | 1.00 |
| O30 | O | 0.35884 | 0.49819 | 0.07679 | 0.00000 | Uiso | 1.00 |
| O31 | O | 0.83615 | 0.80232 | 0.30376 | 0.00000 | Uiso | 1.00 |
| O32 | O | 0.80177 | 0.96990 | 0.40695 | 0.00000 | Uiso | 1.00 |
| O33 | O | 0.49549 | 0.13675 | 0.63792 | 0.00000 | Uiso | 1.00 |
| O34 | O | 0.47140 | 0.30520 | 0.74252 | 0.00000 | Uiso | 1.00 |
| O35 | O | 0.16184 | 0.47095 | 0.97244 | 0.00000 | Uiso | 1.00 |
| O36 | O | 0.13820 | 0.64192 | 0.07619 | 0.00000 | Uiso | 1.00 |
| O37 | O | 0.82839 | 0.02618 | 0.80534 | 0.00000 | Uiso | 1.00 |
| O38 | O | 0.80461 | 0.83445 | 0.90990 | 0.00000 | Uiso | 1.00 |
| O39 | O | 0.49846 | 0.36175 | 0.13807 | 0.00000 | Uiso | 1.00 |
| O4 | O | 0.86525 | 0.49895 | 0.42136 | 0.00000 | Uiso | 1.00 |
| O40 | O | 0.47084 | 0.16289 | 0.24084 | 0.00000 | Uiso | 1.00 |
| O41 | O | 0.16592 | 0.69089 | 0.46899 | 0.00000 | Uiso | 1.00 |
| O42 | O | 0.13776 | 0.50059 | 0.57553 | 0.00000 | Uiso | 1.00 |
| O43 | O | 0.97448 | 0.80331 | 0.80537 | 0.00000 | Uiso | 1.00 |
| O44 | O | 0.16705 | 0.97233 | 0.90976 | 0.00000 | Uiso | 1.00 |
| O45 | O | 0.64473 | 0.14032 | 0.13856 | 0.00000 | Uiso | 1.00 |
| O46 | O | 0.84015 | 0.31484 | 0.24482 | 0.00000 | Uiso | 1.00 |
| O47 | O | 0.31024 | 0.46931 | 0.47243 | 0.00000 | Uiso | 1.00 |
| O48 | O | 0.50149 | 0.63979 | 0.57608 | 0.00000 | Uiso | 1.00 |
| O49 | O | 0.19742 | 0.17230 | 0.80543 | 0.00000 | Uiso | 1.00 |
| O5 | O | 0.50530 | 0.64115 | 0.86117 | 0.00000 | Uiso | 1.00 |
| O50 | O | 0.02810 | 0.19640 | 0.90988 | 0.00000 | Uiso | 1.00 |
| O51 | O | 0.87124 | 0.51149 | 0.14232 | 0.00000 | Uiso | 1.00 |
| O52 | O | 0.68917 | 0.53095 | 0.24245 | 0.00000 | Uiso | 1.00 |
| O53 | O | 0.53361 | 0.83800 | 0.47179 | 0.00000 | Uiso | 1.00 |
| O54 | O | 0.36098 | 0.86331 | 0.57489 | 0.00000 | Uiso | 1.00 |
| O55 | O | 0.02667 | 0.82909 | 0.69392 | 0.00000 | Uiso | 1.00 |
| O56 | O | 0.83451 | 0.80476 | 0.58969 | 0.00000 | Uiso | 1.00 |
| O57 | O | 0.69074 | 0.15980 | 0.02705 | 0.00000 | Uiso | 1.00 |
| O58 | O | 0.50011 | 0.13962 | 0.92339 | 0.00000 | Uiso | 1.00 |
| O59 | O | 0.35529 | 0.50033 | 0.36061 | 0.00000 | Uiso | 1.00 |
| O6 | O | 0.52912 | 0.83388 | 0.75685 | 0.00000 | Uiso | 1.00 |
| O60 | O | 0.17106 | 0.48000 | 0.25911 | 0.00000 | Uiso | 1.00 |
| O61 | O | 0.80274 | 0.97443 | 0.69388 | 0.00000 | Uiso | 1.00 |
| O62 | O | 0.97149 | 0.16722 | 0.58945 | 0.00000 | Uiso | 1.00 |
| O63 | O | 0.46434 | 0.30475 | 0.02720 | 0.00000 | Uiso | 1.00 |
| O64 | O | 0.63956 | 0.50242 | 0.92451 | 0.00000 | Uiso | 1.00 |
| O65 | O | 0.13857 | 0.64333 | 0.35669 | 0.00000 | Uiso | 1.00 |
| O66 | O | 0.30589 | 0.83014 | 0.25881 | 0.00000 | Uiso | 1.00 |
| O67 | O | 0.17156 | 0.19809 | 0.69403 | 0.00000 | Uiso | 1.00 |
| O68 | O | 0.19626 | 0.03088 | 0.58962 | 0.00000 | Uiso | 1.00 |
| O69 | O | 0.83357 | 0.52724 | 0.03063 | 0.00000 | Uiso | 1.00 |
| O7 | O | 0.02865 | 0.20211 | 0.19552 | 0.00000 | Uiso | 1.00 |
| O70 | O | 0.86259 | 0.36419 | 0.92425 | 0.00000 | Uiso | 1.00 |
| O71 | O | 0.49671 | 0.85768 | 0.36053 | 0.00000 | Uiso | 1.00 |
| O72 | O | 0.52165 | 0.69576 | 0.26312 | 0.00000 | Uiso | 1.00 |
| O8 | O | 0.83547 | 0.03173 | 0.09048 | 0.00000 | Uiso | 1.00 |
| O9 | O | 0.69331 | 0.53220 | 0.52626 | 0.00000 | Uiso | 1.00 |
| P73 | P | 0.28753 | -0.00406 | 0.25176 | 0.00000 | Uiso | 1.00 |
| P74 | P | 0.95334 | 0.33415 | 0.58214 | 0.00000 | Uiso | 1.00 |
| P75 | P | 0.61956 | 0.66842 | 0.91690 | 0.00000 | Uiso | 1.00 |
| P76 | P | 0.00793 | 0.29417 | 0.25184 | 0.00000 | Uiso | 1.00 |
| P77 | P | 0.66728 | 0.62011 | 0.58222 | 0.00000 | Uiso | 1.00 |
| P78 | P | 0.33381 | 0.95410 | 0.91647 | 0.00000 | Uiso | 1.00 |
| P79 | P | 0.70374 | 0.71604 | 0.25185 | 0.00000 | Uiso | 1.00 |
| P80 | P | 0.38136 | 0.04881 | 0.58190 | 0.00000 | Uiso | 1.00 |
| P81 | P | 0.04747 | 0.38166 | 0.91664 | 0.00000 | Uiso | 1.00 |
| P82 | P | 0.71410 | 0.00066 | 0.74949 | 0.00000 | Uiso | 1.00 |
| P83 | P | 0.37829 | 0.33233 | 0.08316 | 0.00000 | Uiso | 1.00 |
| P84 | P | 0.05158 | 0.66517 | 0.41306 | 0.00000 | Uiso | 1.00 |
| P85 | P | -0.00004 | 0.71421 | 0.74949 | 0.00000 | Uiso | 1.00 |
| P86 | P | 0.66761 | 0.04788 | 0.08284 | 0.00000 | Uiso | 1.00 |

```
P87    P     0.33524   0.38708   0.41462   0.00000  Uiso   1.00
P88    P     0.28639   0.28646   0.74955   0.00000  Uiso   1.00
P89    P     0.95266   0.62126   0.08494   0.00000  Uiso   1.00
P90    P     0.61334   0.94779   0.41426   0.00000  Uiso   1.00
Ti100  Ti   -0.00463   0.00077   0.14835   0.00000  Uiso   1.00
Ti101  Ti    0.66866   0.33513   0.47727   0.00000  Uiso   1.00
Ti102  Ti    0.33373   0.66723   0.81205   0.00000  Uiso   1.00
Ti103  Ti    0.00323   0.00494   0.35631   0.00000  Uiso   1.00
Ti104  Ti    0.66683   0.33353   0.68582   0.00000  Uiso   1.00
Ti105  Ti    0.33167   0.66770   0.02069   0.00000  Uiso   1.00
Ti106  Ti   -0.00011   0.00067   0.85339   0.00000  Uiso   1.00
Ti107  Ti    0.66192   0.33061   0.18657   0.00000  Uiso   1.00
Ti108  Ti    0.33467   0.66727   0.51874   0.00000  Uiso   1.00
natp2_opt.cif
data_natp2
_audit_creation_date              2014-10-29
_audit_creation_method            'Materials Studio'
_symmetry_space_group_name_H-M    'P1'
_symmetry_Int_Tables_number       1
_symmetry_cell_setting            triclinic
loop_
_symmetry_equiv_pos_as_xyz
  x,y,z
_cell_length_a                    8.5476
_cell_length_b                    8.5538
_cell_length_c                    22.1189
_cell_angle_alpha                 90.0675
_cell_angle_beta                  89.8754
_cell_angle_gamma                 119.8539
loop_
_atom_site_label
_atom_site_type_symbol
_atom_site_fract_x
_atom_site_fract_y
_atom_site_fract_z
_atom_site_U_iso_or_equiv
_atom_site_adp_type
_atom_site_occupancy
Ti108  Ti   -0.00032   0.00194   0.64485   0.00000  Uiso   1.00
Al109  Al    0.67380   0.33655   0.98310   0.00000  Uiso   1.00
Al110  Al    0.34271   0.66995   0.31546   0.00000  Uiso   1.00
Na91   Na   -0.03334  -0.03392   0.00153   0.00000  Uiso   1.00
Na92   Na    0.62881   0.29863   0.33308   0.00000  Uiso   1.00
Na93   Na    0.33540   0.66924   0.66558   0.00000  Uiso   1.00
Na94   Na    0.00185   0.00095   0.49877   0.00000  Uiso   1.00
Na95   Na    0.69125   0.33740   0.83549   0.00000  Uiso   1.00
Na96   Na    0.36078   0.67097   0.16954   0.00000  Uiso   1.00
Na97   Na   -0.00346   0.63538   0.25908   0.00000  Uiso   1.00
Na98   Na    0.32951   0.30222   0.92534   0.00000  Uiso   1.00
O1     O     0.16026   0.96328   0.19692   0.00000  Uiso   1.00
O10    O     0.50150   0.36569   0.42085   0.00000  Uiso   1.00
O11    O     0.36142   0.86833   0.86244   0.00000  Uiso   1.00
O12    O     0.16867   0.69871   0.75720   0.00000  Uiso   1.00
O13    O     0.78785   0.83340   0.19338   0.00000  Uiso   1.00
O14    O     0.96567   0.80300   0.09055   0.00000  Uiso   1.00
O15    O     0.46891   0.16587   0.52618   0.00000  Uiso   1.00
O16    O     0.63706   0.13696   0.42227   0.00000  Uiso   1.00
O17    O     0.11962   0.49685   0.86155   0.00000  Uiso   1.00
O18    O     0.30142   0.47235   0.75872   0.00000  Uiso   1.00
O19    O     0.98036   0.17537   0.30463   0.00000  Uiso   1.00
O2     O     0.19855   0.16408   0.09031   0.00000  Uiso   1.00
O20    O     0.16889   0.19997   0.40850   0.00000  Uiso   1.00
O21    O     0.64385   0.50826   0.63824   0.00000  Uiso   1.00
O22    O     0.83465   0.52943   0.74411   0.00000  Uiso   1.00
O23    O     0.31316   0.84414   0.97341   0.00000  Uiso   1.00
O24    O     0.50256   0.86559   0.07718   0.00000  Uiso   1.00
O25    O     0.20750   0.03116   0.30686   0.00000  Uiso   1.00
O26    O     0.02981   0.83307   0.40776   0.00000  Uiso   1.00
O27    O     0.86389   0.36058   0.63890   0.00000  Uiso   1.00
O28    O     0.69062   0.16452   0.74302   0.00000  Uiso   1.00
O29    O     0.54014   0.69872   0.97533   0.00000  Uiso   1.00
O3     O     0.83816   0.30815   0.52696   0.00000  Uiso   1.00
O30    O     0.35878   0.49684   0.07557   0.00000  Uiso   1.00
O31    O     0.83807   0.80424   0.30242   0.00000  Uiso   1.00
O32    O     0.80225   0.97051   0.40610   0.00000  Uiso   1.00
O33    O     0.49597   0.13919   0.63787   0.00000  Uiso   1.00
O34    O     0.47114   0.30901   0.74245   0.00000  Uiso   1.00
O35    O     0.16961   0.47126   0.97069   0.00000  Uiso   1.00
O36    O     0.13511   0.64037   0.07369   0.00000  Uiso   1.00
O37    O     0.83228   0.03014   0.80446   0.00000  Uiso   1.00
O38    O     0.80369   0.83032   0.90782   0.00000  Uiso   1.00
O39    O     0.50239   0.36082   0.13481   0.00000  Uiso   1.00
O4     O     0.86607   0.50071   0.42127   0.00000  Uiso   1.00
O40    O     0.47080   0.16505   0.23974   0.00000  Uiso   1.00
O41    O     0.16719   0.69267   0.46841   0.00000  Uiso   1.00
O42    O     0.13781   0.50301   0.57574   0.00000  Uiso   1.00
O43    O     0.97730   0.80779   0.80500   0.00000  Uiso   1.00
O44    O     0.17245   0.98170   0.91181   0.00000  Uiso   1.00
O45    O     0.64724   0.14038   0.13872   0.00000  Uiso   1.00
O46    O     0.84107   0.31636   0.24432   0.00000  Uiso   1.00
O47    O     0.31057   0.47017   0.47217   0.00000  Uiso   1.00
O48    O     0.50144   0.64082   0.57609   0.00000  Uiso   1.00
O49    O     0.20541   0.17910   0.80951   0.00000  Uiso   1.00
O5     O     0.49248   0.62701   0.86569   0.00000  Uiso   1.00
O50    O     0.02088   0.19829   0.90983   0.00000  Uiso   1.00
```

```
O51    O     0.87389   0.51168   0.14189   0.00000  Uiso   1.00
O52    O     0.68911   0.53311   0.24092   0.00000  Uiso   1.00
O53    O     0.53423   0.83907   0.47135   0.00000  Uiso   1.00
O54    O     0.36148   0.86517   0.57491   0.00000  Uiso   1.00
O55    O     0.02496   0.82682   0.69322   0.00000  Uiso   1.00
O56    O     0.83488   0.80742   0.58931   0.00000  Uiso   1.00
O57    O     0.68598   0.16606   0.02658   0.00000  Uiso   1.00
O58    O     0.50427   0.14824   0.92504   0.00000  Uiso   1.00
O59    O     0.35761   0.50325   0.36011   0.00000  Uiso   1.00
O6     O     0.52914   0.83162   0.75995   0.00000  Uiso   1.00
O60    O     0.17279   0.48142   0.25802   0.00000  Uiso   1.00
O61    O     0.79619   0.97008   0.69341   0.00000  Uiso   1.00
O62    O     0.97344   0.17015   0.59056   0.00000  Uiso   1.00
O63    O     0.46712   0.30764   0.02294   0.00000  Uiso   1.00
O64    O     0.64038   0.49828   0.92713   0.00000  Uiso   1.00
O65    O     0.13852   0.64362   0.35584   0.00000  Uiso   1.00
O66    O     0.30668   0.83147   0.25772   0.00000  Uiso   1.00
O67    O     0.16509   0.19386   0.69752   0.00000  Uiso   1.00
O68    O     0.19633   0.03309   0.59028   0.00000  Uiso   1.00
O69    O     0.82544   0.52175   0.03019   0.00000  Uiso   1.00
O7     O     0.03035   0.20475   0.19363   0.00000  Uiso   1.00
O70    O     0.85499   0.36570   0.93061   0.00000  Uiso   1.00
O71    O     0.49640   0.86037   0.35997   0.00000  Uiso   1.00
O72    O     0.52302   0.69890   0.26251   0.00000  Uiso   1.00
O8     O     0.83617   0.03495   0.08766   0.00000  Uiso   1.00
O9     O     0.69458   0.53376   0.52624   0.00000  Uiso   1.00
P73    P     0.28814  -0.00251   0.25030   0.00000  Uiso   1.00
P74    P     0.95348   0.33665   0.58263   0.00000  Uiso   1.00
P75    P     0.62046   0.66344   0.91915   0.00000  Uiso   1.00
P76    P     0.00915   0.29578   0.25036   0.00000  Uiso   1.00
P77    P     0.66792   0.62193   0.58202   0.00000  Uiso   1.00
P78    P     0.34114   0.96195   0.91844   0.00000  Uiso   1.00
P79    P     0.70493   0.71863   0.25037   0.00000  Uiso   1.00
P80    P     0.38168   0.05067   0.58170   0.00000  Uiso   1.00
P81    P     0.03628   0.38394   0.91881   0.00000  Uiso   1.00
P82    P     0.71121  -0.00071   0.74942   0.00000  Uiso   1.00
P83    P     0.38221   0.33122   0.07959   0.00000  Uiso   1.00
P84    P     0.05254   0.66700   0.41234   0.00000  Uiso   1.00
P85    P     0.00034   0.71474   0.74913   0.00000  Uiso   1.00
P86    P     0.66881   0.05442   0.08071   0.00000  Uiso   1.00
P87    P     0.33576   0.38838   0.41397   0.00000  Uiso   1.00
P88    P     0.28547   0.28790   0.75155   0.00000  Uiso   1.00
P89    P     0.94637   0.61664   0.08268   0.00000  Uiso   1.00
P90    P     0.61429   0.94915   0.41364   0.00000  Uiso   1.00
Ti99   Ti   -0.00301   0.00315   0.14588   0.00000  Uiso   1.00
Ti100  Ti    0.66954   0.33657   0.47643   0.00000  Uiso   1.00
Ti101  Ti    0.32812   0.66724   0.81455   0.00000  Uiso   1.00
Ti102  Ti    0.00418   0.00616   0.35461   0.00000  Uiso   1.00
Ti103  Ti    0.66476   0.33479   0.68618   0.00000  Uiso   1.00
Ti104  Ti    0.33444   0.67246   0.02311   0.00000  Uiso   1.00
Ti105  Ti   -0.00593  -0.00258   0.85305   0.00000  Uiso   1.00
Ti106  Ti    0.66317   0.33162   0.18486   0.00000  Uiso   1.00
Ti107  Ti    0.33507   0.66854   0.51819   0.00000  Uiso   1.00
natp3_opt.cif
data_NATP_1
_audit_creation_date              2014-10-29
_audit_creation_method            'Materials Studio'
_symmetry_space_group_name_H-M    'P1'
_symmetry_Int_Tables_number       1
_symmetry_cell_setting            triclinic
loop_
_symmetry_equiv_pos_as_xyz
  x,y,z
_cell_length_a                    8.5325
_cell_length_b                    8.5378
_cell_length_c                    22.0397
_cell_angle_alpha                 89.9922
_cell_angle_beta                  89.8458
_cell_angle_gamma                 119.7499
loop_
_atom_site_label
_atom_site_type_symbol
_atom_site_fract_x
_atom_site_fract_y
_atom_site_fract_z
_atom_site_U_iso_or_equiv
_atom_site_adp_type
_atom_site_occupancy
Al109  Al    0.00775   0.00446   0.64885   0.00000  Uiso   1.00
Al110  Al    0.67439   0.33780   0.98219   0.00000  Uiso   1.00
Al111  Al    0.34110   0.67115   0.31548   0.00000  Uiso   1.00
Na91   Na   -0.03004  -0.02967   0.00032   0.00000  Uiso   1.00
Na92   Na    0.63663   0.30368   0.33353   0.00000  Uiso   1.00
Na93   Na    0.30344   0.63714   0.66696   0.00000  Uiso   1.00
Na94   Na    0.02967   0.00383   0.50199   0.00000  Uiso   1.00
Na95   Na    0.69640   0.33717   0.83545   0.00000  Uiso   1.00
Na96   Na    0.36307   0.67050   0.16871   0.00000  Uiso   1.00
Na97   Na   -0.00170   0.63632   0.25423   0.00000  Uiso   1.00
Na98   Na    0.66497  -0.03036   0.58750   0.00000  Uiso   1.00
Na99   Na    0.33157   0.30290   0.92088   0.00000  Uiso   1.00
O1     O     0.16035   0.96005   0.19755   0.00000  Uiso   1.00
O10    O     0.50340   0.36884   0.42113   0.00000  Uiso   1.00
O11    O     0.36423   0.87028   0.86080   0.00000  Uiso   1.00
O12    O     0.17017   0.70231   0.75452   0.00000  Uiso   1.00
O13    O     0.78773   0.83279   0.19373   0.00000  Uiso   1.00
```

| | | | | | | | |
|---|---|---|---|---|---|---|---|
| O14 | O | 0.96567 | 0.80340 | 0.09059 | 0.00000 | Uiso | 1.00 |
| O15 | O | 0.45456 | 0.16604 | 0.52705 | 0.00000 | Uiso | 1.00 |
| O16 | O | 0.63243 | 0.13679 | 0.42387 | 0.00000 | Uiso | 1.00 |
| O17 | O | 0.12116 | 0.49947 | 0.86045 | 0.00000 | Uiso | 1.00 |
| O18 | O | 0.29889 | 0.47004 | 0.75727 | 0.00000 | Uiso | 1.00 |
| O19 | O | 0.98065 | 0.17757 | 0.30558 | 0.00000 | Uiso | 1.00 |
| O2 | O | 0.19893 | 0.16337 | 0.09043 | 0.00000 | Uiso | 1.00 |
| O20 | O | 0.16978 | 0.19882 | 0.41003 | 0.00000 | Uiso | 1.00 |
| O21 | O | 0.64751 | 0.51099 | 0.63894 | 0.00000 | Uiso | 1.00 |
| O22 | O | 0.83650 | 0.53215 | 0.74342 | 0.00000 | Uiso | 1.00 |
| O23 | O | 0.31409 | 0.84432 | 0.97231 | 0.00000 | Uiso | 1.00 |
| O24 | O | 0.50319 | 0.86547 | 0.07672 | 0.00000 | Uiso | 1.00 |
| O25 | O | 0.20575 | 0.03147 | 0.30770 | 0.00000 | Uiso | 1.00 |
| O26 | O | 0.02482 | 0.82987 | 0.40867 | 0.00000 | Uiso | 1.00 |
| O27 | O | 0.87237 | 0.36463 | 0.64107 | 0.00000 | Uiso | 1.00 |
| O28 | O | 0.69136 | 0.16312 | 0.74207 | 0.00000 | Uiso | 1.00 |
| O29 | O | 0.53914 | 0.69803 | 0.97444 | 0.00000 | Uiso | 1.00 |
| O3 | O | 0.82713 | 0.29358 | 0.53088 | 0.00000 | Uiso | 1.00 |
| O30 | O | 0.35803 | 0.49645 | 0.07538 | 0.00000 | Uiso | 1.00 |
| O31 | O | 0.83675 | 0.80484 | 0.30331 | 0.00000 | Uiso | 1.00 |
| O32 | O | 0.80182 | 0.97379 | 0.40688 | 0.00000 | Uiso | 1.00 |
| O33 | O | 0.50332 | 0.13825 | 0.63668 | 0.00000 | Uiso | 1.00 |
| O34 | O | 0.46844 | 0.30716 | 0.74024 | 0.00000 | Uiso | 1.00 |
| O35 | O | 0.17000 | 0.47150 | 0.97004 | 0.00000 | Uiso | 1.00 |
| O36 | O | 0.13513 | 0.64049 | 0.07356 | 0.00000 | Uiso | 1.00 |
| O37 | O | 0.83535 | 0.02752 | 0.80149 | 0.00000 | Uiso | 1.00 |
| O38 | O | 0.80393 | 0.83155 | 0.90742 | 0.00000 | Uiso | 1.00 |
| O39 | O | 0.50210 | 0.36098 | 0.13478 | 0.00000 | Uiso | 1.00 |
| O4 | O | 0.86561 | 0.49680 | 0.42371 | 0.00000 | Uiso | 1.00 |
| O40 | O | 0.47056 | 0.16495 | 0.24068 | 0.00000 | Uiso | 1.00 |
| O41 | O | 0.16878 | 0.69428 | 0.46809 | 0.00000 | Uiso | 1.00 |
| O42 | O | 0.13725 | 0.49832 | 0.57405 | 0.00000 | Uiso | 1.00 |
| O43 | O | 0.97924 | 0.80664 | 0.80522 | 0.00000 | Uiso | 1.00 |
| O44 | O | 0.17375 | 0.98248 | 0.91099 | 0.00000 | Uiso | 1.00 |
| O45 | O | 0.64578 | 0.13990 | 0.13853 | 0.00000 | Uiso | 1.00 |
| O46 | O | 0.84037 | 0.31579 | 0.24425 | 0.00000 | Uiso | 1.00 |
| O47 | O | 0.31250 | 0.47330 | 0.47183 | 0.00000 | Uiso | 1.00 |
| O48 | O | 0.50703 | 0.64905 | 0.57761 | 0.00000 | Uiso | 1.00 |
| O49 | O | 0.20695 | 0.17871 | 0.80890 | 0.00000 | Uiso | 1.00 |
| O5 | O | 0.49366 | 0.62673 | 0.86429 | 0.00000 | Uiso | 1.00 |
| O50 | O | 0.02114 | 0.19926 | 0.90871 | 0.00000 | Uiso | 1.00 |
| O51 | O | 0.87365 | 0.51197 | 0.14219 | 0.00000 | Uiso | 1.00 |
| O52 | O | 0.68788 | 0.53262 | 0.24197 | 0.00000 | Uiso | 1.00 |
| O53 | O | 0.54032 | 0.84536 | 0.47551 | 0.00000 | Uiso | 1.00 |
| O54 | O | 0.35460 | 0.86597 | 0.57534 | 0.00000 | Uiso | 1.00 |
| O55 | O | 0.02063 | 0.83357 | 0.69269 | 0.00000 | Uiso | 1.00 |
| O56 | O | 0.83895 | 0.81580 | 0.59065 | 0.00000 | Uiso | 1.00 |
| O57 | O | 0.68736 | 0.16694 | 0.02602 | 0.00000 | Uiso | 1.00 |
| O58 | O | 0.50566 | 0.14913 | 0.92404 | 0.00000 | Uiso | 1.00 |
| O59 | O | 0.35389 | 0.50017 | 0.35931 | 0.00000 | Uiso | 1.00 |
| O6 | O | 0.53227 | 0.83000 | 0.75712 | 0.00000 | Uiso | 1.00 |
| O60 | O | 0.17224 | 0.48251 | 0.25731 | 0.00000 | Uiso | 1.00 |
| O61 | O | 0.80027 | 0.97384 | 0.68910 | 0.00000 | Uiso | 1.00 |
| O62 | O | 0.97501 | 0.16552 | 0.59316 | 0.00000 | Uiso | 1.00 |
| O63 | O | 0.46690 | 0.30710 | 0.02242 | 0.00000 | Uiso | 1.00 |
| O64 | O | 0.64164 | 0.49878 | 0.92656 | 0.00000 | Uiso | 1.00 |
| O65 | O | 0.13365 | 0.64057 | 0.35572 | 0.00000 | Uiso | 1.00 |
| O66 | O | 0.30841 | 0.83223 | 0.25984 | 0.00000 | Uiso | 1.00 |
| O67 | O | 0.15828 | 0.18881 | 0.69673 | 0.00000 | Uiso | 1.00 |
| O68 | O | 0.18941 | 0.03399 | 0.59644 | 0.00000 | Uiso | 1.00 |
| O69 | O | 0.82495 | 0.52225 | 0.03004 | 0.00000 | Uiso | 1.00 |
| O7 | O | 0.03094 | 0.20370 | 0.19407 | 0.00000 | Uiso | 1.00 |
| O70 | O | 0.85608 | 0.36735 | 0.92983 | 0.00000 | Uiso | 1.00 |
| O71 | O | 0.49173 | 0.85556 | 0.36335 | 0.00000 | Uiso | 1.00 |
| O72 | O | 0.52269 | 0.70056 | 0.26312 | 0.00000 | Uiso | 1.00 |
| O8 | O | 0.83680 | 0.03559 | 0.08785 | 0.00000 | Uiso | 1.00 |
| O9 | O | 0.69747 | 0.53691 | 0.52741 | 0.00000 | Uiso | 1.00 |
| P73 | P | 0.28795 | -0.00261 | 0.25137 | 0.00000 | Uiso | 1.00 |
| P74 | P | 0.95462 | 0.33074 | 0.58473 | 0.00000 | Uiso | 1.00 |
| P75 | P | 0.62127 | 0.66401 | 0.91812 | 0.00000 | Uiso | 1.00 |
| P76 | P | 0.00915 | 0.29620 | 0.25034 | 0.00000 | Uiso | 1.00 |
| P77 | P | 0.67583 | 0.62949 | 0.58370 | 0.00000 | Uiso | 1.00 |
| P78 | P | 0.34252 | 0.96286 | 0.91709 | 0.00000 | Uiso | 1.00 |
| P79 | P | 0.70407 | 0.71849 | 0.25089 | 0.00000 | Uiso | 1.00 |
| P80 | P | 0.37077 | 0.05185 | 0.58425 | 0.00000 | Uiso | 1.00 |
| P81 | P | 0.03740 | 0.38518 | 0.91764 | 0.00000 | Uiso | 1.00 |
| P82 | P | 0.71545 | -0.00216 | 0.74585 | 0.00000 | Uiso | 1.00 |
| P83 | P | 0.38211 | 0.33119 | 0.07916 | 0.00000 | Uiso | 1.00 |
| P84 | P | 0.04885 | 0.66457 | 0.41245 | 0.00000 | Uiso | 1.00 |
| P85 | P | 0.00192 | 0.72094 | 0.74697 | 0.00000 | Uiso | 1.00 |
| P86 | P | 0.66859 | 0.05423 | 0.08029 | 0.00000 | Uiso | 1.00 |
| P87 | P | 0.33521 | 0.38758 | 0.41357 | 0.00000 | Uiso | 1.00 |
| P88 | P | 0.27970 | 0.28325 | 0.74924 | 0.00000 | Uiso | 1.00 |
| P89 | P | 0.94642 | 0.61661 | 0.08255 | 0.00000 | Uiso | 1.00 |
| P90 | P | 0.61310 | 0.94992 | 0.41584 | 0.00000 | Uiso | 1.00 |
| Ti100 | Ti | -0.00285 | 0.00256 | 0.14550 | 0.00000 | Uiso | 1.00 |
| Ti101 | Ti | 0.66382 | 0.33590 | 0.47880 | 0.00000 | Uiso | 1.00 |
| Ti102 | Ti | 0.33045 | 0.66922 | 0.81219 | 0.00000 | Uiso | 1.00 |
| Ti103 | Ti | 0.00098 | 0.00552 | 0.35505 | 0.00000 | Uiso | 1.00 |
| Ti104 | Ti | 0.66764 | 0.33884 | 0.68844 | 0.00000 | Uiso | 1.00 |
| Ti105 | Ti | 0.33432 | 0.67218 | 0.02176 | 0.00000 | Uiso | 1.00 |
| Ti106 | Ti | -0.00519 | -0.00296 | 0.85151 | 0.00000 | Uiso | 1.00 |
| Ti107 | Ti | 0.66151 | 0.33040 | 0.18479 | 0.00000 | Uiso | 1.00 |
| Ti108 | Ti | 0.32816 | 0.66371 | 0.51813 | 0.00000 | Uiso | 1.00 |

NMP STRUCTURES

```
------------------------------------------------------------
NaGe2PO4_3_hex.cif
data_NaGe2PO4_3ext\(2)_hex
_audit_creation_date              2014-09-09
_audit_creation_method            'Materials Studio'
_symmetry_space_group_name_H-M    'R-3C'
_symmetry_Int_Tables_number       167
_symmetry_cell_setting            trigonal
loop_
_symmetry_equiv_pos_as_xyz
  x,y,z
  -y,x-y,z
  -x+y,-x,z
  y,x,-z+1/2
  x-y,-y,-z+1/2
  -x,-x+y,-z+1/2
  -x,-y,-z
  y,-x+y,-z
  x-y,x,-z
  -y,-x,z+1/2
  -x+y,y,z+1/2
  x,x-y,z+1/2
  x+2/3,y+1/3,z+1/3
  -y+2/3,x-y+1/3,z+1/3
  -x+y+2/3,-x+1/3,z+1/3
  y+2/3,x+1/3,-z+5/6
  x-y+2/3,-y+1/3,-z+5/6
  -x+2/3,-x+y+1/3,-z+5/6
  -x+2/3,-y+1/3,-z+1/3
  y+2/3,-x+y+1/3,-z+1/3
  x-y+2/3,x+1/3,-z+1/3
  -y+2/3,-x+1/3,z+5/6
  -x+y+2/3,y+1/3,z+5/6
  x+2/3,x-y+1/3,z+5/6
  x+1/3,y+2/3,z+2/3
  -y+1/3,x-y+2/3,z+2/3
  -x+y+1/3,-x+2/3,z+2/3
  y+1/3,x+2/3,-z+1/6
  x-y+1/3,-y+2/3,-z+1/6
  -x+1/3,-x+y+2/3,-z+1/6
  -x+1/3,-y+2/3,-z+2/3
  y+1/3,-x+y+2/3,-z+2/3
  x-y+1/3,x+2/3,-z+2/3
  -y+1/3,-x+2/3,z+1/6
  -x+y+1/3,y+2/3,z+1/6
  x+1/3,x-y+2/3,z+1/6
_cell_length_a                    8.3423
_cell_length_b                    8.3423
_cell_length_c                    21.8897
_cell_angle_alpha                 90.0000
_cell_angle_beta                  90.0000
_cell_angle_gamma                 120.0000
loop_
_atom_site_label
_atom_site_type_symbol
_atom_site_fract_x
_atom_site_fract_y
_atom_site_fract_z
_atom_site_U_iso_or_equiv
_atom_site_adp_type
_atom_site_occupancy
O1     O     0.16814   0.96194   0.69326   0.01146  Uani   1.00
O2     O     0.19037   0.16188   0.59117   0.00873  Uani   1.00
Na1    Na    0.33333   0.66667   0.16667   0.03191  Uani   1.00
Ge1    Ge    0.33333   0.66667   0.31341   0.00343  Uani   1.00
P1     P     0.28621   1.00000   0.75000   0.00398  Uani   1.00
NaGe2PO4_3_LDA_hex.cif
data_NaGe2PO4_3ext\(2)_hex
_audit_creation_date              2014-09-09
_audit_creation_method            'Materials Studio'
_symmetry_space_group_name_H-M    'R-3C'
_symmetry_Int_Tables_number       167
_symmetry_cell_setting            trigonal
loop_
_symmetry_equiv_pos_as_xyz
  x,y,z
  -y,x-y,z
  -x+y,-x,z
  y,x,-z+1/2
  x-y,-y,-z+1/2
  -x,-x+y,-z+1/2
  -x,-y,-z
  y,-x+y,-z
  x-y,x,-z
  -y,-x,z+1/2
  -x+y,y,z+1/2
  x,x-y,z+1/2
  x+2/3,y+1/3,z+1/3
  -y+2/3,x-y+1/3,z+1/3
  -x+y+2/3,-x+1/3,z+1/3
  y+2/3,x+1/3,-z+5/6
  x-y+2/3,-y+1/3,-z+5/6
  -x+2/3,-x+y+1/3,-z+5/6
  -x+2/3,-y+1/3,-z+1/3
```

```
  y+2/3,-x+y+1/3,-z+1/3
  x-y+2/3,x+1/3,-z+1/3
  -y+2/3,-x+1/3,z+5/6
  -x+y+2/3,y+1/3,z+5/6
  x+2/3,x-y+1/3,z+5/6
  x+1/3,y+2/3,z+2/3
  -y+1/3,x-y+2/3,z+2/3
  -x+y+1/3,-x+2/3,z+2/3
  y+1/3,x+2/3,-z+1/6
  x-y+1/3,-y+2/3,-z+1/6
  -x+1/3,-x+y+2/3,-z+1/6
  -x+1/3,-y+2/3,-z+2/3
  y+1/3,-x+y+2/3,-z+2/3
  x-y+1/3,x+2/3,-z+2/3
  -y+1/3,-x+2/3,z+1/6
  -x+y+1/3,y+2/3,z+1/6
  x+1/3,x-y+2/3,z+1/6
_cell_length_a                    8.1530
_cell_length_b                    8.1530
_cell_length_c                    21.2751
_cell_angle_alpha                 90.0000
_cell_angle_beta                  90.0000
_cell_angle_gamma                 120.0000
loop_
_atom_site_label
_atom_site_type_symbol
_atom_site_fract_x
_atom_site_fract_y
_atom_site_fract_z
_atom_site_U_iso_or_equiv
_atom_site_adp_type
_atom_site_occupancy
O1     O     0.16577   0.96314   0.69268   0.01090  Uani   1.00
O2     O     0.19004   0.16074   0.59126   0.00830  Uani   1.00
Na1    Na    0.33333   0.66667   0.16667   0.03043  Uani   1.00
Ge1    Ge    0.33333   0.66667   0.31234   0.00326  Uani   1.00
P1     P     0.28498   1.00000   0.75000   0.00378  Uani   1.00
NaSi2PO4_3_hex.cif
data_NaSi2PO4_3_hex
_audit_creation_date              2014-09-09
_audit_creation_method            'Materials Studio'
_symmetry_space_group_name_H-M    'R-3C'
_symmetry_Int_Tables_number       167
_symmetry_cell_setting            trigonal
loop_
_symmetry_equiv_pos_as_xyz
  x,y,z
  -y,x-y,z
  -x+y,-x,z
  y,x,-z+1/2
  x-y,-y,-z+1/2
  -x,-x+y,-z+1/2
  -x,-y,-z
  y,-x+y,-z
  x-y,x,-z
  -y,-x,z+1/2
  -x+y,y,z+1/2
  x,x-y,z+1/2
  x+2/3,y+1/3,z+1/3
  -y+2/3,x-y+1/3,z+1/3
  -x+y+2/3,-x+1/3,z+1/3
  y+2/3,x+1/3,-z+5/6
  x-y+2/3,-y+1/3,-z+5/6
  -x+2/3,-x+y+1/3,-z+5/6
  -x+2/3,-y+1/3,-z+1/3
  y+2/3,-x+y+1/3,-z+1/3
  x-y+2/3,x+1/3,-z+1/3
  -y+2/3,-x+1/3,z+5/6
  -x+y+2/3,y+1/3,z+5/6
  x+2/3,x-y+1/3,z+5/6
  x+1/3,y+2/3,z+2/3
  -y+1/3,x-y+2/3,z+2/3
  -x+y+1/3,-x+2/3,z+2/3
  y+1/3,x+2/3,-z+1/6
  x-y+1/3,-y+2/3,-z+1/6
  -x+1/3,-x+y+2/3,-z+1/6
  -x+1/3,-y+2/3,-z+2/3
  y+1/3,-x+y+2/3,-z+2/3
  x-y+1/3,x+2/3,-z+2/3
  -y+1/3,-x+2/3,z+1/6
  -x+y+1/3,y+2/3,z+1/6
  x+1/3,x-y+2/3,z+1/6
_cell_length_a                    8.1021
_cell_length_b                    8.1021
_cell_length_c                    21.2172
_cell_angle_alpha                 90.0000
_cell_angle_beta                  90.0000
_cell_angle_gamma                 120.0000
loop_
_atom_site_label
_atom_site_type_symbol
_atom_site_fract_x
_atom_site_fract_y
_atom_site_fract_z
_atom_site_U_iso_or_equiv
```

```
_atom_site_adp_type
_atom_site_occupancy
O1     O     0.15942   0.95986   0.69187   0.01079  Uani   1.00
O2     O     0.18959   0.15761   0.59287   0.00822  Uani   1.00
Na1    Na    0.33333   0.66667   0.16667   0.03008  Uani   1.00
Si1    Si    0.33333   0.66667   0.31252   0.00323  Uani   1.00
P1     P     0.28229   1.00000   0.75000   0.00375  Uani   1.00
NaSn2PO4_3_hex.cif
data_NaSn2PO4_3_hex
_audit_creation_date              2014-09-09
_audit_creation_method            'Materials Studio'
_symmetry_space_group_name_H-M    'R-3C'
_symmetry_Int_Tables_number       167
_symmetry_cell_setting            trigonal
loop_
_symmetry_equiv_pos_as_xyz
  x,y,z
  -y,x-y,z
  -x+y,-x,z
  y,x,-z+1/2
  x-y,-y,-z+1/2
  -x,-x+y,-z+1/2
  -x,-y,-z
  y,-x+y,-z
  x-y,x,-z
  -y,-x,z+1/2
  -x+y,y,z+1/2
  x,x-y,z+1/2
  x+2/3,y+1/3,z+1/3
  -y+2/3,x-y+1/3,z+1/3
  -x+y+2/3,-x+1/3,z+1/3
  y+2/3,x+1/3,-z+5/6
  x-y+2/3,-y+1/3,-z+5/6
  -x+2/3,-x+y+1/3,-z+5/6
  -x+2/3,-y+1/3,-z+1/3
  y+2/3,-x+y+1/3,-z+1/3
  x-y+2/3,x+1/3,-z+1/3
  -y+2/3,-x+1/3,z+5/6
  -x+y+2/3,y+1/3,z+5/6
  x+2/3,x-y+1/3,z+5/6
  x+1/3,y+2/3,z+2/3
  -y+1/3,x-y+2/3,z+2/3
  -x+y+1/3,-x+2/3,z+2/3
  y+1/3,x+2/3,-z+1/6
  x-y+1/3,-y+2/3,-z+1/6
  -x+1/3,-x+y+2/3,-z+1/6
  -x+1/3,-y+2/3,-z+2/3
  y+1/3,-x+y+2/3,-z+2/3
  x-y+1/3,x+2/3,-z+2/3
  -y+1/3,-x+2/3,z+1/6
  -x+y+1/3,y+2/3,z+1/6
  x+1/3,x-y+2/3,z+1/6
_cell_length_a                    8.4397
_cell_length_b                    8.4397
_cell_length_c                    22.1985
_cell_angle_alpha                 90.0000
_cell_angle_beta                  90.0000
_cell_angle_gamma                 120.0000
loop_
_atom_site_label
_atom_site_type_symbol
_atom_site_fract_x
_atom_site_fract_y
_atom_site_fract_z
_atom_site_U_iso_or_equiv
_atom_site_adp_type
_atom_site_occupancy
O1     O     0.17033   0.96678   0.69450   0.01175  Uani   1.00
O2     O     0.19146   0.16499   0.59183   0.00895  Uani   1.00
Na1    Na    0.33333   0.66667   0.16667   0.03268  Uani   1.00
Sn1    Sn    0.33333   0.66667   0.31324   0.00352  Uani   1.00
P1     P     0.28706   1.00000   0.75000   0.00408  Uani   1.00
NaTi2PO4_3_hex.cif
data_NaTi2(PO4)_3_hex
_audit_creation_date              2014-09-09
_audit_creation_method            'Materials Studio'
_symmetry_space_group_name_H-M    'R-3C'
_symmetry_Int_Tables_number       167
_symmetry_cell_setting            trigonal
loop_
_symmetry_equiv_pos_as_xyz
  x,y,z
  -y,x-y,z
  -x+y,-x,z
  y,x,-z+1/2
  x-y,-y,-z+1/2
  -x,-x+y,-z+1/2
  -x,-y,-z
  y,-x+y,-z
  x-y,x,-z
  -y,-x,z+1/2
  -x+y,y,z+1/2
  x,x-y,z+1/2
  x+2/3,y+1/3,z+1/3
  -y+2/3,x-y+1/3,z+1/3
```

```
  -x+y+2/3,-x+1/3,z+1/3
  y+2/3,x+1/3,-z+5/6
  x-y+2/3,-y+1/3,-z+5/6
  -x+2/3,-x+y+1/3,-z+5/6
  -x+2/3,-y+1/3,-z+1/3
  y+2/3,-x+y+1/3,-z+1/3
  x-y+2/3,x+1/3,-z+1/3
  -y+2/3,-x+1/3,z+5/6
  -x+y+2/3,y+1/3,z+5/6
  x+2/3,x-y+1/3,z+5/6
  x+1/3,y+2/3,z+2/3
  -y+1/3,x-y+2/3,z+2/3
  -x+y+1/3,-x+2/3,z+2/3
  y+1/3,x+2/3,-z+1/6
  x-y+1/3,-y+2/3,-z+1/6
  -x+1/3,-x+y+2/3,-z+1/6
  -x+1/3,-y+2/3,-z+2/3
  y+1/3,-x+y+2/3,-z+2/3
  x-y+1/3,x+2/3,-z+2/3
  -y+1/3,-x+2/3,z+1/6
  -x+y+1/3,y+2/3,z+1/6
  x+1/3,x-y+2/3,z+1/6
_cell_length_a                    8.5817
_cell_length_b                    8.5817
_cell_length_c                    22.2259
_cell_angle_alpha                 90.0000
_cell_angle_beta                  90.0000
_cell_angle_gamma                 120.0000
loop_
_atom_site_label
_atom_site_type_symbol
_atom_site_fract_x
_atom_site_fract_y
_atom_site_fract_z
_atom_site_U_iso_or_equiv
_atom_site_adp_type
_atom_site_occupancy
O1     O     0.53059   0.69257  -0.02776   0.01191  Uiso   1.00
O2     O     0.69537   1.16624  -0.25708   0.00874  Uiso   1.00
Na1    Na    0.33333   0.66667   0.16667   0.06079  Uiso   1.00
Ti1    Ti    0.33333   0.66667   0.02068   0.00887  Uiso   1.00
P1     P    -0.04754   0.33333  -0.41667   0.00633  Uiso   1.00
NaTi2PO4_3_LDA_hex.cif
data_NaTi2(PO4)_3_lda_hex
_audit_creation_date              2014-09-09
_audit_creation_method            'Materials Studio'
_symmetry_space_group_name_H-M    'R-3C'
_symmetry_Int_Tables_number       167
_symmetry_cell_setting            trigonal
loop_
_symmetry_equiv_pos_as_xyz
  x,y,z
  -y,x-y,z
  -x+y,-x,z
  y,x,-z+1/2
  x-y,-y,-z+1/2
  -x,-x+y,-z+1/2
  -x,-y,-z
  y,-x+y,-z
  x-y,x,-z
  -y,-x,z+1/2
  -x+y,y,z+1/2
  x,x-y,z+1/2
  x+2/3,y+1/3,z+1/3
  -y+2/3,x-y+1/3,z+1/3
  -x+y+2/3,-x+1/3,z+1/3
  y+2/3,x+1/3,-z+5/6
  x-y+2/3,-y+1/3,-z+5/6
  -x+2/3,-x+y+1/3,-z+5/6
  -x+2/3,-y+1/3,-z+1/3
  y+2/3,-x+y+1/3,-z+1/3
  x-y+2/3,x+1/3,-z+1/3
  -y+2/3,-x+1/3,z+5/6
  -x+y+2/3,y+1/3,z+5/6
  x+2/3,x-y+1/3,z+5/6
  x+1/3,y+2/3,z+2/3
  -y+1/3,x-y+2/3,z+2/3
  -x+y+1/3,-x+2/3,z+2/3
  y+1/3,x+2/3,-z+1/6
  x-y+1/3,-y+2/3,-z+1/6
  -x+1/3,-x+y+2/3,-z+1/6
  -x+1/3,-y+2/3,-z+2/3
  y+1/3,-x+y+2/3,-z+2/3
  x-y+1/3,x+2/3,-z+2/3
  -y+1/3,-x+2/3,z+1/6
  -x+y+1/3,y+2/3,z+1/6
  x+1/3,x-y+2/3,z+1/6
_cell_length_a                    8.4460
_cell_length_b                    8.4460
_cell_length_c                    21.6971
_cell_angle_alpha                 90.0000
_cell_angle_beta                  90.0000
_cell_angle_gamma                 120.0000
loop_
_atom_site_label
```

```
_atom_site_type_symbol
_atom_site_fract_x
_atom_site_fract_y
_atom_site_fract_z
_atom_site_U_iso_or_equiv
_atom_site_adp_type
_atom_site_occupancy
O1     O     0.17401   0.97882   0.19375   0.01191  Uiso   1.00
O2     O     0.19457   0.16657   0.08909   0.00874  Uiso   1.00
Na1    Na    0.00000   0.00000   0.00000   0.06079  Uiso   1.00
Ti1    Ti    0.00000   0.00000   0.14464   0.00887  Uiso   1.00
P1     P     0.28624   0.00000   0.25000   0.00633  Uiso   1.00
NaZr2PO4_3_hex.cif
data_NaZr2(PO4)_3_hex
_audit_creation_date              2014-09-09
_audit_creation_method            'Materials Studio'
_symmetry_space_group_name_H-M    'R-3C'
_symmetry_Int_Tables_number       167
_symmetry_cell_setting            trigonal
loop_
_symmetry_equiv_pos_as_xyz
  x,y,z
  -y,x-y,z
  -x+y,-x,z
  y,x,-z+1/2
  x-y,-y,-z+1/2
  -x,-x+y,-z+1/2
  -x,-y,-z
  y,-x+y,-z
  x-y,x,-z
  -y,-x,z+1/2
  -x+y,y,z+1/2
  x,x-y,z+1/2
  x+2/3,y+1/3,z+1/3
  -y+2/3,x-y+1/3,z+1/3
  -x+y+2/3,-x+1/3,z+1/3
  y+2/3,x+1/3,-z+5/6
  x-y+2/3,-y+1/3,-z+5/6
  -x+2/3,-x+y+1/3,-z+5/6
  -x+2/3,-y+1/3,-z+1/3
  y+2/3,-x+y+1/3,-z+1/3
  x-y+2/3,x+1/3,-z+1/3
  -y+2/3,-x+1/3,z+5/6
  -x+y+2/3,y+1/3,z+5/6
  x+2/3,x-y+1/3,z+5/6
  x+1/3,y+2/3,z+2/3
  -y+1/3,x-y+2/3,z+2/3
  -x+y+1/3,-x+2/3,z+2/3
  y+1/3,x+2/3,-z+1/6
  x-y+1/3,-y+2/3,-z+1/6
  -x+1/3,-x+y+2/3,-z+1/6
  -x+1/3,-y+2/3,-z+2/3
  y+1/3,-x+y+2/3,-z+2/3
  x-y+1/3,x+2/3,-z+2/3
  -y+1/3,-x+2/3,z+1/6
  -x+y+1/3,y+2/3,z+1/6
  x+1/3,x-y+2/3,z+1/6
_cell_length_a                    8.9236
_cell_length_b                    8.9236
_cell_length_c                    23.3138
_cell_angle_alpha                 90.0000
_cell_angle_beta                  90.0000
_cell_angle_gamma                 120.0000
loop_
_atom_site_label
_atom_site_type_symbol
_atom_site_fract_x
_atom_site_fract_y
_atom_site_fract_z
_atom_site_U_iso_or_equiv
_atom_site_adp_type
_atom_site_occupancy
O1     O     0.18105   0.97970   0.19696   0.01191  Uiso   1.00
O2     O     0.19872   0.17305   0.08927   0.00874  Uiso   1.00
Na1    Na    0.00000   0.00000   0.00000   0.06079  Uiso   1.00
Zr1    Zr    0.00000   0.00000   0.14639   0.00887  Uiso   1.00
P1     P     0.28912  -0.00000   0.25000   0.00633  Uiso   1.00
```